\documentclass[a4paper,fleqn,usenatbib]{mnras}

\usepackage{newtxtext,newtxmath}

\usepackage[T1]{fontenc}
\usepackage{ae,aecompl}

\usepackage{graphicx}	
\setkeys{Gin}{draft=false}
\usepackage{amsmath}	
\usepackage{amssymb}	
\usepackage{subfig}
\usepackage{threeparttable}

\graphicspath{{Figures/}}

\def\Ha{H$\alpha$}
\def\/{$^{-1}$}
\def\MH2{M$_{\text{H}_2}$}
\def\arcsec{$^{\prime\prime}$}
\def\atlas3d{ATLAS$^{3\text{D}}$}
\def\SFdens{$\Sigma_{\text{SFR}}$}
\def\H2dens{$\Sigma_{{\text{H}_2}}$}
\def\HI{H{\sc i}}
\def\KS{$\Sigma_{{\text{H}_2}} - \Sigma_{\text{SFR}}$}
\def\DT{$t_\text{dep}$}
\newcommand{\farc}{\mbox{\ensuremath{.\!\!^{\prime\prime}}}}
\newcommand\change[1]{\textcolor{black}{#1}}
\newcommand{\tHa}{$t_{\rm H\alpha}$}   
\newcommand{\tHaref}{$t_{\rm H\alpha, ref}$}   
\newcommand{\tCO}{$t_{\rm CO}$}   
\newcommand{\tover}{$t_{\rm fb}$}   
\newcommand{\Msun}{\mbox{M$_\odot$}}
\newcommand{\msun}{\mbox{M$_\odot$}}
\newcommand{\kms}{\mbox{${\rm km}~{\rm s}^{-1}$}}
\newcommand{\pc}{\mbox{${\rm pc}$}}
\newcommand{\myr}{\mbox{${\rm Myr}$}}
\newcommand\refrep[1]{\textcolor{black}{#1}}
\newcommand\secround[1]{\textcolor{black}{#1}}

\title[AlFoCS + F3D: resolved star formation]{\change{AlFoCS + Fornax3D: resolved star formation in the Fornax cluster with ALMA and MUSE}}

\author[N. Zabel et al.]{Nikki Zabel,$^{1}$\thanks{E-mail: ZabelNJ@cardiff.ac.uk}
Timothy A. Davis,$^{1}$
Marc Sarzi,$^{2}$
Boris Nedelchev,$^{2}$
M\'elanie Chevance,$^{3}$ \newauthor
J. M. Diederik Kruijssen,$^{3}$
Enrichetta Iodice,$^{4}$ 
Maarten Baes,$^{5}$
George J. Bendo,$^{6}$ \newauthor
Enrico Maria Corsini,$^{7, 8}$ 
Ilse De Looze,$^{5,9}$ 
P. Tim de Zeeuw,$^{10, 11}$ 
Dimitri A. Gadotti,$^{12}$  \newauthor
Marco Grossi,$^{13}$
Reynier Peletier,$^{14}$ 
Francesca Pinna,$^{15}$ 
Paolo Serra,$^{16}$ \newauthor
Freeke van de Voort,$^{1, 17}$
Aku Venhola,$^{18}$ 
S\'ebastien Viaene,$^{5}$ 
Catherine Vlahakis$^{19}$ \\
$^{1}$School of Physics and Astronomy, Cardiff University, Queen's Building, The Parade, Cardiff, CF24 3AA, Wales, UK \\
$^{2}$Armagh Observatory and Planetarium, College Hill, Armagh, BT61 9DG, UK \\
$^{3}$Astronomisches Rechen-Institut, Zentrum f{\"u}r Astronomie der Universit{\"a}t Heidelberg, M{\"o}nchhofstra{\ss}e 12-14, D-69120 Heidelberg, Germany \\
$^{4}$INAF-Osservatorio Astronomico di Capodimonte, via Moiariello 16, 80131, Napoli, Italy \\
$^{5}$Sterrenkundig Observatorium, Universiteit Gent, Krijgslaan 281 S9, B-9000 Gent, Belgium\\
$^{6}$UK ALMA Regional Centre Node, Jodrell Bank Centre for Astrophysics, School of Physics and Astronomy, The University of Manchester, \\ 
Oxford Road, Manchester M13 9PL, UK \\
$^{7}$Dipartimento di Fisica e Astronomia ``G. Galilei'', Universit\`a di Padova, vicolo dell'Osservatorio 3, 35122, Padova, Italy \\
$^{8}$INAF-Osservatorio Astronomico di Padova, vicolo dell'Osservatorio 5, 35122, Padova, Italy \\
$^{9}$Department of Physics and Astronomy, University College London, Gower Street, London WC1E 6BT, UK \\
$^{10}$Sterrewacht Leiden, Leiden University, Postbus 9513, 2300 RA, Leiden, The Netherlands \\
$^{11}$Max-Planck-Institut f\"ur extraterrestrische Physik, Giessenbachstra{\ss}e, 85741, Garching bei Muenchen, Germany \\
$^{12}$European Southern Observatory, Karl-Schwarzschild-Stra{\ss}e 2, 85748, Garching bei M{\"u}nchen, Germany \\
$^{13}$Observatório do Valongo, Universidade Federal do Rio de Janeiro, Ladeira Pedro Antônio 43, 20080-090 Rio de Janeiro, RJ, Brazil\\
$^{14}$Kapteyn Astronomical Institute, University of Groningen, PO Box 72, 9700 AB Groningen, The Netherlands\\
$^{15}$Max-Planck-Institut f\"ur Astronomie, K\"onigstuhl 17, 69117, Heidelberg, Germany \\
$^{16}$INAF - Osservatorio Astronomico di Cagliari, Via della Scienza 5, I-09047 Selargius (CA), Italy\\
$^{17}$Max Planck Institute for Astrophysics, Karl-Schwarzschild-Stra{\ss}e 1,
85748, Garching, Germany\\
$^{18}$Astronomy Research Unit, University of Oulu, 90014, Oulu, Finland\\
$^{19}$National Radio Astronomy Observatory, 520 Edgemont Road, Charlottesville, VA 22903-2475, USA
}

\date{Accepted 2020 May 27. Received 2020 May 21; in original form 2020 January 17}

\pubyear{2020}

\begin{document}
\label{firstpage}
\pagerange{\pageref{firstpage}--\pageref{lastpage}}
\maketitle
\begin{abstract}
We combine data from ALMA and MUSE to study the resolved ($\sim$300 pc scale) star formation relation (star formation rate vs. molecular gas surface density) in cluster galaxies. Our sample consists of 9 Fornax cluster galaxies, including spirals, ellipticals, and dwarfs, covering a stellar mass range of $\sim10^{8.8} - 10^{11}$ M$_\odot$. CO(1-0) and extinction corrected \Ha\ were used as tracers for the molecular gas mass and star formation rate, respectively. We compare our results with \citet{Kennicutt1998} and \citet{Bigiel2008}. Furthermore, we create depletion time maps to reveal small-scale variations in individual galaxies. \change{We explore these further in FCC290, using the `uncertainty principle for star formation' \citep{Kruijssen2014} to estimate molecular cloud lifetimes, which we find to be short ($<$10 Myr) in this galaxy.} Galaxy-averaged depletion times are compared with other parameters such as stellar mass and cluster-centric distance. We find that the star formation relation in the Fornax cluster is close to those from \citet{Kennicutt1998} and \citet{Bigiel2008}, \refrep{but overlaps mostly with the shortest depletion times predicted by \citet{Bigiel2008}}. This slight decrease in depletion time is mostly driven by dwarf galaxies with disturbed molecular gas reservoirs close to the virial radius. In FCC90, a dwarf galaxy with a molecular gas tail, we find that depletion times are a factor $\gtrsim10$ higher in its tail than in its stellar body.
\end{abstract}




\section{Introduction}
\label{sec:introduction}
Galaxy clusters are known to harbour relatively many passive, elliptical galaxies compared to the field \citep[e.g.][]{Oemler1974,Dressler1980}. This suggests that clusters are extreme environments that are capable of quenching the star formation in galaxies. It has been known for a few decades that atomic hydrogen is affected by these environments \refrep{\citep{Haynes1984,Cayatte1990,Solanes2001,Gavazzi2005,Jaffe2015,Scott2018}} through processes such as ram pressure stripping (RPS, \citealt{Gunn1972}), galaxy-galaxy interactions \citep{Moore1996}, and starvation \citep{Larson1980}. It was not until much more recently that evidence started to accumulate that the molecular gas in cluster galaxies can also be directly affected by the cluster environment \citep[e.g.][]{Vollmer2008,Fumagalli2009,Boselli2014, Zabel2019}. Since molecular gas is the direct fuel for star formation, this could have serious consequences for the star formation in cluster galaxies, and it begs the question whether they still follow ``traditional'' star formation relations (i.e. the ones found by \citet{Schmidt1959, Kennicutt1998}, and \citet{Bigiel2008} (hereafter K98 and B08, respectively).

\refrep{The star formation relation links the observed star formation rate surface density \SFdens\ to the total gas surface density $\Sigma{_{{\text{H{\sc i}}} + {\text{H}_2}}}$. It was first shown by \citet{Schmidt1959} and \citet{Kennicutt1998}, who found that, on galaxy scales, the two are related by the power law $\Sigma_{\text{SFR}} \propto \left( \Sigma{_{{\text{H{\sc i}}} + {\text{H}_2}}} \right) ^n$ where $n = 1.4 \pm 0.15$. They used integrated measurements for entire spiral and starburst galaxies, $\sim$40\% of which are located in the Virgo cluster, and they used \Ha\ to trace their star formation.} Since this discovery, many others have studied this relation in a variety of contexts. It has, for example, been shown that a similar, possibly even stronger correlation exists between the molecular gas surface density \H2dens\ and \SFdens\ (e.g. \citealt{Wong2002, Leroy2008, Bigiel2008, Schruba2011, Reyes2019}). \citet{Bigiel2008} studied a more diverse sample of galaxies on sub-kpc scales, and found a linear relation between both surface densities on these smaller scales. Other studies focused on (variations within) individual galaxies (e.g. \citealt{Ford2013, Utomo2017}), or compared results for individual molecular clouds to those for entire galaxies (e.g. \citealt{Lada2012}). The Physics at High Angular Resolution in Nearby GalaxieS survey (including PHANGS-ALMA, PI: E. Schinnerer, e.g. Leroy et al., in prep., and PHANGS-MUSE, PI: E. Schinnerer, e.g. \citealt{Kreckel2019}) will study the star formation relation in nearby, face-on spiral galaxies with unprecedented resolution ($<$100 pc). \change{The K98 relation has recently been revisited by \citet{Reyes2019}, who showed that this relation is still a good approximation for the star formation relation (they find a power law with $n = 1.41$, with similar dispersion). Moreover, they obtain this relation from spiral galaxies only, not including starbursting galaxies. They also find that \SFdens\ scales roughly linearly with H$_2$, while it only weakly depends on \HI. This is in agreement with the resolved studies described above.}

\refrep{While the initial sample from which the star formation relation was derived mainly consisted of spiral galaxies and galaxies located in the field, it has been shown to hold for samples consisting of cluster spiral galaxies exclusively.} For example, \citet{Vollmer2012} investigated the influence of environment on the gas surface density and depletion time in 12 spiral galaxies in the Virgo cluster on a pixel-by-pixel \change{(of the order kpc in size)} basis. The depletion time (\DT) is the ratio of the molecular gas surface density and the star formation rate (the star formation efficiency, SFE, is the inverse of this ratio). While they find that depletion times are mostly unaffected, they do report increased depletion times in the extraplanar gas of three spiral galaxies, likely caused by the loss of gravitational confinement of stripped gas and the associated pressure of the disk. Their work was an extension of the work by \citet{Fumagalli2008}, who showed that the bulk of the star formation in spiral galaxies is fuelled by a molecular interstellar medium (ISM), but that the atomic gas is important for the star formation activity in the outer parts of the disks. 

The star formation relation has been studied in several Virgo cluster galaxies, e.g. NGC4501, a galaxy undergoing RPS, and NGC4567/68, an interacting pair, by \citet{Nehlig2016}, and NGC4330, NGC4402, and NGC4522, all spirals undergoing RPS, by \citet{Lee2017}. While no significant deviations from the K98 relations are found, some regions with modified depletion times were observed in compressed parts of RPS galaxies and the interacting pair. Furthermore, they found locally increased SFRs on the sides of the galaxies where the intracluster medium pressure acts, and the opposite on the other side.

The GAs Stripping Phenomena in galaxies (GASP) with the Multi Unit Spectroscopic Explorer (MUSE) survey targeted 114 RPS galaxies, specifically to study gas removal processes in galaxies in different environments \citep{Poggianti2017}. In this context, studies by \citet{Vulcani2018}, \citet{Ramatsoku2019}, and \citet{Moretti2018} have found that RPS can enhance the SFR in galaxy disks by 0.2 dex, and that depletion times in the disks of RPS galaxies are decreased, while they are increased in their tails (with a difference of a factor $\sim$10).

So far, studies of the star formation relation in cluster galaxies have been performed almost exclusively on spiral galaxies in the Virgo cluster. In this paper, we will study the star formation relation in 9 Fornax cluster galaxies, in the context of the ALMA Fornax Cluster Survey (AlFoCS). The aim of AlFoCS is to study the effects of the cluster environment on galaxy evolution (in particular the cold molecular ISM), by targeting 30 Fornax galaxies with the Atacama Large Millimeter/submillimeter Array (ALMA, see \citealt{Zabel2019}, hereafter Paper I). The Fornax cluster is the smaller sibling of the Virgo cluster, at a similar distance ($\sim$20 Mpc) but $\sim$1/10 times the mass ($\sim7 \times 10^{13} M_\odot$, \citealt{Drinkwater2001a, Jordan2007}) and $\sim$1/6 times the number of galaxies. Despite this, the galaxy number density is 2-3 times higher in Fornax than in Virgo, and it is more \change{symmetric} and dynamically evolved. \refrep{Being a relatively poor and evolved cluster, Fornax is likely representative of the type of environment many galaxies in the local universe reside in. According to e.g. \citet{Robotham2011}, $\sim$40\% of galaxies in the local universe are located in groups and poor clusters. This means that the study of galaxy evolution in such clusters is important if we aim to understand galaxy evolution in the universe as a whole.}

Our sample is diverse, containing both spirals and ellipticals, as well as dwarf galaxies. In Paper I we used ALMA observations of $^{12}$CO(1-0) to create maps of the molecular gas surface density at a resolution of $\sim$3\arcsec\ ($\sim$0.3 kpc at the distance of the Fornax cluster). In this paper we combine these with \Ha\ observations from MUSE \citep{Bacon2010}, from the Fornax3D project \change{(F3D, \citealt{Sarzi2018, Iodice2019}), binned to the same resolution as the maps from ALMA}, to create maps of the star formation rate surface density. F3D targeted all galaxies from the Fornax Cluster Catalogue \citep{Ferguson1989} brighter than $m_B = 15$ within or close to the virial radius ($R_{\text{vir}} = 0.7$ Mpc; \citealt{Drinkwater2001a}). Both \refrep{are} combined \refrep{here} to study the star formation relation in these galaxies on a pixel-by-pixel basis, and to create maps of their depletion times. We compare the outcome with both the integrated K98 relation, and the one derived for kpc sized regions by B08. Since the sample, resolution, and use of molecular gas makes this study more similar to the one by B08, we expect our results to be close to this relation, should the cluster environment have no or little effect on the depletion times. We compare the galaxy-average depletion times with other parameters, i.e. stellar mass, stellar mass surface brightness, and (projected) distance from the cluster centre. Where possible, we compare the results to a field control sample.

\change{For the purpose of consistency with F3D publications, we adopt the distance to the Fornax cluster (20 Mpc, \citealt{Blakeslee2009}) as a common distance to all galaxies. The cluster has a virial radius of 0.7 Mpc \citep{Drinkwater2001a}}. 

This paper is structured as follows: in \S \ref{sec:observations} the sample, observations, and data reduction are described. In \S \ref{sec:methods} we describe how surface density maps and \KS\ relations are obtained from the data. The results are shown in \S \ref{sec:results}. These include the derived star \KS\ relations and depletion time maps, as well as the relations between depletion times and various other parameters, such as stellar mass. 
In \S \ref{sec:cloud_lifetimes} we apply the `uncertainty principle of star formation' to the face-on, flocculent spiral FCC290 to estimate its molecular cloud lifetimes and compare them to those in field galaxies. The results are discussed in \S \ref{sec:discussion}, and finally summarised in \S \ref{sec:conclusions}.

\begin{table*}
	\centering
	\begin{threeparttable}
	\caption{Key properties of the galaxies in the sample.}
	\label{tab:targets}
	\setlength{\tabcolsep}{1.5mm}
	\begin{tabular}{rrrrrrrrrr}
	\hline
	FCC & Common name & $M_\star$ & Type & $D$ & Gas distribution & M$_{\text{H}_2}$ & H$_2$ def. & SFR & $\mu_{\bigstar, e}$ \\
	- & - & ($10^{10} M_{\odot}$) & - & (kpc) & - & (log $M_\odot$) & (dex) & ($M_\odot$ yr$^{-1}$) & (log $M_\odot$ kpc$^{-2}$) \\
	(1) & (2) & (3) & (4) & (5) & (6) & (7) & (8) & (9) & (10) \\
	\hline
	90 & MGC-06-08-024 & 0.08$^\dagger$ & E4 pec & 595 & D & 6.97 $\pm$ 0.07 & -1.21 & 0.035 & $^{\star\star}$5.0 $\pm\ 0.6 \times 10^8$ \\
	167 & NGC1380 & 9.85$^\dagger$ & S0/a & 219 & R & 7.67 $\pm$ 0.06 & -1.56 & 0.000 & $^\star$8.0 $ \times\ 10^8$  \\
	179 & NGC1386 & 1.58$^\dagger$ & Sa & 226 & R & 8.37 $\pm$ 0.04 & -0.70 & 0.155 & 1.20 $\pm\ 0.04 \times 10^9$ \\	
	184 & NGC1387 & 4.70$^\dagger$ & SB0 & 111 & R & 8.33 $\pm$ 0.04 & -0.85 & 0.008 & $^\star$2.21 $\times 10^9$  \\
	207 & FCC207 & \change{0.06*} & - & 113 & D & 6.54 $\pm$ 0.22 & -1.43 & 0.004 & $^{\star\star}$2.2 $\pm\ 0.4 \times 10^8$  \\
	263 & PGC013571 & 0.04$^\dagger$ & SBcdIII & 293 & D & 7.22 $\pm$ 0.05 & -1.07 & 0.282 & $^{\star\star}$2.0 $\pm\ 0.3\ \times 10^8$  \\
	290 & NGC1436 & 0.64$^\dagger$ & ScII & 389 & R & 8.44 $\pm$ 0.05 & -0.53 & 0.127 & 1.82 $\pm\ 0.02 \times 10^8$ \\
	308 & NGC1437B & 0.04$^\dagger$ & Sd & 611 & D & 7.76 $\pm$ 0.04 & -0.64 & 0.167 & 6.1 $\pm\ 0.3 \times 10^7$ \\
	312 & ESO358-G063 & 1.48$^\dagger$ & Scd & 584 & R & 8.57 $\pm$ 0.05 & -0.38 & 0.751 & 3.1 $\pm\ 0.5 \times 10^7$ \\
	\hline
	\end{tabular}
	\textit{Notes:} \textbf{1}: Fornax Cluster Catalogue \citep{Ferguson1989} number of the galaxy; \textbf{2}: Common name of the galaxy; \textbf{3}: \change{$^\dagger$Stellar mass in the MUSE FoV from \citet{Iodice2019}}; \change{*Stellar mass obtained from 3.6 $\mu$m image. The uncertainty is $^{+0.04}_{-0.05}$}; \textbf{4}: Morphological type from \citet{Sarzi2018}; \textbf{5}: Distance from brightest cluster galaxy NGC1399; \textbf{6}: Whether the molecular gas in the galaxy is regular (R) or disturbed (D) as classified in Paper I; \textbf{7}: H$_2$ mass from Paper I; \textbf{8}: H$_2$ deficiency, defined as log$(M_{\text{H}_2, \text{observed}}) - \text{log}(M_{\text{H}_2, \text{expected}})$ \change{(where $M_{\text{H}_2, \text{expected}}$ is the expected gas fraction based on field galaxies with similar stellar masses, see Paper I)}. Uncertainties are 0.01 dex for each galaxy. \textbf{9}: Total star formation rate in the MUSE FoV from \citet[except FCC207, which is from this work]{Iodice2019}; \textbf{10}: Stellar mass surface density: half the stellar mass divided by the area within the effective radius $R_{\rm e}$. $^\star$Effective radii from (\citealt{Iodice2019}, no uncertainties given), $^{\star\star}$effective radii from \cite{Venhola2018}, remaining effective radii are from \citet{Raj2019}. 
	\end{threeparttable}
\end{table*}

\section{Observations and data reduction}
\label{sec:observations}

\subsection{The sample}
\label{sub:sample}
Our sample consists of all Fornax cluster galaxies for which both CO and \Ha\ maps are available, from AlFoCS and F3D, respectively. AlFoCS targeted all Fornax cluster galaxies that show signs of the presence of an ISM, as indicated by either far-infrared observations from the \textit{Herschel} Space Observatory \citep{Pilbratt2010}, \HI\ observations (from \citealt{Waugh2002} and Loni et al. in prep., based on Australia Telescope Compact Array data), or both. There are 9 overlapping galaxies in these surveys. This means that 6 galaxies detected in Paper I are not included here: the large spiral FCC121 (NGC1365), the edge-on spiral FCC67 (NGC1351A), and four dwarf galaxies with disturbed molecular gas reservoirs. 

\refrep{As F3D targeted only the brightest galaxies ($m_B < 15$), and AlFoCS is mostly IR selected, the sample is likely biased towards higher-mass, star forming galaxies. As we have seen in Paper I, the lower-mass galaxies are the ones that are affected most by the cluster environment, and that are most gas deficient. Including more such low mass galaxies in our analysis, or including galaxies with less molecular gas, would be required to obtain a complete census of the star formation activity in the Fornax cluster. Possible effects on our conclusions are discussed in more detail in \S \ref{sec:discussion}.}

The 9 galaxies in the sample are listed in Table \ref{tab:targets}, along with their key properties. \change{Column 3 lists the total stellar mass in the MUSE field of view from \citet{Iodice2019}. Since FCC207 is not included in F3D, but was taken from the archive, its stellar mass was obtained from aperture photometry on its archival Wide-field Infrared Survey Explorer (WISE, \citealt{Wright2010}) band 1 (3.4 $\mu$m) image, assuming a mass-to-light ratio of 1. The aperture was chosen using the effective radius determined by \citet{Venhola2018}. The uncertainty in this case \change{(see Table caption)} is a combination of the uncertainty in the effective radius and the rms in the image.} \refrep{Column 6 indicates whether the galaxy has a regular (R) or disturbed (D) molecular gas reservoir (as determined in Paper I), and molecular gas masses and deficiencies are listed in column 8 and 9, respectively (both also from Paper I). The latter is defined as log($M_{\text{H}_2, \text{observed}}) - \text{log}(M_{\text{H}_2, \text{expected}})$, where $\text{log}(M_{\text{H}_2, \text{expected}})$ is the estimated molecular gas mass of a field galaxy of similar stellar mass (from interpolation of field samples).} \change{The total SFR in the MUSE field of view from \citet{Iodice2019} is listed in column 10. \refrep{These were calculated through the conversion SFR (M$_\odot$ y$^{-1}$) = L$_{\text{H}\alpha}$ (erg s$^{-1}$) / 1.82 $\times 10^{41}$ ($M_\odot$ yr$^{-1}$ erg$^{-1}$ s$^{-1}$), provided by \citep{Calzetti2012}.} Stellar mass surface densities, defined here as half the stellar mass divided by the effective radius $R_{\rm e}$ (from \citealt{Venhola2018}, \citealt{Iodice2019}, or \citealt{Raj2019}, see Table caption) are listed in column 11.}

\subsection{CO data}
\label{subsec:CO_data}
The molecular gas data used are $^{12}$CO(1-0) integrated intensity (moment zero) maps from AlFoCS. A detailed description of the observations and data reduction can be found in Paper I, \refrep{in which these CO data are published.} Some important details are summarised below.

ALMA Band 3 observations were carried out between the 7th and 12th of January 2016 \change{under project ID 2015.1.00497.S (PI: T. Davis), using the main (12m) array in the C36-1 configuration.} The data were calibrated manually, \textsc{clean}ed (interactively, using a natural weighting scheme), and continuum subtracted using the Common Astronomy Software Applications package (CASA, version 5.1.1, \citealt{McMullin2007}). The spaxel sizes in the final data cubes (and hence in the moment zero maps used here) are 0\farc 5 ($\sim$50 pc at the distance of the Fornax cluster), \change{and typical beam sizes are around 3\arcsec ($\sim$0.3 kpc at the distance of Fornax). Channel widths are 10 \kms\ for most galaxies, and 2 \kms\ for the dwarf FCC207, because of its narrow linewidth.} Typical rms noise levels are $\sim$3 mJy/beam. \change{The cleaned data cubes were used to produce primary beam corrected moment maps of the CO(1-0) line emission using the masked moment method \citep{Dame2011}.}

\subsection{\Ha\ data}
\label{subsec:Ha_data}
\Ha\ maps are from F3D, with the exception of FCC207 (described below). A detailed description of the survey and data reduction can be found in \citet{Sarzi2018} and some important details are summarised here. 

Integral-field spectroscopic observations were carried out with MUSE in Wide Field Mode \citep{Bacon2010} between July 2016 and December 2017. It was mounted on the Yepun Unit Telescope 4 at the ESO Very Large Telescope (VLT). A field of 1 $\times$ 1 square arcminutes was covered, with 0.2 $\times$ 0.2 square arcsecond spatial sampling. \change{For some of the more extended galaxies, this is smaller than their optical discs. In these cases two or three pointings were used to map the whole galaxy. An exception is FCC290, which was only observed partially (including the centre and most of the outskirts, see Figure \ref{fig:FCC290} in Appendix \ref{app:SFandDT}). The MUSE pointings of all F3D galaxies can be found in \citet{Sarzi2018}. } The observations cover a wavelength range of 4650-9300 \AA, with a spectral resolution of 2.5 \AA\ (at the full width at half maximum, FWHM) at 7000 \AA\ and spectral sampling of 1.25 \AA\ pixel\/. 

Data reduction was performed using the MUSE pipeline (version 1.6.2, \citealt{Weilbacher2012, Weilbacher2016}) under the ESOREFLEX environment \citep{Freudling2013}. In summary, the data reduction involved bias and overscan subtraction, flat fielding, wavelength calibration, determination of the line spread function, illumination correction with twilight flats (to account for large-scale variation of the illumination of the detectors) and similar with lamp flats (to correct for edge effects between the integral-field units).

\refrep{Extinction corrected \Ha\ maps were obtained as described in \citet{Sarzi2018} and \citet{Iodice2019}: the Gas and Absorption Line Fitting code (GandALF, \citealt{Sarzi2006, Barroso2006}) was used to perform a spaxel-by-spaxel fit, simultaneously for both the stellar and ionised gas contributions. The full extent of the MILES library was used to decrease the impact of template-mismatch on the emission line measurement. As detailed also in \citet{Oh2011} and in \citet{Sarzi2018}, the GandALF fit included two reddening components, a first affecting the entire spectrum and representing dust diffusion everywhere throughout the target galaxy, and a second which affects only the nebular emission and can account for dust more localized around the emission-line regions. Differentiating the two allows for a more accurate estimation of the effects of dust mixed with gas in star forming regions. A Calzetti \citep{Calzetti1994} extinction law was used in both cases.}

\begin{figure}
	\includegraphics[width=0.47\textwidth]{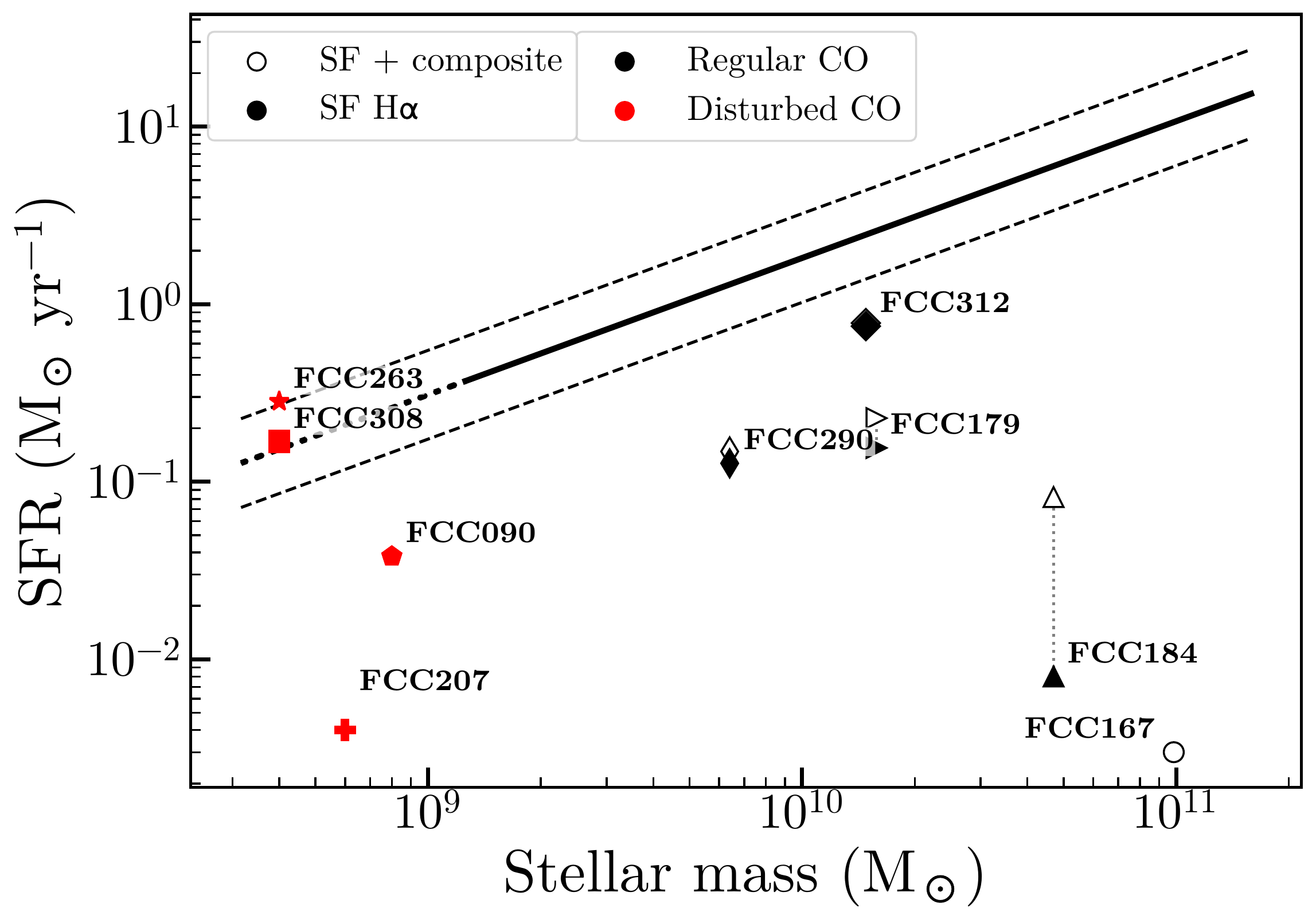}
	\caption{\change{Location of the galaxies in the sample in the $M_\star$ - SFR plane (values are from Table \ref{tab:targets}), compared to the star formation main sequence from \citealt{Elbaz2007} (solid line, the dashed lines represent the 1 $\sigma$ confidence interval, and the dotted line is an extrapolation towards lower masses). Filled markers indicate data points for which only the star forming \Ha\ was used and open markers indicate data points for which composite \Ha\ was considered also. Both are plotted for each galaxy with the same marker shape, unless it contains only non-star forming \Ha\ (FCC167). Galaxies' FCC numbers are indicated. Galaxies with regular molecular gas reservoirs are shown in black and galaxies with disturbed molecular gas reservoirs in red (\S \ref{sub:sample}, Table \ref{tab:targets}). With few exceptions, galaxies in the sample lie below the star formation main sequence.}}
	\label{fig:SFMS}
\end{figure}

The \change{MUSE cube} of FCC207 was taken from the ESO archive, Program IDs 098.B-0239, 094.B-0576, 097.B-0761, and 096.B-0063 \change{(PI: E. Emsellem)}. Observations were carried out between October 2014 and January 2017, similarly to the observations described above. A field of 1.2 arcmin$^2$ was covered, with 0.2 $\times$ 0.2 square arcsecond spatial sampling. The spectral range covered is 4650-9351 \AA, with a spectral resolution of 2.3 \AA\ at 7000 \AA. \change{The \Ha\ map was obtained in the same way as the ones from F3D.}

\section{Methods}
\label{sec:methods}

\subsection{Obtaining surface density maps}
\label{sub:SDmaps}

\subsubsection{Star formation rate surface densities}
\label{subsub:SFR_SD}

\begin{figure*}

	\centering
	
	\hspace{-2mm} \subfloat[\label{subfig:KS_FCC312}]
	{\includegraphics[height=0.43\textwidth, width=0.46\textwidth]{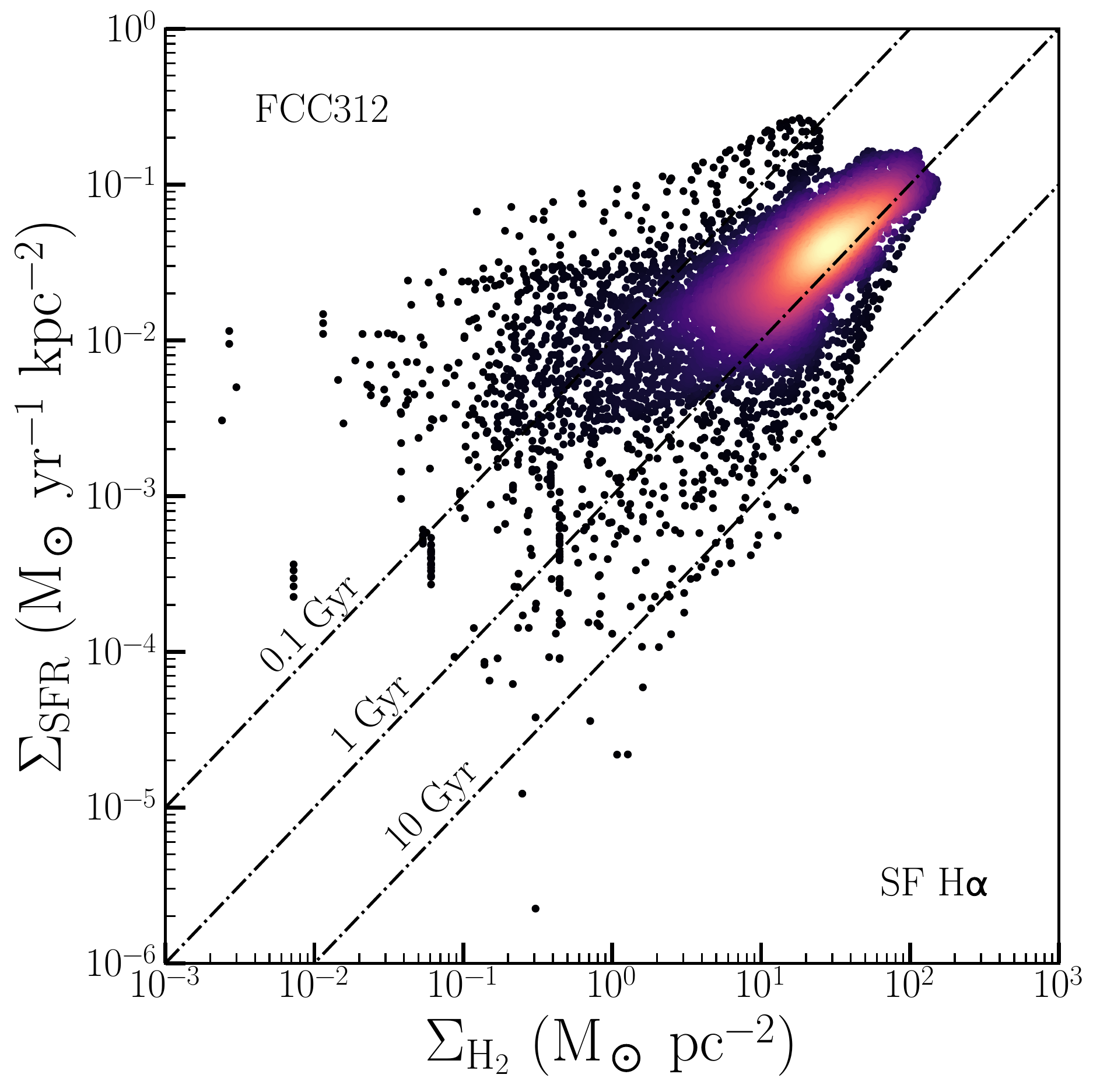}}		
	\hspace{12mm}
	\subfloat[\label{subfig:KS_FCC090}]
	{\includegraphics[height=0.425\textwidth, width=0.415\textwidth]{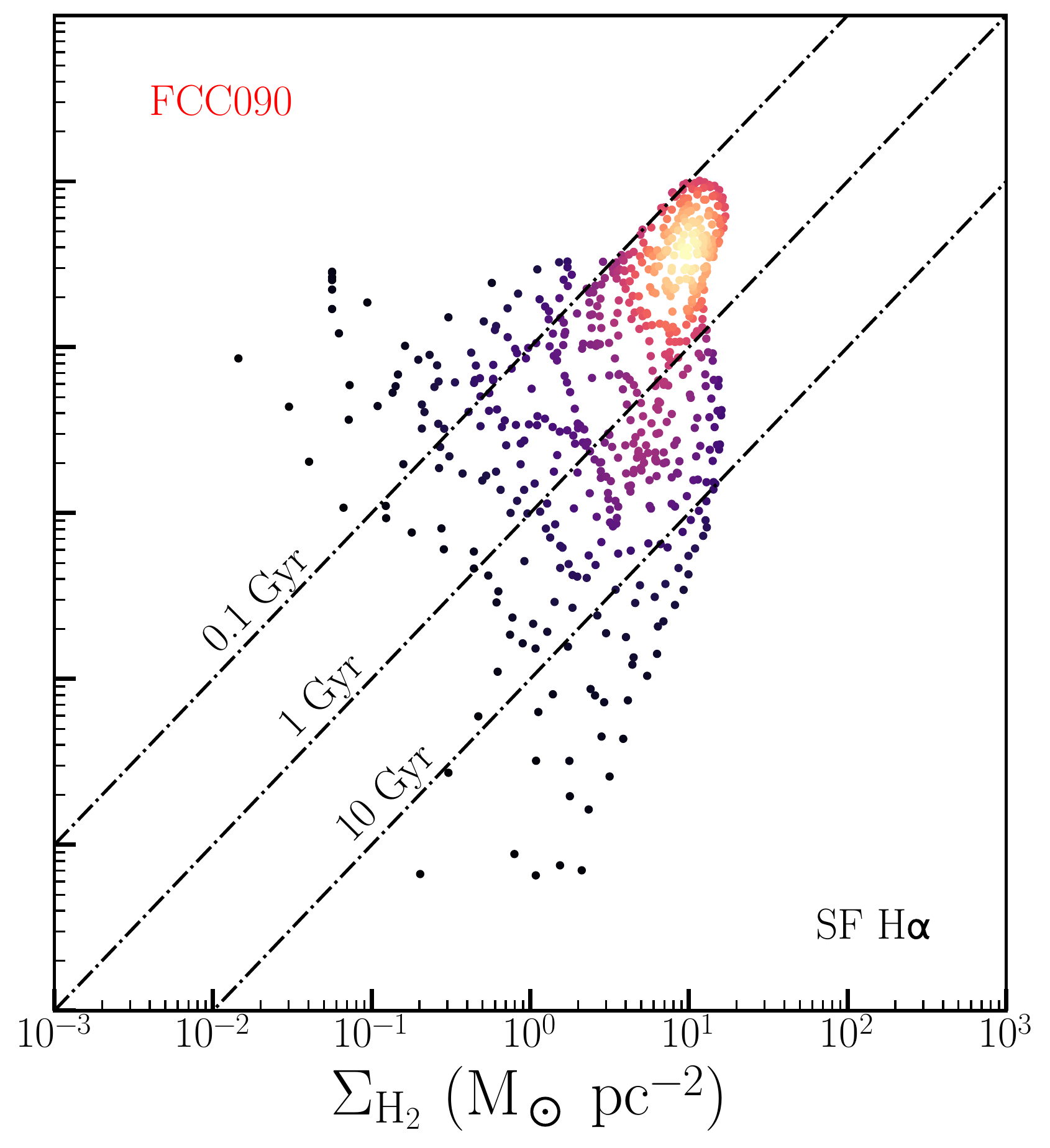}}
	\hspace{-5mm}\subfloat[\label{subfig:DT_FCC312}]
	{\includegraphics[height=0.44\textwidth, width=0.5\textwidth]{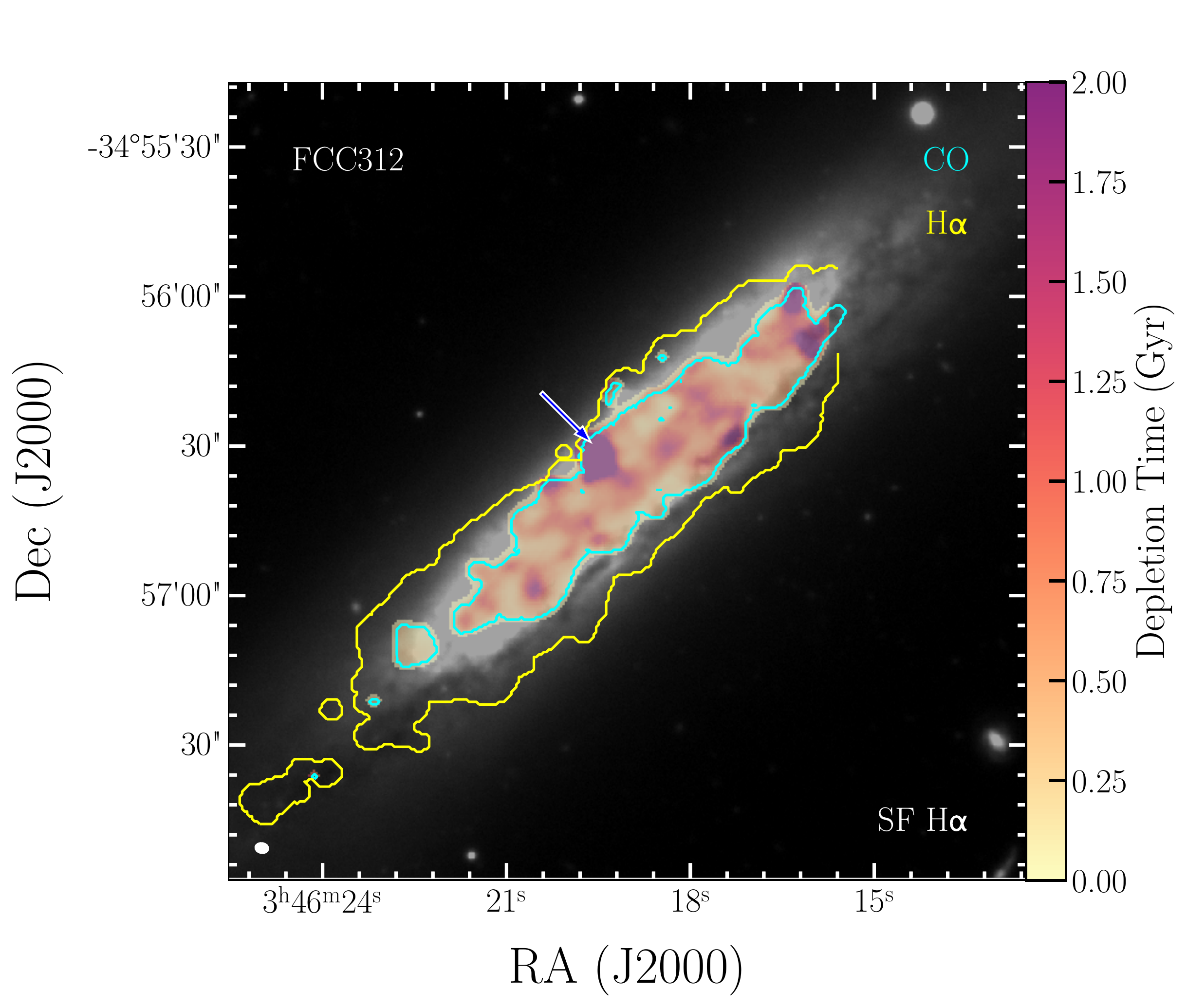}}
	\subfloat[\label{subfig:DT_FCC090}]
	{\includegraphics[height=0.44\textwidth, width=0.49\textwidth]{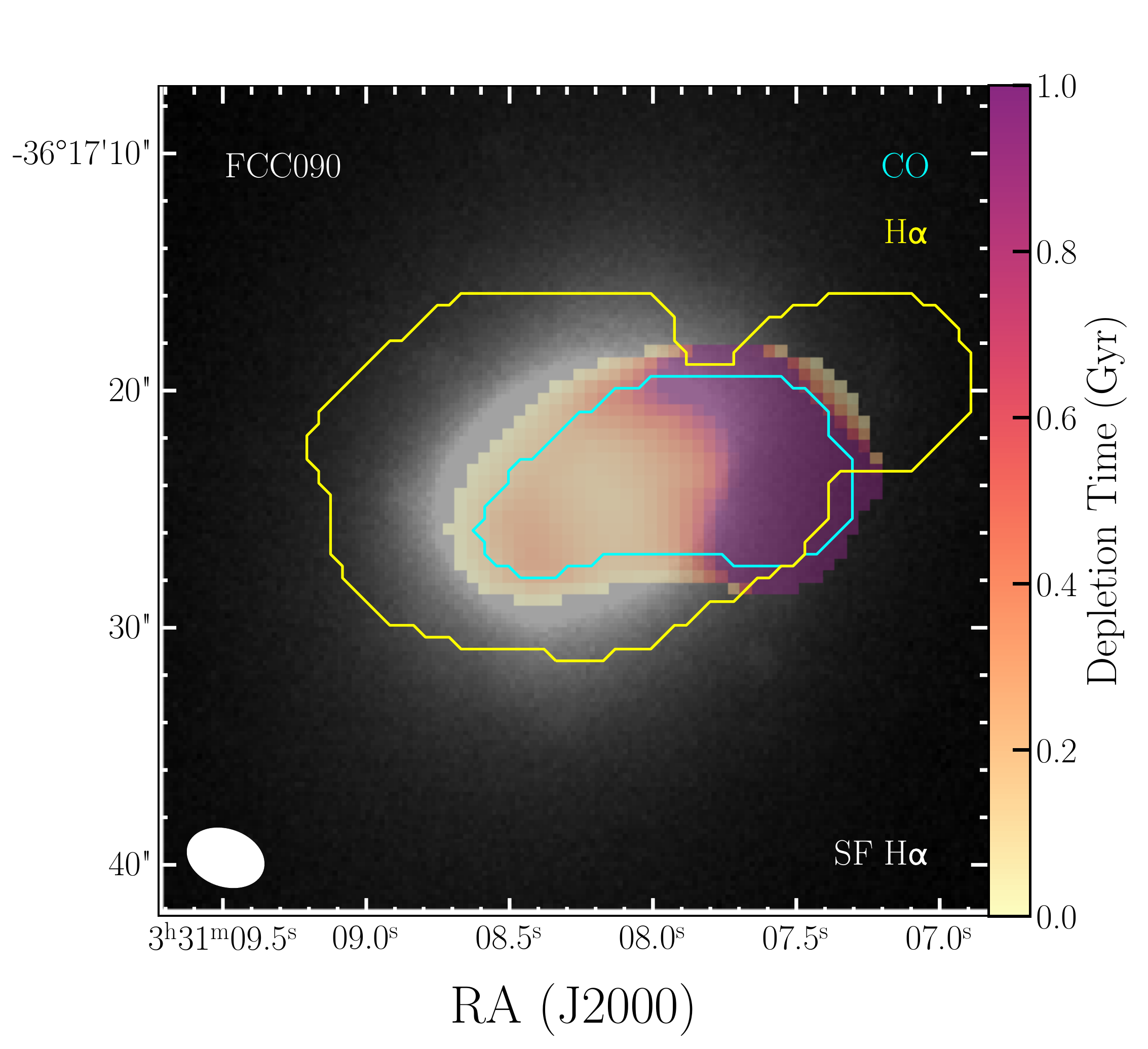}}
	
		\caption{\textbf{Top row}: Relation between the molecular gas and star formation rate surface densities for FCC312 (left) and FCC90 (right). The colours of the markers indicate the density around the data point, determined using a Gaussian kernel density estimation. Lines of constant depletion times (0.1, 1, and 10 Gyr) are shown. The galaxy names are indicated in the upper left corners, in black if the galaxy has a regular molecular gas reservoir, and in red if it is disturbed (see \S \ref{sub:sample}, Table \ref{tab:targets}).
	\textbf{Bottom row}: depletion time maps of FCC312 (left) and FCC90 (right). The maps are overplotted on optical (\textit{g}-band) images from the Fornax Deep Survey (FDS, \citealt{Iodice2016,Iodice2017,Venhola2017,Venhola2018}, Peletier et al., in prep.). The extent of the CO emission is indicated with a cyan contour, and that of the \Ha\ emission with a yellow contour. The beam of the ALMA observations is shown in the lower left corner. In these examples, only the pixels dominated by star forming \Ha\ are used (maps where composite \Ha\ is considered can be found in Appendix \ref{app:SFandDT}).
Most regions in FCC312 have depletion times shorter than the ``standard'' depletion time of 1-2 Gyr, and they vary in the galaxy from region to region. The long depletion time region indicated with the blue/white arrow is due to mosaicking effects in the MUSE data\refrep{: a small fraction of the galaxy was not covered by the mosaic, resulting in an overestimation of the depletion time here}. FCC90 deviates from the ``standard'' depletion time of 1-2 Gyr by a factor 5-10. Depletion times are short in the galaxy's body ($<$ 0.5 Gyr), but long in the gas tail ($\gg$2 Gyr).}
	
	\label{fig:examples}
	
\end{figure*}

SFR maps are obtained by calculating star formation rates from the Balmer decrement corrected \Ha\ maps as follows: 
\begin{equation}
\label{eq:SFR}
\text{SFR} \left( M_\odot\ \text{yr}^{-1} \right) = \frac{L \left( \text{H}_\alpha \right)}{1.86 \times 10^{41}},
\end{equation}
where $1.86 \times 10^{41} M_\odot\ \text{yr}^{-1} \text{erg}^{-1} \text{s}^{-1}$ was adopted from \citet{Hao2011} and \citet{Murphy2011}, and $L \left( \text{H}_\alpha \right)$ is the \Ha\ luminosity in erg s$^{-1}$. The latter is obtained from the MUSE images via
\begin{equation}
L \left( \text{H}_\alpha \right) = 4 \pi D^2 F \left( \text{H}_\alpha \right),   
\end{equation}
where $F \left( \text{H}_\alpha \right)$ is the measured \Ha\ flux in erg s\/ cm$^{-2}$ and $D$ the distance to the galaxy in cm. The SFR is then divided by the spaxel area in kpc$^2$ to obtain the SFR surface density \SFdens\ in $M_\odot\ \text{yr}^{-1}$ kpc$^{-2}$. \change{While some authors opt to correct for inclination in order to obtain a measure of the intrinsic surface density (assuming the gas is distributed in a flat disk), here we do not apply this correction, both due to the disturbed nature of some of the sources, and the fact that we are only comparing surface density maps within individual galaxies, in which case this correction cancels out.} Finally, the image is convolved to match the ALMA beam and regridded to the resolution of the ALMA images.

\refrep{Two sets of \SFdens\ images were created, one where only the \Ha\ emission powered purely by star formation is considered, and another including also other so-called ``composite'' emission regions, where other sources of ionisation (such as shocks or active galactic nuclei, AGN) in addition to O-stars could contribute to the observed \Ha\ emission.} Only spaxels with \Ha\ detections (S/N $>$ 3) are considered. Which spaxels are dominated purely by star forming \Ha, and which by composite ionisation, was determined using Baldwin, Phillips \& Terlevich (BPT, \citealt{Baldwin1981}) diagrams, using the [OIII]$\lambda5007$/H$\beta$ and [NII]$\lambda 6583$/H$\alpha$ line ratios. \change{Only spaxels with sufficient S/N ($>$3) in these lines to reliably determine the nature of the \Ha\ emission were included in the maps}). The boundary line from \citet{Kauffmann2003} was used as a cut-off to define which pixels contain star-forming \Ha, and the one from \citet{Kewley2001} as an upper limit for composites. 

\refrep{The line ratios used in BPT diagrams have a certain degree of sensitivity to conditions like gas density and shocks, and relative abundances. Moreover, there is a level of uncertainty in the exact dividing line between spaxels dominated by ionisation as a result of star formation, and other ionising sources (such as AGN, shocks or old stars). The \citet{Kewley2001} line is a conservative lower limit for the true number of spaxels dominated by these alternative ionising mechnisms. \citet{Kauffmann2003}, on the other hand, posit that star forming galaxies tend to follow a relation on the BPT diagram with relatively little scatter, and thus all spaxels above the scatter in this relation, defined by their dividing line, can be considered as dominated by other sources. Therefore, although these are not perfect dividing lines, both are conservative in what they consider purely AGN/other sources or purely star forming spaxels, and both are widely used to disentangle between contribution from star formation and AGN/other sources to H$\alpha$ emission.}

\refrep{Since emission from composite regions is only partly due to star formation, including these regions allows us to set an upper limit to the total star formation rate of each system. A corresponding lower limit can be set by removing these spaxels entirely. For this reason, we consider two cases throughout this paper: one where only the spaxels strictly dominated by star formation are taken into account, and one where spaxels classified as ``composite'' are included also. We remove all spaxels which are dominated purely by ionisation from other sources (such as AGN or shocks).}

\change{BPT diagrams for all galaxies in the sample (except FCC207) galaxy are presented in \citet{Iodice2019}. Which pixels are excluded from the \Ha\ maps in the \Ha-only analysis, and how this affects the resulting star formation relations, can be seen in the maps in Appendix \ref{app:SFandDT}.} 

\subsubsection{Molecular gas surface densities}
\label{subsub:H2_dens}

\H2dens maps are obtained by first deriving H$_2$ column densities from the CO maps as follows: 
\begin{equation}
	\label{eq:NH2}
	N_{\text{H}_2} \left( \text{mol.\ cm}^{-2} \right) = X_{\text{CO}}\ \frac{\lambda^2}{2 k_{\text{B}}} \int S_\nu\  d\nu,  
\end{equation}
where $X_{\text{CO}}$ is the CO-to-H$_2$ mass conversion factor, $\lambda$ the rest wavelength of the line observed, $k_\text{B}$ is the Boltzmann constant, and $\int S_\nu\ d\nu$ the total flux of the line observed \refrep{in Jy km s$^{-1}$}. 

\change{\refrep{For the purpose of consistency with Paper I}, we use the metallicity-dependent mass conversion factor from eqn. 25 from \citet[see \S4.3 in Paper I]{Accurso2017}. Although \citet{Bigiel2008} used a fixed conversion factor, our results change only marginally when a different conversion factor is adopted, not affecting a comparison with their result.}
The relation used calls for a distance from the star formation main sequence (SFMS, $\text{log}\ \Delta \left( \text{MS} \right)$). To calculate this, we use the SFMS from \citet{Elbaz2007}, and the star formation rates and stellar masses listed in Table \ref{tab:targets}. Where our targets lie in the $M_\star$ - SFR plane compared to the SFMS is shown in Figure \ref{fig:SFMS}. The SFMS is shown as a solid line, the dotted line is an extrapolation from it towards lower stellar masses. The 1 $\sigma$ confidence interval is indicated with the dashed lines. Filled markers indicate data points for which only the star forming \Ha\ was used and open markers indicate data points for which composite \Ha\ was considered also. Both are plotted for each galaxy with the same marker shape, unless it only contains \Ha\ that is not dominated by star formation (FCC167). Galaxies' FCC numbers are annotated. The majority of our galaxies lie well below the SFMS. This is not surprising, as they are molecular gas deficient cluster galaxies (Paper I).

In order to obtain the gas mass surface density \H2dens in $M_\odot\ \text{pc}^{-2}$, equation \ref{eq:NH2} is multiplied by twice the mass of a hydrogen atom ($2\ m_\text{H}$, in $M_\odot$) and converted from cm$^{-2}$ to pc$^{-2}$. 

\subsection{Obtaining the \KS\ relation and depletion times}
\label{sub:SFplots}

\begin{figure*}

	\centering
	
	\subfloat[\label{subfig:KDE_all}]
	{\includegraphics[height=0.47\textwidth, width=0.5\textwidth]{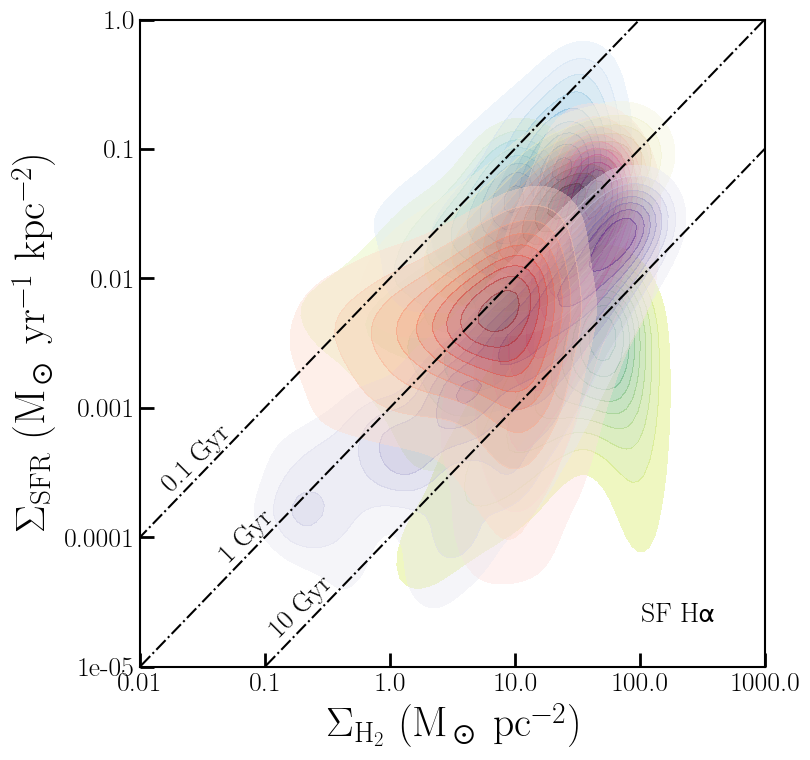}}	\hspace{5mm}
	\subfloat[\label{subfig:KDE_all_allHa}]
	{\includegraphics[height=0.465\textwidth, width=0.46\textwidth]{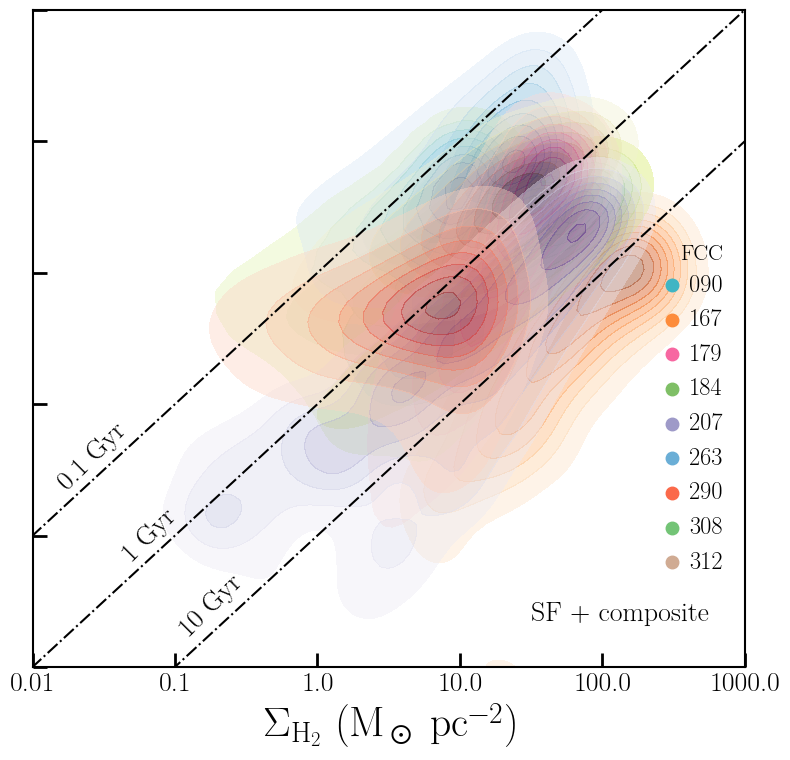}}	
	
	\subfloat[\label{subfig:KDE_one}]
	{\includegraphics[height=0.47\textwidth, width=0.5\textwidth]{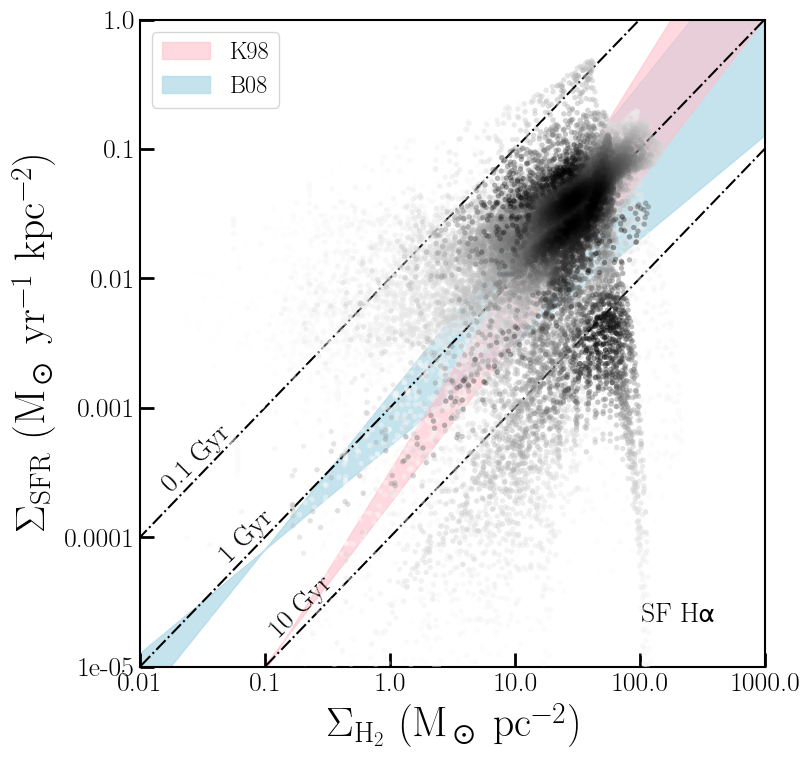}} \hspace{5mm}
	\subfloat[\label{subfig:KDE_one_allHa}]
	{\includegraphics[height=0.465\textwidth, width=0.46\textwidth]{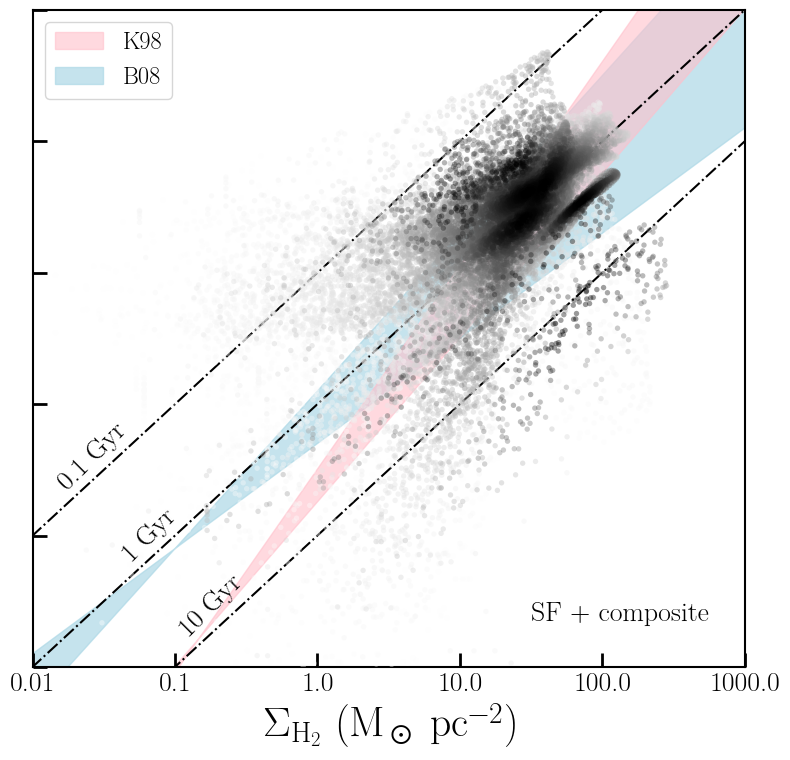}}
	
	\caption{\refrep{Combined \KS\ relation for all galaxies in the sample, both considering the star forming \Ha\ only (left column) and including composite \Ha\ also (right column). In the top row separate contour plots (based on kernel density estimates) are shown for each galaxy, whose FCC numbers are indicated in the right-hand panel. The bottom row shows the same data, represented as points coloured by galaxies' individual KDE values, similar to the top panels (a darker shade indicates a higher KDE value), to allow for a an easier comparison between the Fornax cluster and the K98 and B08 relations, shown in pink and blue, respectively. Lines of constant depletion time (0.1, 1, and 10 Gyr) are shown in each panel. The bulk of the resolution elements in the Fornax cluster have depletion times close to what is predicted by K98 and B08. However, many have slightly decreased depletion times compared to the B08 relation, lying towards its short-depletion time side. A few galaxies have significantly longer depletion times. The \KS\ relation is tighter and looks more physical when composite \Ha\ is taken into account, suggesting that the contribution of AGN and other sources in ionising \Ha\ in these galaxies is low.}}
	\label{fig:combinedKS}
	
\end{figure*}

The actual resolution of the \H2dens\ map is constrained by the ALMA beam size. To avoid over-interpreting the data by studying these maps on smaller scales, both images are binned to the size of the ALMA beam. When doing so, the starting point of the first bin, and therefore the location of the resulting grid, is arbitrary. As a result, the binned map is slightly different depending on the chosen starting point, which is especially noticeable in galaxies with relatively small angular sizes. To account for this, we create a separate map for each possible starting point, effectively shifting the grid up/down and sideways until each possibility is covered. \change{We then show all resulting star formation relations in one figure. This way, the denser areas of the star formation relation will represent the more robust pixel values, whereas sparse regions and scatter are the result of small-scale variations and edge-effects.}
Depletion times are then calculated by dividing both surface density maps by each other. Because we have a different map for each possible grid for \change{both the gas and SFR surface densities}, we take the median of these maps in each pixel before dividing both maps by each other. This way we end up with a map at our initial resolution with the most likely depletion time in each pixel (one version where composite \Ha\ is included, and one where only star forming \Ha\ is considered).

\section{Results}
\label{sec:results}

\begin{figure*}
	\centering
	\includegraphics[width=0.7\textwidth]{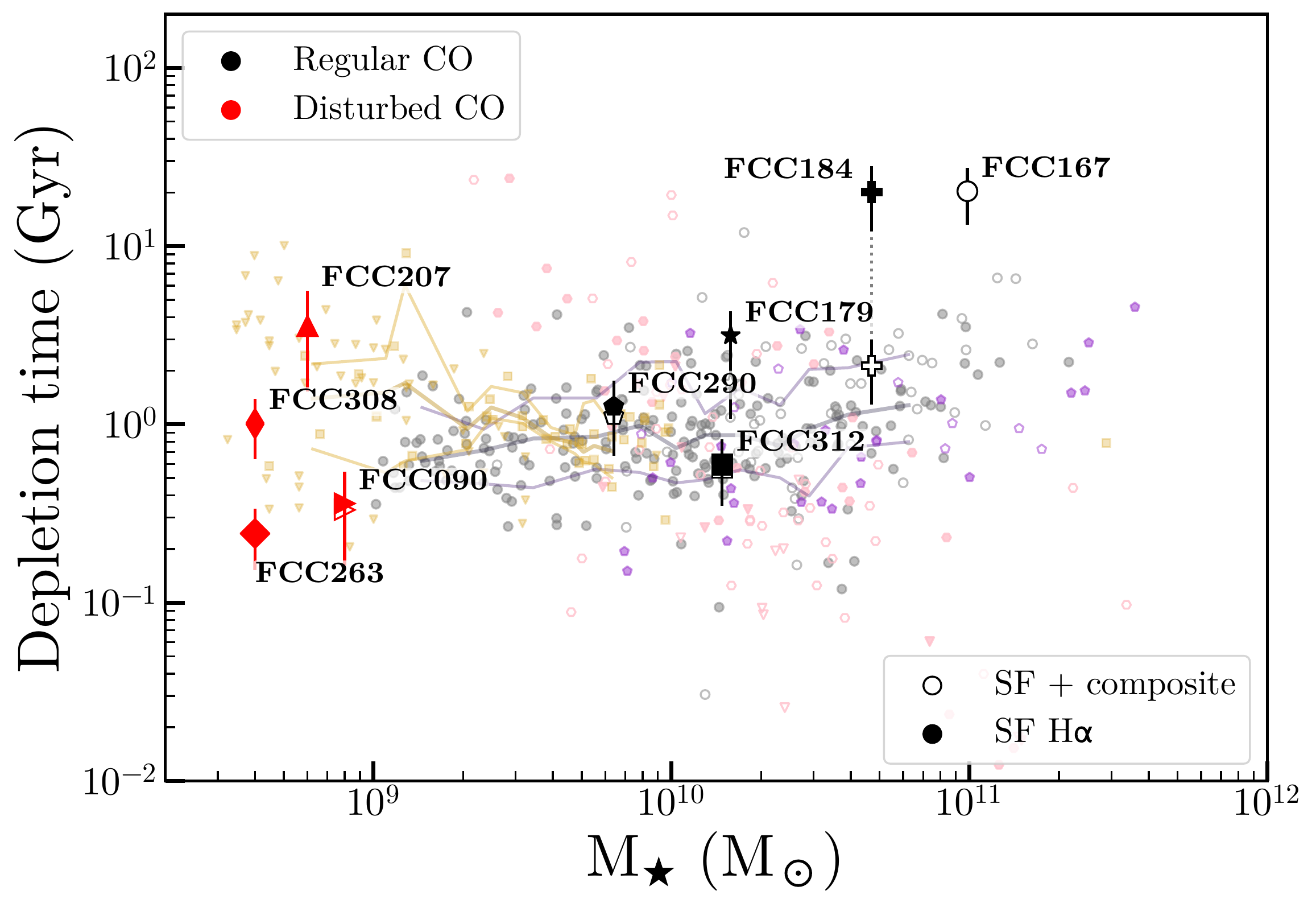}
	\caption{\refrep{Depletion times of the galaxies in the sample as a function of their stellar mass. Markers are the same as in Figure \ref{fig:SFMS}. The underlying field sample consists of xCOLD GASS (\citealt{Saintonge2017}, shown as grey markers), \atlas3d\ (\citealt{Cappellari2011, Young2011, Cappellari2013}, shown as purple markers), ALLSMOG (\citealt{Cicone2017}, shown as yellow markers, triangles indicate upper limits), and HRS (\citealt{Boselli2010,Boselli2014b}, shown as pink markers). Filled purple markers indicate field ellipticals from \atlas3d, open markers are ellipticals residing in the Virgo cluster. Similarly, pink open markers are HRS galaxies located in the Virgo cluster, and filled markers are field galaxies. Grey open markers indicate galaxies in the xCOLD GASS sample that were classified as ``composite'', while filled ones are classified as star forming. Upper limits are indicated with down-pointing triangles for each sample. The median of the xCOLD GASS sample, as well as the 16$^{\text{th}}$ and 84$^{\text{th}}$ percentiles, are indicated with solid lines, and similarly for ALLSMOG in yellow. Our targets mostly follow the weak relation between stellar mass and depletion time seen in the xCOLD GASS sample.}}
	\label{fig:DT_SM}
\end{figure*}

Examples of \KS\ relations and depletion time maps are shown in Figure \ref{fig:examples} for FCC312 (left) and FCC090 (right) \refrep{for the case where only star forming spaxels are considered}. The full set of similar figures (including the versions of these examples where composite spaxels are included also, see below) for all the sample can be found in Appendix \ref{app:SFandDT}. FCC312 was \change{chosen because of its large angular size, which allows it to nicely demonstrate the resulting star formation relation}. It is a relatively massive (log($M_\star / M_\odot$) = 10.04), edge-on spiral with a regular molecular gas reservoir (see Table \ref{tab:targets}). FCC090 was \change{chosen because it has a stripped molecular gas tail}. It is a dwarf (log($M_\star / M_\odot$) = 8.98) elliptical with a ``disturbed'' molecular gas reservoir. Only the pixels dominated by star forming \Ha\ were used in these examples, although they would not look significantly different if composite \Ha\ was considered (Figures \ref{fig:FCC90} and \ref{fig:FCC312}).

In the \KS\ relations (Figure \ref{fig:examples}, top row), the colour of the marker indicates the density of the data around the point \change{in log space}. \refrep{Thus, the lighter the colour, the more resolution elements exist in the galaxy that lie on a similar location on the \KS\ plane. Darker colours, on the other hand, correspond to locations that are only occupied by a few pixels, which usually means they lie at the edge of the detected emission.} These densities were calculated using a Gaussian kernel density estimation (KDE). Furthermore, lines of constant depletion time (0.1, 1, and 10 Gyr, from top to bottom) are shown. The galaxy name is indicated in the upper left corner, in red if the galaxy has a disturbed molecular gas reservoir, and in black if it is regular (see Table \ref{tab:targets}).

The \DT\ maps (Figure \ref{fig:examples}, bottom row) are overplotted on optical (\textit{g}-band) images from the Fornax Deep Survey (FDS, \citealt{Iodice2016,Iodice2017}, Peletier et al., in prep., \citealt{Venhola2017,Venhola2018}). The extent of the CO emission is indicated with a cyan contour, and the extent of the \Ha\ emission with a yellow contour (as indicated in the top right corner).

FCC312 has slightly decreased depletion times compared to the ``standard'' 1-2 Gyr for spiral galaxies, with the majority of the regions having depletion times lower than 1 Gyr. Both CO and \Ha\ are present throughout the galaxy. There is some small-scale variation in depletion time, mostly towards depletion times shorter than the bulk of the resolution elements in the galaxy, shown by the scatter in the SF plot and the variations in the \DT\ map. The relatively high-depletion time area in the middle of the disk on the north side of the galaxy \change{(indicated with the blue/white arrow)} is due to mosaicking effects in the MUSE data\refrep{: a small fraction of the galaxy was not covered by the mosaic, resulting in an overestimation of the depletion time here.}

FCC90 also has relatively short depletion times in its stellar body, as can be seen in the depletion time map and corresponding points in the \KS\ relation. In the molecular gas tail depletion times are $>$10 times longer, with a sharp transition between the stellar body and this tail. In the \KS\ relation, the large ``wing'' towards longer depletion times corresponds to this area. This is discussed further in \S \ref{sec:discussion}.

Similar figures for the remaining galaxies, along with their discussions, are provided in Appendix \ref{app:SFandDT}. General trends in these figures are discussed in \S \ref{sec:discussion}.

\subsection{Combined \KS\ relations}
\label{sub:combined_relations}
In order to obtain a better understanding of whether galaxies in the Fornax cluster follow the classic star formation relations, it is useful to show the \KS\ relations for each galaxy in one figure. This is done in Figure \ref{fig:combinedKS}. The top row shows the KDE plots for all the sample, similar to the top rows in Figure \ref{fig:examples} and Appendix \ref{app:SFandDT}, each in a different colour. FCC numbers are indicated in the upper right-hand panel. \refrep{In the bottom panels the same data are shown, but represented with points coloured by their KDE value. This is to better visualise how the \KS\ relation in the Fornax cluster relates to the K98 and B08 relations, which are shown in pink and blue, respectively.} The left column of the figure shows the case in which only star forming \Ha\ is included, and the right column shows the case in which composite \Ha\ is considered also (this is indicated in the lower right corners of the panels). FCC167 is only present in the right column (where composite \Ha\ is considered as well as star forming \Ha), as its \Ha\ image does not contain any pixels that are dominated purely by star formation according to a BPT classification. 

\refrep{The \KS\ relation in the left panel, where only \Ha\ from regions dominated purely by star formation is considered, looks unphysical for some galaxies. In particular, FCC184 shows a strong downturn in its star formation efficiency at high H$_2$ densities (see also Figure \ref{fig:FCC184}), resulting in unrealistically long depletion times. This effect arises near the centre of this object, where other sources of ionisation begin to become important. This results in a large area consisting of ``composite'' spaxels in this area. Including the \Ha\ emission from this region moves these spaxels into agreement with a normal \KS\ relation (see the right-hand panels in Figure \ref{fig:combinedKS} and \ref{fig:FCC184}). This implies that star formation contributes significantly to the ``composite'' spaxels in this galaxy. We therefore consider the situation where \Ha\ originating from composite regions is taken into account, in this case, more realistic. This means that the contribution from AGN (and/or old stars/shocks) to the ionisation of the gas in composite regions is probably low in this galaxy. This Figure is discussed further in \S \ref{sec:discussion}}.

\refrep{Together with the dwarf galaxy FCC207, which has depletion times of $\sim$2 - $>$10 Gyr, the lenticular FCC167 is the only galaxy that has depletion times significantly longer than the usual 1-2 Gyr, almost every spaxel has a deplition time significantly longer than 10 Gyr. Depletion times in the other galaxies overlap mostly with those predicted by the K98 and B08 relations, but lie towards the short-depletion time side of the B08 relation.}

\begin{figure*}
	\centering
	\includegraphics[width=0.7\textwidth]{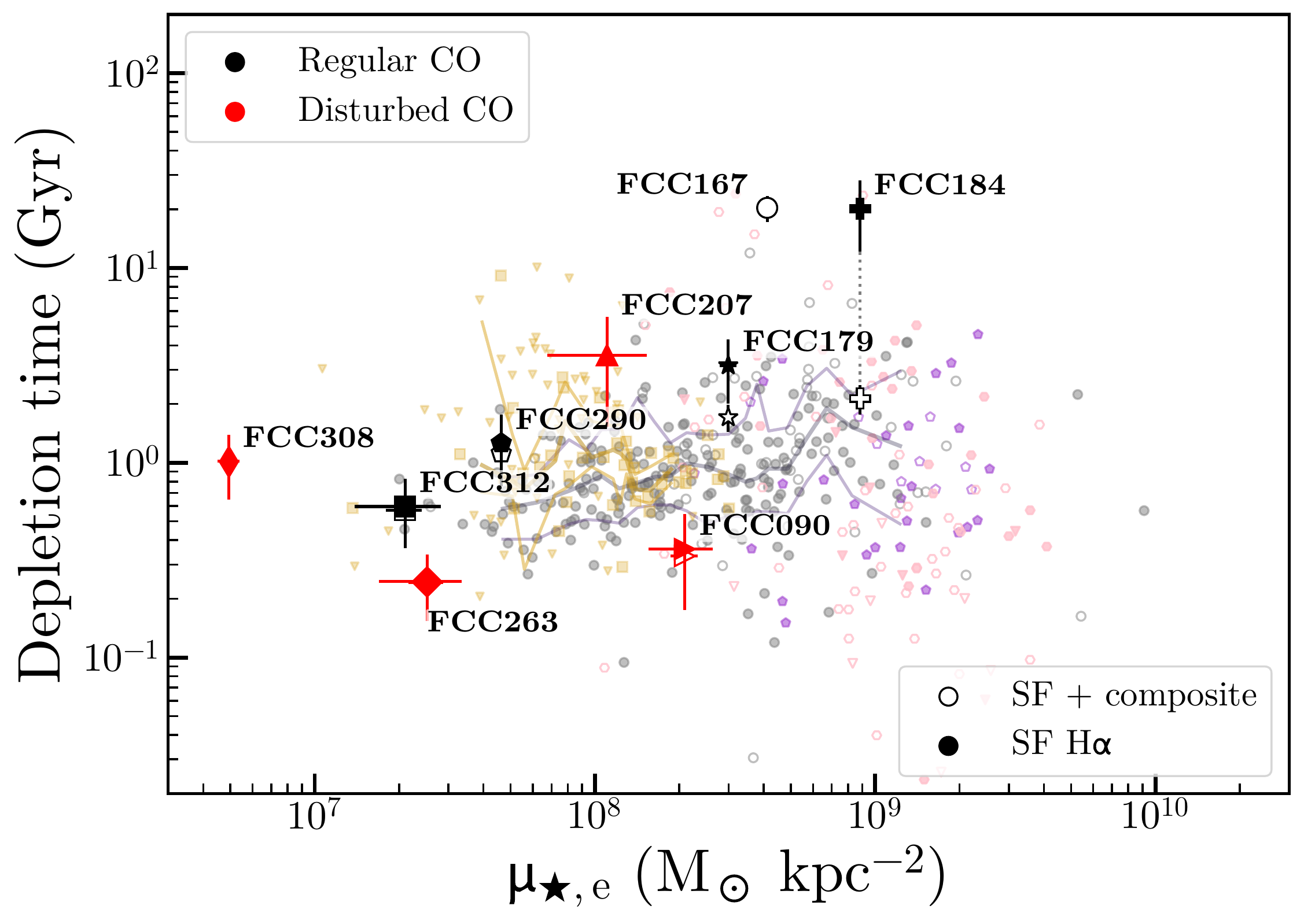}
	\caption{Similar to Figure \ref{fig:DT_SM}, but with depletion times as a function of stellar mass surface density. \refrep{A weak relation between depletion time and stellar mass surface density can be seen in the xCOLD GASS sample. The Fornax galaxies mostly follow this weak trend.} Dwarf galaxies with disturbed molecular gas are widely scattered around this relation.}
	\label{fig:DT_SB}
\end{figure*}\

\subsection{Relations with other parameters}
\label{sub:relations}
Above we have seen that depletion times can vary significantly between galaxies in the sample. Here we test whether this variation depends on other variables, i.e. stellar mass, morphology, and (projected) location within the cluster. We compare our sample to field galaxies where possible.

Figure \ref{fig:DT_SM} shows galaxy-average depletion times as a function of stellar mass. Filled markers represent median depletion times with only the star formation dominated pixels in the \Ha\ image taken into account, and empty markers represent the case where composite pixels in the \Ha\ image are also taken into account. With the exception of FCC167, whose \Ha\ image contains no pixels dominated by star formation according to our classification, there are thus two measurements for each galaxy, represented by markers of the same shape, and connected by a dotted line. Galaxies with a disturbed molecular gas reservoir are shown with red markers, and galaxies with a regular molecular gas reservoirs in black. Uncertainties are a combination of the error on the molecular gas surface density and the error on the star formation rate surface density (in both cases the rms error and a 10\% calibration error). The smaller markers in the background represent the \change{control} sample. It consists of data from the extended CO Legacy Database for \textit{GALEX} Arecibo SDSS Survey (xCOLD GASS, \citealt{Saintonge2017}, grey), \change{which is a field sample containing a broad range of galaxy types, data from the APEX low-redshift legacy survey for molecular gas (ALLSMOG, \citealt{Cicone2017}, yellow), which spans galaxies with stellar masses from 10$^{8.5}$ to 10$^{10} M_\odot$}, \refrep{data from the Herschel Reference Survey\footnote{\refrep{Herschel \citep{Pilbratt2010} is an ESA space observatory with science instruments provided by European-led Principal Investigator consortia and with important participation from NASA.}} (HRS, \citealt{Boselli2010,Boselli2014b}, in pink),} and early-type galaxies from \atlas3d\ (\citealt{Cappellari2011, Young2011, Cappellari2013}, purple). \atlas3d\ was added because the xCOLD GASS sample might not detect the gas in more massive ellipticals, as more sensitive observations are often required to detect CO in these galaxies. This sample was thus added to complete the higher mass end of the relation, and put the massive early-type galaxy FCC184 into more relevant context. \secround{From xCOLD GASS we only include star forming galaxies (shown as filled markers) and composite galaxies (shown as open markers), as classified by their position in the BPT diagram. Upper limits are indicated with down-pointing triangles for all samples.} \secround{Error bars are omitted for visibility purposes. Uncertainties in the depletion times are typically slightly over 10\% for xCOLD-GASS, 30\% for ALLSMOG and 40\% for the HRS.}

\secround{Where necessary, we recalibrate the literature data to share the same X$_\text{CO}$ prescription we use to estimate M$_{\text{H}_2}$ in our Fornax sample (see \S \ref{subsub:H2_dens} and Paper I). xCOLD GASS already uses this prescription. From ALLSMOG we directly use CO luminosities, allowing us to use the same recipe to derive H$_2$ masses. To derive X$_\text{CO}$ we use their gas-phase metallicities calculated using the O3N2 calibration from \cite{Pettini2004}, and a distance from the main sequence from \citet{Elbaz2007}, as with our Fornax measurements. To recalibrate the \atlas3d\ values, we use stellar masses derived from $r$-band luminosities and M/L from \citet{Cappellari2013} to estimate metallicities. Distances from the main sequence are derived using the SFRs from \citet{Davis2014} compared to the main sequence from \cite{Elbaz2007}. We then derive a correction factor from the X$_\text{CO}$ we derive using our method and these parameters, compared to the X$_\text{CO}$ used to derive the molecular gas masses in this sample. From the HRS we use total CO fluxes, and derive M$_{\text{H}_2}$ following the same prescription as for the Fornax, xCOLD GASS, and ALLSMOG samples. We estimate metallicities from the O3N2 line ratio, using the [N{\sc i}{\sc i}]$\lambda$6584, [O{\sc i}{\sc i}{\sc i}]$\lambda$5007, and H$\beta$ line fluxes (all normalised to the H$\alpha$ line flux) from \citet{Boselli2013}, and the calibration from \citet{Pettini2004}.}

The \atlas3d\ and HRS samples are split into field galaxies (filled markers) and galaxies located in the Virgo cluster (open markers). \refrep{To better visualise how Fornax galaxies compare to those in Virgo, this figure is duplicated in Appendix \ref{app:virgo}, but with only data from Virgo galaxies shown.} \change{The ALLSMOG sample was added to put the lower-mass galaxies into better context.} \refrep{To calculate depletion times for the HRS sample, we use star formation rates from \citet{Boselli2015}, and molecular gas masses from \citet{Boselli2014b}.} The solid grey lines indicate the median and 16$^{\text{th}}$ and 84$^{\text{th}}$ quantiles of the xCOLD GASS data, indicating the 1$\sigma$ spread in the data. To calculate these, running bins were used, where the data is divided into horizontal bins, each containing an equal number of \change{(33)} galaxies, where each bin is shifted by half a bin size with respect to the previous one. Similarly, the yellow lines show the median and 1$\sigma$ spread in the ALLSMOG data. 

Most of our cluster galaxies lie within or close to the 1$\sigma$ confidence interval of the field relations.

\begin{figure*}
	\centering
	\includegraphics[width=0.7\textwidth]{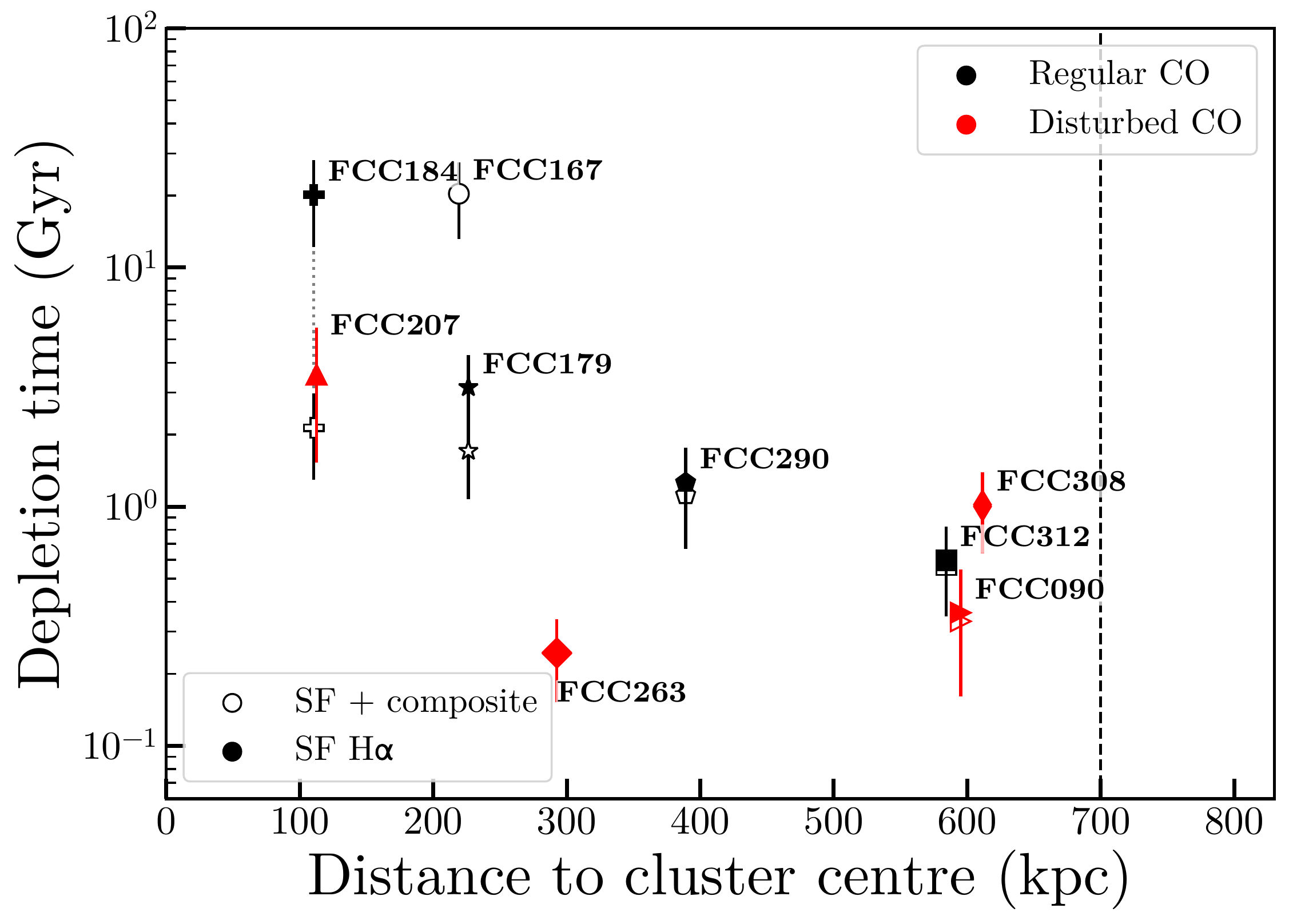}
	\caption{Depletion time as a function of (projected) distance to the cluster centre (i.e. BCG NGC1399). Markers are the same as in Figures \ref{fig:SFMS}, \ref{fig:DT_SM} and \ref{fig:DT_SB}. The virial radius is indicated with a dashed line. A relation between depletion time and cluster-centric distance seems to exist, although it is likely partly driven by galaxy mass/morphology.}
	\label{fig:DT_distance}
\end{figure*}

Figure \ref{fig:DT_SB} shows the depletion times of the galaxies (obtained from the median values of the resolution elements of their molecular gas and star formation rate surface densities) in the sample as a function of their stellar mass surface density (SMSD), here defined as half the stellar mass divided by the area enclosed by the effective radius $R_e$. A galaxy's SMSD is a crude measure of its radial mass profile, and can therefore be interpreted as a proxy for its morphology. Effective radii are taken from \citet{Venhola2018} for dwarf galaxies, and from \citet{Raj2019} and \citet{Iodice2019} for more massive galaxies. Markers and lines are the same as in Figure \ref{fig:DT_SM}. Most galaxies have depletion times within 1$\sigma$ of what is expected from their surface densities as predicted by the field samples.

Figure \ref{fig:DT_distance} shows depletion times as a function of distance from the cluster centre, defined here as the brightest cluster galaxy (BCG) NGC1399 (FCC213). Markers are the same as in Figures \ref{fig:DT_SM} and \ref{fig:DT_SB}. The virial radius (at 0.7 Mpc, \citealt{Drinkwater2001a}) is indicated with a dashed line. \change{There is a clear trend between cluster-centric distance and depletion time. Although this is likely partly driven by galaxy mass/morphology (massive ellipticals, which often have low SFEs \citep{Saintonge2011,Davis2014} are in the cluster centre, whereas less massive spiral and dwarf galaxies tend to be located further out), there is possibly an additional correlation between depletion time and cluster-centric distance.} The figures described above are discussed further in \S \ref{sec:discussion}.

\section{The molecular cloud lifetime in FCC290}
\label{sec:cloud_lifetimes}

\change{FCC290 is a nearly face-on, flocculent spiral (Appendix~A7, Paper~I). It is therefore an ideal candidate to study the small-scale relation between molecular gas and star formation in cluster galaxies. We have measured the molecular cloud lifetime in FCC290 by applying the statistical method (named the `uncertainty principle for star formation') developed by \citet{KL14} and \citet{Kruijssen2018}.}

\subsection{Uncertainty principle for star formation}
\label{sub:uncert_princ}
\change{Contrary to the tight correlation defining the K98 relation observed on galaxy scales, a spatial de-correlation between CO and \Ha\ clumps is commonly observed in nearby galaxies \citep[e.g.][]{Schruba2010, Kreckel2018, Chevance2020, Hygate2019b, Kruijssen2019}. This spatial de-correlation, which is responsible for the observed scatter around the K98 relation below $\sim 1$\,kpc scales (Figure~2 and Appendix~A; see also e.g.\ \citealt{Bigiel2008, Blanc2009, Leroy2013}), can be explained by the fact that individual regions in galaxies follow independent evolutionary lifecycles where cloud assemble, collapse, form stars, and get disrupted by stellar feedback \citep[e.g.][]{Feldmann2011,KL14}. On small scales, each individual region is observed at a specific time of this cycle and it is only when averaging over many regions, sampling all phases of this cycle, that the galactic K98 relation can be retrieved.}

\change{The small-scale de-correlation between CO and \Ha\ clumps can be directly linked to the underlying timeline of this evolutionary cycle, and represents a fundamental probe of stellar feedback physics in observations and numerical simulations \citep[e.g.][]{Fujimoto2019}. In this paper, we follow the specific implementation of this methodology presented by \cite{Kruijssen2018} and formalised in the \textsc{Heisenberg} code to characterise the cycle of star formation and feedback in the cluster galaxy FCC290, from cloud formation (as traced by CO emission), to star formation (as traced by \Ha\ emission), and to cloud destruction.}

\change{In practice, we measure the CO-to-\Ha\ flux ratio in apertures of certain sizes centred on peaks of CO or \Ha\ emission. The relative change of the CO-to-\Ha\ flux ratio compared to the galactic CO-to-\Ha\ ratio as a function of the aperture size (see Figure~\ref{fig:tuningfork}), is governed by the relative timescales of the different phases of the evolutionary cycle of cloud evolution and star formation. These observations can be fitted by the model of \cite{Kruijssen2018}, which depends on three independent parameters: the duration of the cloud lifetime relative to the isolated stellar phase (\tCO/\tHaref), the timescale during which CO and \Ha\ coexist in a region relative to the isolated stellar phase (which is the time during which stellar feedback can affect the cloud, \tover/\tHaref) and the characteristic separation length $\lambda$ between individual regions (i.e. molecular clouds and star forming regions). To convert these relative timescales into absolute ones, we use the duration of the isolated stellar phase traced by \Ha\ only (i.e.\ after the complete disruption of the CO cloud, \tHaref). This time-scale has been calibrated by \cite{Haydon2019} and is a function of the metallicity \footnote{We note that there this \Ha\ emission timescale can in principle be affected by extinction. However, \citet{Haydon2020} demonstrate that this only happens for kpc-scale gas surface densities $>20~\msun~\pc^{-2}$, much larger than the typical surface density of FCC290, which is $\sim3~\msun~\pc^{-2}$} (Table \ref{tab:input}).}

\begin{table*}
	\centering
	\begin{threeparttable}
	\caption{Main input parameters of and quantities constrained (output) by the analysis of FCC290, along with their descriptions from Tables 2 and 4 in \citet{Kruijssen2018}. The other input parameters use the default values as listed in Table~2 of \citet{Kruijssen2018}.}
	\label{tab:input}
	\begin{tabular}{lll}	
	\hline
	
Input quantity    &   Value   &  Description  \\
\hline
$D$ [Mpc]          &  $17.7$  & Distance to galaxy \\
$i$ [\degr]        &   47  &  Inclination angle \\
$l_{\rm ap,min}$ [pc]   & 215  & Minimum inclination-corrected aperture size (i.e. diameter) to convolve the input maps to \\
$l_{\rm ap,max}$ [pc]  & 6400 & Maximum inclination-corrected aperture size (i.e. diameter) to convolve the input maps to \\
$N_{\rm ap}$            & 15 & Number of aperture sizes used to create logarithmically-spaced aperture size array in the range ($l_{\rm ap,min}$, $l_{\rm ap,max}$) \\
$N_{\rm pix,min}$       & 10  & Minimum number of pixels for a valid peak (use $N_{\rm pix,min}$ = 1 to allow single points to be identified as peaks) \\
$\Delta \log_{10} \mathcal{F}_{\rm star}$$^{\textit{a}}$ & 2.00   &  Logarithmic range below flux maximum covered by flux contour levels for stellar peak identification  \\
$\delta \log_{10} \mathcal{F}_{\rm star}$$^{\textit{a}}$ & 0.03    & Logarithmic interval between flux contour levels for stellar peak identification  \\
$\Delta \log_{10} \mathcal{F}_{\rm gas}$$^{\textit{a}} $ & 0.6     &  Logarithmic range below flux maximum covered by flux contour levels for gas peak identification \\
$\delta \log_{10} \mathcal{F}_{\rm gas}$$^{\textit{a}}$  & 0.06    & Logarithmic interval between flux contour levels for stellar peak identification \\
$t_{\text{\Ha ,ref}}$ [Myr]                  & 4.29 &  Reference time-scale spanned by star formation tracer (in our case \Ha) \\
$\sigma(t_{\text{\Ha ,ref}})$ [Myr]$^{\textit{b}} $     & 0.16  & Uncertainty on reference time-scale \\
SFR [\Msun~yr$^{-1}$]$^{\textit{c}}$     & 0.127  & Total star formation rate in the MUSE FoV \\
log$_{10}X_{\rm gas}$$^{\textit{d}}$ & 0.64 & Logarithm of conversion factor from map pixel value to an absolute gas mass in $M_\odot$ \\
$\sigma_{\rm rel}(X_{\rm gas})$$^{\textit{b}}$         & 0.34  & Relative uncertainty (i.e. $\sigma_x$/x) of X$_\text{gas}$ \\
$n_{\lambda \rm,iter}$   & 13  & Characteristic width used for the Gaussian filter to filter out diffuse emission in Fourier space (see text) \\
\hline
\hline
Output quantity & Value & Description \\ 
\hline
$\chi^2_{\rm red}$     & 0.84 & Goodness-of-fit statistic \\
$t_{\rm CO}$ [Myr]     & $9.3_{-2.2}^{+2.6}$ & Best-fitting gas tracer lifetime (e.g. the cloud lifetime, in our case CO) \\
$t_{\rm fb}$ [Myr]     & $1.7_{-0.7}^{+0.9}$ & Best-fitting overlap lifetime (e.g. the feedback time-scale, like in our case) \\
$\lambda$ [pc]         & $327_{-74}^{+133}$ & Best-fitting mean separation length of independent regions (e.g. the fragmentation length)  \\
$r_{\rm star}$ [pc]    & $100_{-15}^{+13}$ & Disc radius or Gaussian dispersion radius of star formation tracer peaks  \\
$r_{\rm gas}$ [pc]     & $97_{-13}^{+17}$ & Disc radius or Gaussian dispersion radius of gas tracer peaks \\
$\zeta_{\rm star}$     & $0.61_{-0.11}^{+0.06}$ & Star formation tracer peak concentration parameter \\
$\zeta_{\rm gas}$      & $0.59_{-0.09}^{+0.07}$ & Gas tracer peak concentration parameter  \\
$n_{\rm peak,star}$    & 49 & Number of peaks in the star formation tracer map (H{\sc i}{\sc i} regions) \\
$n_{\rm peak,gas}$     & 52 & Number of peaks in the gas tracer map (giant molecular clouds) \\
\hline

	\end{tabular}
\textit{Notes:} $^\textit{a}$The parameters for the peak identification listed here are valid for the diffuse-emission filtered maps (\S \ref{sub:uncert_princ}). Different values are used for the first iteration during which emission peaks are identified in unfiltered maps, but we have verified that the choice of these initial parameter values does not significantly affect our results.\\
$^\textit{b}$Standard error. The subscript `rel' indicates a relative error.\\
$^\textit{c}$This is the SFR measured across the MUSE field of view considered, rather than of the entire galaxy.\\
$^\textit{d}$The gas conversion factor corresponds to $\alpha_{\rm CO(1-0)}$ in $\Msun~({\rm K}~\kms~\pc^2)^{-1}$.
	\end{threeparttable}
\end{table*}

\refrep{The above measurements are potentially affected by the presence of large-scale diffuse emission, which is not associated with individual molecular clouds or star forming regions and is therefore not participating in the evolutionary cycle of the emission peaks identified. It is therefore necessary to filter the diffuse emission on scales larger than $\lambda$ to ensure an unbiased measurement of the molecular cloud lifecycle. We use the method of \cite{Hygate2019b} to filter out diffuse emission in Fourier space with a Gaussian filter, using a characteristic width of 13$\lambda$. This multiplicative factor of $\lambda$ has been chosen to ensure the best compromise between a selective filter to filter a maximum of the diffuse emission and a broader filter to limit the appearance of artefacts around compact structures (see \citealt{Hygate2019b} for details). During the analysis, a first estimation of $\lambda$ is made for the original, unfiltered maps. The maps are then transformed to Fourier space and spatial wavelengths larger than 13$\lambda$ are masked with the Gaussian filter, before transforming the maps back to normal space. We then estimate $\lambda$ again using these new, filtered maps, and iterate over this process until convergence is reached (defined as obtaining the same value of $\lambda$ to within 5~per~cent for four successive iterations).}

\subsection{Application to FCC290}
\begin{figure*}
\begin{minipage}{0.40\linewidth}
\includegraphics[trim=10mm 2mm 18mm 80mm, clip=true, width=\linewidth]{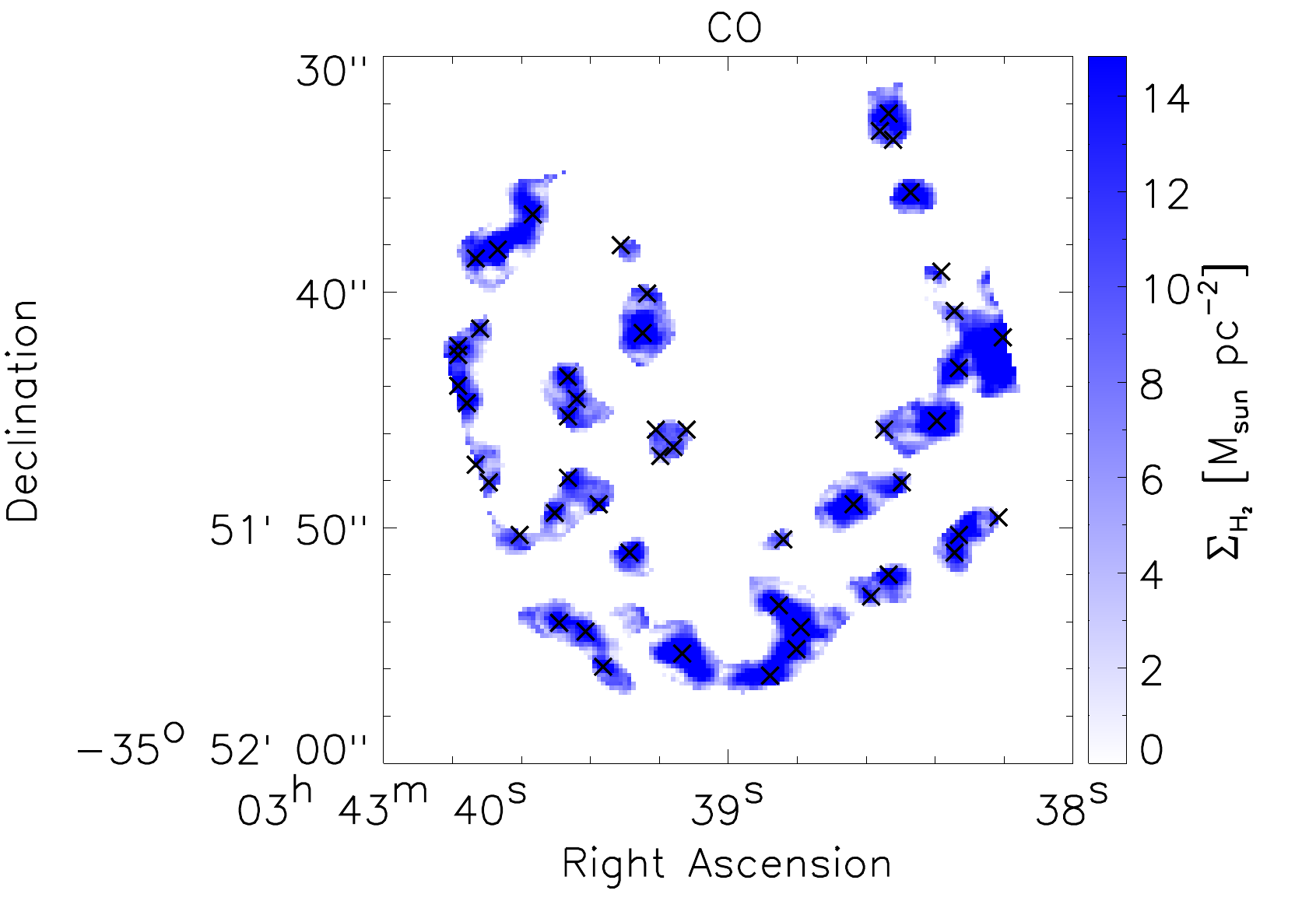}
\includegraphics[trim=10mm 2mm 18mm 80mm, clip=true,width=\linewidth]{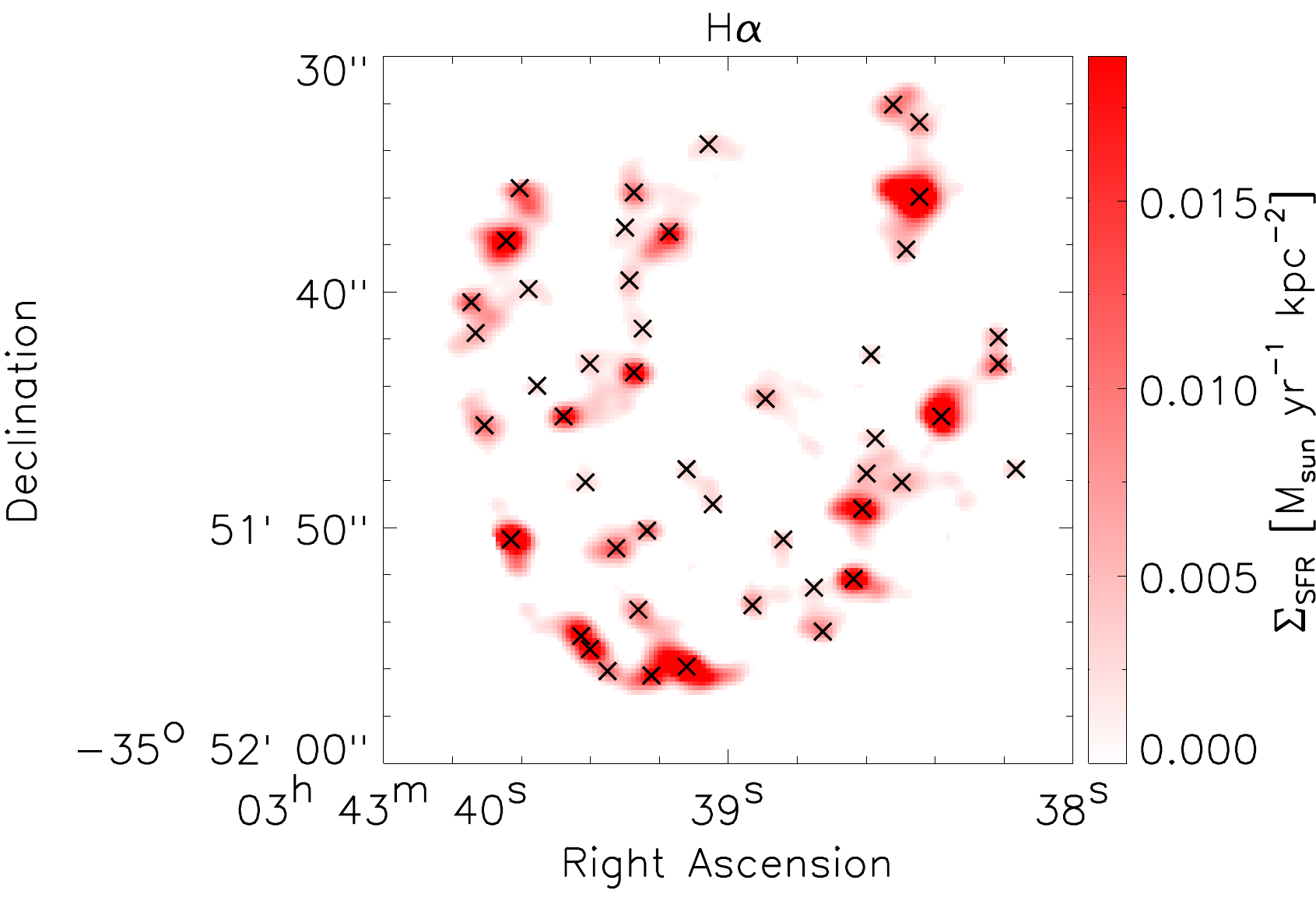}
  \end{minipage}
   \hspace{1mm} 
    \begin{minipage}{0.58\linewidth}
\includegraphics[width=\linewidth]{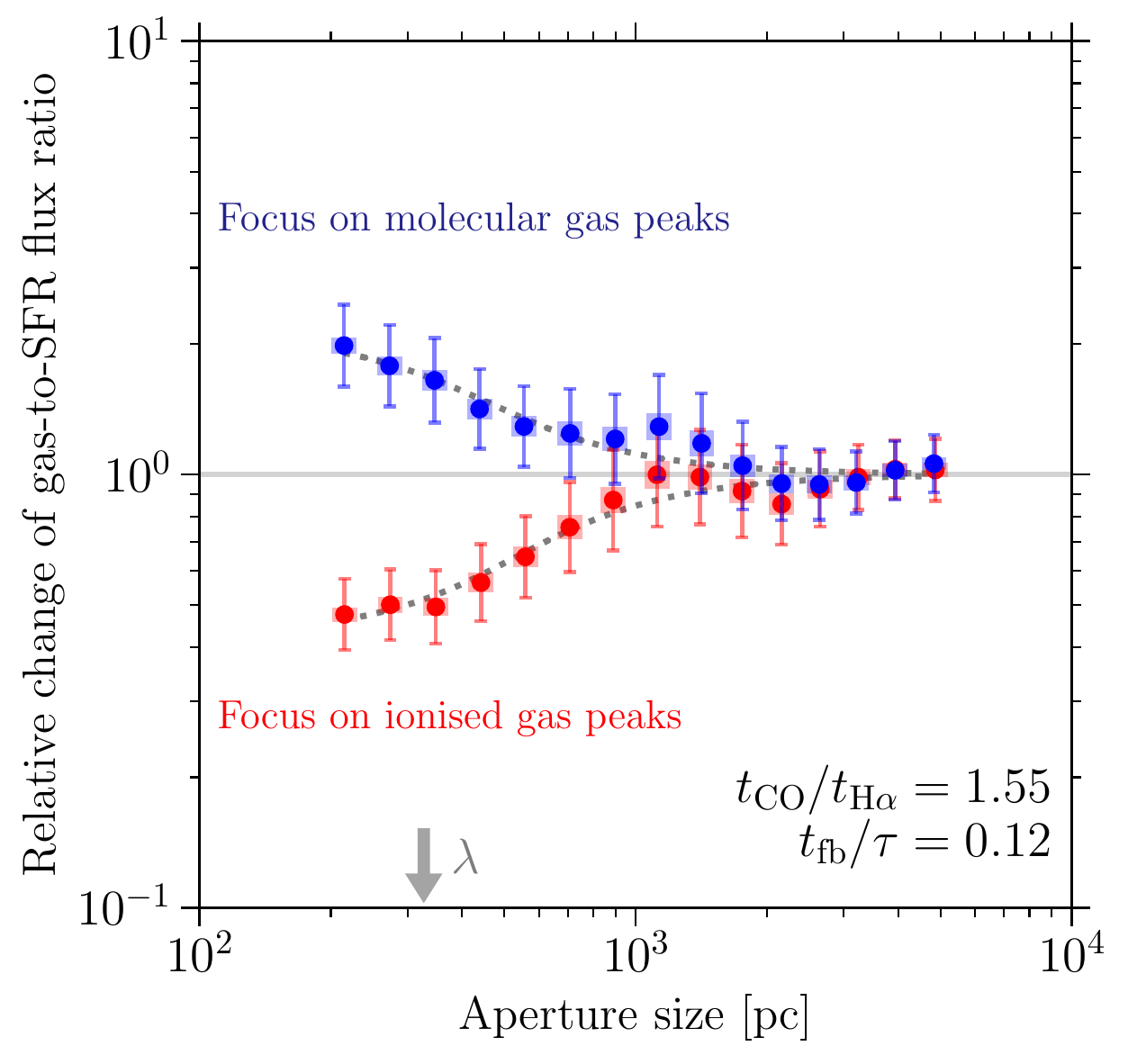}
  \end{minipage}
\caption{Left panels: CO (top) and \Ha\ (bottom) maps of FCC290 after removing the diffuse emission (\S \ref{sub:uncert_princ}). The identified emission peaks in each map are indicated with black crosses. Right panel: Relative change of the CO-to-\Ha\ flux ratio compared to the galactic average as a function of the aperture size, for apertures placed on CO emission peaks (blue) and \Ha\ emission peaks (red). \refrep{The error bars indicate the $1\sigma$ uncertainty on each individual data point, whereas the shaded areas indicate the effective $1\sigma$ uncertainty range that accounts for the covariance between the data points and should be used when visually assessing the quality of the fit.} The horizontal solid line indicates the galactic average and the dotted line is the best-fitting model \citep{Kruijssen2018}, which allows us to constrain the molecular cloud lifecycle. The arrow indicates the best-fitting value of the region separation length $\lambda$. The ratios \tCO/\tHa\ (controlling the asymmetry between the two branches) and \tover/$\tau$ (controlling the flattening of the branches) are indicated in the bottom-right corner. Also compare to \citet[Fig.~1]{Kruijssen2019}, \citet[Fig.~2]{Chevance2020}, and \citet[Fig.~2]{Hygate2019b}.}
\label{fig:tuningfork}
\end{figure*}

\change{We apply the above analysis to the CO and \Ha\ maps of FCC290, limited to the field of view of the MUSE map, in order to ensure the best possible spatial resolution (see Figure~\ref{fig:tuningfork}). The input parameters we use for the \textsc{Heisenberg} code, as well as the main output quantities, are summarised in Table~\ref{tab:input}. Descriptions are from Table 2 and 4 from \citet{Kruijssen2018}. Other input parameters not listed are set to their default values, which can be found in the same tables in \citet{Kruijssen2018} along with their descriptions.}

\change{We measure a relatively short molecular cloud lifetime of $t_{\rm CO}=9.3_{-2.2}^{+2.6}~\myr$ and a characteristic (deprojected) separation length between regions of $\lambda = 327_{-74}^{+133}~\pc$. Comparison with previous measurements of the molecular cloud lifetime using the same method \citep{Chevance2020, Hygate2019b, Kruijssen2019} reveals similarities, but also some differences with the values found in nearby spirals. They might be linked to the cluster environment of FCC290. Indeed, short molecular cloud lifetimes of $\sim 10$\,Myr are commonly measured in low mass galaxies like FCC290 (e.g. NGC300, NGC5068, and M33, which have stellar masses of $\log{(M_\star/\msun)} \leq 9.8$, FCC290 has a stellar mass of $\log{(M_\star/\msun)} = 9.8$), with low molecular gas surface densities ($\sim 3~\msun~\pc^{-2}$ for NGC300 and NGC5068, between 1 and $10~\msun~\pc^{-2}$ for M33). Despite a somewhat larger total gas mass, the average molecular gas surface density in FCC290 is $\Sigma_{\rm H_2} \sim 3~\msun~\pc^{-2}$, very similar to NGC300 and NGC5068. The short molecular cloud lifetime of $\sim10~\myr$ in FCC290 is consistent with the conclusions of \cite{Chevance2020}, who show that large-scale surface density is an important factor in driving the cloud lifecycle. At kpc-scale molecular gas surface densities of $\Sigma_{\rm H_2} \leq 8~\msun~\pc^{-2}$, the cloud lifetime is regulated by internal dynamical processes and clouds live for a short time, comparable to one cloud free-fall time or cloud crossing time, with typical values of $\sim10~\myr$. At larger molecular gas surface densities, the cloud lifetime is instead regulated by galactic dynamics, often resulting in cloud lifetimes of $20{-}30~\myr$.}

\change{A notable difference between FCC290 and nearby star-forming galaxies in the field is the large characteristic distance ($\lambda\sim300~\pc$) between individual regions. By contrast, typical values measured by \cite{Chevance2020} are in the range 100--250\,pc, where galaxies with low molecular gas surface densities are typically located on the lower end of this range. This could be a result of the different environment.}

\change{Despite the relatively low number of identified regions ($n_{\rm peak,star} = 49$ in the \Ha\ map and $n_{\rm peak,gas} = 52$ in the CO map), the main output quantities (\tCO , \tover\ and $\lambda$) are well-constrained. As tested in \cite{Kruijssen2018}, a minimum of 35 peaks in each map ensures a precision better than 50\%. In addition, the region separation length $\lambda$ should be well resolved to guarantee a good accuracy of the feedback time-scale \tover, i.e.\ the time-scale over which the CO and \Ha\ emission co-exist. This time-scale represents the time it takes feedback from massive stars to disperse the parent molecular cloud. We measure a ratio of $\lambda$ to the minimum aperture size ($l_{\rm ap,min}$) of 1.52. This is sufficient to bring a strong constraint on \tCO , but marginal for \tover\ and $\lambda$, which therefore might be slightly biased because of the insufficient spatial resolution \citep{Kruijssen2019}. Indeed, the relative filling factor of the emission peaks ($\zeta_{\rm star} = 2 r_{\rm star}/\lambda$ and $\zeta_{\rm gas} = 2 r_{\rm gas}/\lambda$, where $r_{\rm star}$ and $r_{\rm gas}$ are the average radii of the emission peaks in the \Ha\ and the CO maps respectively) is high ($\sim 0.6$).}

\change{The above indications of region blending imply that our measurement of \tover\ may be overestimated (see sect.~4.4 of \citealt{Kruijssen2018}). Given that $t_{\rm fb}=1.7_{-0.7}^{+0.9}~\myr$ is already considerably shorter than the minimum timescale for supernova feedback \citep[$\sim4~\myr$, e.g.][]{Leitherer2014}, the fact that it is an upper limit implies that cloud destruction is likely driven by early feedback mechanisms in FCC290, such as photoionisation or stellar winds. This is consistent with the results found in other galaxies with similarly low molecular gas surface densities, such as NGC300 \citep[$t_{\rm fb}=1.5_{-0.2}^{+0.2}~\myr$,][]{Kruijssen2019} and NGC5068 \citep[$t_{\rm fb}=1.0_{-0.3}^{+0.4}~\myr$,][]{Chevance2020}.}

\section{Discussion}
\label{sec:discussion}

\subsection{The star formation relation in the Fornax cluster}
From Figure \ref{fig:examples}, Figure \ref{fig:combinedKS}, and the figures in Appendix \ref{app:SFandDT}, we can conclude that depletion times of most galaxies in the sample lie close to the K98 and B08 relations. The star formation relation found in the Fornax cluster mostly overlaps with those found by K98 and B08, \refrep{but is, with few exceptions, shifted towards the short-depletion time end of the B08 relation. This is noteworthy, as the study by B08 is more similar to ours than that by K98 in terms of sample, resolution, and use of molecular gas (see \S \ref{sec:introduction}).} There are some observational differences that can contribute to this difference. First, we use CO(1-0) as a tracer for the molecular gas rather than CO(2-1), used for most of the B08 sample. This can affect the molecular gas measurements, for example through the assumption of a line ratio (e.g. \citealt{Rahman2010}). Second, the CO-to-H$_2$ conversion factor can make a difference in the star formation relation. Here we use a conversion factor that varies with stellar mass/metallicity (\S \ref{sub:SDmaps}). However, this mostly results in an increase of the conversion factor for the low-metallicity dwarf galaxies. Therefore, using a constant X$_\text{CO}$ would give a lower \H2dens\ for these dwarfs, only moving them further away from the K98 and B08 relations. Third, we use \Ha\ as a star formation tracer rather than UV + IR, used by B08. This traces the SFR on shorter timescales ($\sim$10 Myr rather than $\sim$100 Myr), which can also affect the star formation relation. Another factor is the increased number of early-type galaxies in our sample compared to the field samples used by B08 (and K98), whose star formation is often dynamically suppressed (e.g. \citealt{Davis2014}). Last, it has been shown by various authors (e.g. \citealt{Schruba2010, Liu2011, Calzetti2012, Williams2018}) that the index of the star formation relation can vary significantly depending on the resolution of the observations. While both B08 and this work examine the relation on sub-kpc scales, the resolution used by B08 is still a factor $\sim$2 worse than the resolution used here, which could play a role (K98 used integrated measurements for entire galaxies only). However, the integrated depletion times in Figures \ref{fig:DT_SM}, \ref{fig:DT_SB}, and \ref{fig:DT_distance} correspond very well with the densest areas in the star formation relations of the individual galaxies (see Appendix \ref{app:SFandDT}), suggesting this does probably not play a significant role.


\refrep{Galaxies that have depletion times significantly shorter than predicted by the K98 and B08 relations are the dwarf galaxies FCC090 and FCC263, and the more massive, edge-on spiral FCC312. Galaxies with significantly longer depletion times are the dwarf FCC207 and the lenticular FCC167. FCC184 has slightly increased depletion times of several Gyrs.} Figure \ref{fig:DT_SM} shows that most of these galaxies still have depletion times within 1 $\sigma$ of what is expected based on their stellar mass. However, this figure is somewhat misleading at the low-mass end due to the low number statistics, and in reality there are probably more dwarf galaxies with increased than decreased depletion times. Indeed, \citet{Reyes2019} have shown that dwarf galaxies in the field have longer depletion times than spirals, and therefore lie below the K98 relation. They also note that second-order correlations with gas fraction, SMSD, and dynamical timescales can be important in dwarfs. In Paper I we have seen that dwarf galaxies in Fornax have molecular gas fractions around an order of magnitude smaller than expected based on observations of field galaxies due to environmental processes, which could play a role here. All three dwarf galaxies have disturbed molecular gas reservoirs, so their star formation rates are likely enhanced by the tidal interactions or ram pressure that cause these disturbances and compress the molecular gas. Ram pressure is very effective at stripping the gas from dwarf galaxies (e.g. \citealt{Venhola2019}). \refrep{The depletion times of these dwarfs are longer than the time it will take them to cross the cluster core (and therefore reach the peak of their gas stripping; e.g. \citealt{Boselli2014}). They are therefore likely to be influenced by RPS before depleting their gas reservoirs.} Tidal disruptions, on the other hand, should be less important (if they are as dark matter dominated as similar galaxies in the field), therefore favouring the latter as an explanation. It could be that these galaxies have unusual chemical features as a result of their bursty star formation histories, as seen for example in \cite{Pinna2019a, Pinna2019b}. This will be explored further in future work.

\change{With the exception of FCC263, which is closer (in projection) to the cluster centre than the other two, galaxies with very short depletion times are located around the virial radius of the cluster (Figure \ref{fig:DT_distance}). This suggests that they are on their first infall into the cluster, their gas being compressed and stripped as they fall in. This explanation is supported by the observation that the galaxies with the shortest depletion times are less H$_2$ deficient than the ones with longer depletion times (Table \ref{tab:targets}), suggesting that they have not yet spent much time in the cluster. There is, however, one other galaxy in the sample with a disturbed molecular gas reservoir, FCC207, that has increased rather than decreased depletion times (\S \ref{sub:FCC207}). In projection (and redshift), FCC207 is the galaxy closest to the BCG in the AlFoCS sample. It is possible that it has passed the slight starburst phase during its first infall. However, although it is more H$_2$ deficient than the other dwarfs, the amount of gas present that has had to have survived the infall in this case (Table \ref{tab:targets}, Table 3 in Paper I), makes this explanation unlikely. The other possibility is that it is at the edge of the cluster along the line of sight, in which case we underestimate its distance from the BCG. In this case, it is possible that the starburst has not yet started. In Figure \ref{fig:FCC207} we can see a strong gradient in depletion times across the galaxy, suggesting that the gas is compressed on one side due to ram pressure stripping or tidal effects. This supports the idea that the galaxy is on its first infall. A future study of the stellar population content from the MUSE data could help distinguish between both scenarios.}

\change{The result that infalling objects in Fornax have relatively short depletion times is somewhat different from what has previously been found in several studies of the Virgo cluster, which concluded that RPS and interacting galaxies here do not have significantly different depletion times \citep{Vollmer2012, Nehlig2016}. Other studies of the Virgo cluster did find, on the other hand, locally increased SFRs in RPS galaxies \citep{Lee2017}. Perhaps the differences with studies of the Virgo cluster originate from a difference in sample, as most studies of the Virgo cluster were performed using observations of higher-mass spiral galaxies, whereas the shorter depletion times in the Fornax cluster are largely driven by dwarf galaxies, which are more susceptible to environmental effects (see also Paper I).}

\subsection{The star formation relation in individual galaxies}
\change{There is significant scatter in the star formation relation of most individual galaxies (Figure \ref{fig:examples}, the figures in Appendix \ref{app:SFandDT}). This is partly due to the statistical nature of these figures (\S \ref{sec:methods}), especially in case of the more extreme scatter and galaxies with small angular sizes. In some cases (e.g. FCC090, \S \ref{sub:FCC90} and FCC207, \S \ref{sub:FCC207}) it is due to larger scale variations within the galaxy, as a result of the stripping and compression of the gas. Other, small scale variations (as can be seen nicely in FCC290, \S \ref{sub:FCC290}, see also \S \ref{sec:cloud_lifetimes}, and FCC312, \S \ref{sub:FCC312}) are also seen in other studies (e.g. \citealt{Bigiel2008, Blanc2009, Onodera2010, Leroy2013}), and are generally contributed to individual regions in galaxies undergoing independent evolutionary lifecycles of cloud assembly, collapse, star formation, and disruption by feedback (e.g. \citealt{Schruba2010, Feldmann2011, KL14}). This small scale scatter can therefore be directly related to the underlying evolutionary timeline of the above processes see also \S \ref{sec:cloud_lifetimes}.}

\subsection{Depletion time as a function of galaxy parameters}
\subsubsection{Relation with stellar mass}
\secround{In the xCOLD GASS sample}, there is a shallow trend between galaxies' depletion times and stellar masses (the Kendall's tau value is $\sim$0.21 with a corresponding p-value of $\sim4.7 \times 10^{-7}$, Figure \ref{fig:DT_SM}). Fornax galaxies follow a similar trend that is possibly slightly steeper, with dwarf galaxies having decreased and more massive galaxies having increased depletion times. However, this partly depends on the true values of the depletion times of FCC179 and FCC184, depending on whether composite \Ha\ or only \Ha\ classified as star forming is taken into account. Moreover, the numbers are small. If FCC184 does indeed have increased depletion times compared to the field, this can be explained by the fact that early-type galaxies in clusters tend to be very regular and relaxed, likely because they have not recently experienced any mergers. Regular and relaxed gas reservoirs are more susceptible to dynamical effects caused by the deep potential, which some studies have suggested suppress star formation \citep{Martig2009, Martig2013, Davis2014, Gensior2019}. A version of this Figure with Fornax and Virgo galaxies shown only is provided in Appendix \ref{app:virgo}.

\subsubsection{Relation with stellar mass surface density}
If we look at depletion times as a function of SMSD rather than stellar mass (Figure \ref{fig:DT_SB}), we can see there is a similar weak trend in the xCOLD GASS data (the Kendall's tau value is $\sim$0.23 with a corresponding p-value of $\sim8.5 \times 10^{-8}$) between these parameters in the field. Most Fornax galaxies are close or within 1 $\sigma$ of this \refrep{field} relation. FCC090, which follows the trend with stellar mass in Figure \ref{fig:DT_SM}, is below 1 $\sigma$ in this relation. This is likely because its elliptical morphology predicts longer depletion times. However, due to its low mass and shallow potential well, it is still more easily affected by the cluster environment. Its depletion time is therefore shorter than predicted by its SMSD. For this reason, galaxy stellar mass appears to be a better predictor of depletion times in the cluster than galaxy morphology as probed by their SMSD.

\subsubsection{Relation with cluster-centric distance}
\change{Figure \ref{fig:DT_distance} shows a clear trend between galaxy depletion times and their distance from the cluster centre. Since massive ellipticals tend to be located closer to the cluster centre than less massive spirals, this trend is likely at least partly driven by galaxy mass/morphology rather than distance from the cluster centre per se. In fact, if we compare this figure with Figure \ref{fig:DT_SM}, the weak correlation between depletion time and stellar mass indeed seems mostly anti-correlated with the trend with radial distance. Galaxies that do not follow this anti-correlation are FCC207, FCC263, and FCC312. The distance to FCC207 is likely underestimated as a result of the projection, as we would not expect it to have any gas left if it was actually only $\sim$100 kpc away from the cluster centre. The same is possibly true for FCC263. FCC263 does show evidence of either a merger or a strong bar, both of which drive gas to the centre and increase the gas (surface) density, thus decreasing depletion times. FCC312 is located close to the virial radius (in projection). Like the dwarf galaxies with short depletion times, it is likely on its first infall into the cluster, undergoing a slight starburst. Based on the location of FCC312, and that of FCC207 if it is accurate, the location of a galaxy within the cluster possibly affects its depletion times in addition to its mass (and morphology), with galaxies close to the virial radius, on their first infall, having relatively short depletion times. However, it is difficult to draw conclusions with the poor statistics and lower limits on the distance from the BCG. Improved statistics and/or simulations are needed in order to break the degeneracy between galaxy morphology and location in the cluster, and see whether this relation holds independent of galaxy morphology.}

\subsection{Star formation in stripped material}
FCC90 is a dwarf elliptical galaxy with a tail of molecular gas extending beyond its stellar body (Figure \ref{fig:examples} and \S \ref{sub:FCC90}). There is a clear difference between depletion times in the galaxy's stellar body and in the tail, with a sharp transition. Depletion times in the galaxy body are low, around 0.5 Gyr. Depletion times in the tail are much higher, $>$10 times that in the stellar body. This is in agreement with what is observed in RPS galaxies, for example by \citet{Ramatsoku2019}, who find that depletion times in the tail of the JO206 RPS galaxy are $\sim$10 times smaller than those in its disk. \change{While this work studies \HI\ rather than H$_2$, and overall \HI\ depletion times are shorter than the H$_2$ depletion times found here, the clear split in \DT\ between the stellar body and the gas tail, differing by a factor $\sim$10 in both works, is in good agreement.} This clear split in depletion times was also observed by \citet{Moretti2018, Moretti2019} in several GASP galaxies. \citet{Jachym2019} find that molecular gas directly stripped from the Norma jellyfish galaxy (ESO 137-001) does not form stars. 

\subsection{Sample bias}
\change{As described in \S \ref{sub:sample}, not all galaxies from Paper I are represented in this work. With the exception of the massive spiral galaxy FCC121 (NGC1365), galaxies that were detected in Paper I, but not studied here, are dwarf galaxies with disturbed molecular gas reservoirs, and a spiral galaxy close to the virial radius. As discussed above, the slightly decreased depletion times observed in the Fornax cluster are driven by disturbed dwarf galaxies and the spiral galaxy FCC312, which is also close to the virial radius and likely experiencing a slight starburst on its infall. If MUSE data were obtained such that all these additional objects from Paper I could be included in this study, this would double the number of disturbed dwarf galaxies and add another spiral galaxy close to the virial radius. If these galaxies are similar to the ones studied here, this would move the star formation relation in the Fornax cluster further away from the K98 and B08 relations. This would strengthen the conclusion that the effects of the cluster environment on the molecular gas content of galaxies result in decreased depletion times compared to field galaxies, likely due to the compression of molecular gas as a result of environmental effects.}

\subsection{\refrep{The inclusion of composite \Ha}}
\refrep{Comparing the cases using \Ha\ from regions dominated by star formation ionisation, and also including \Ha\ from regions with composite ionisation, only significantly changes the star formation relation for a few galaxies. These are FCC179, FCC184 (and FCC167, which has no \Ha\ that is dominated by star formation only, according to a BPT classification). Both FCC179 and FCC184 harbour AGN, which contributes to the ionised gas emission around the galactic centre. In our star forming \Ha\ versus star forming + composite \Ha\ comparison, the star formation relations of these objects both change visibly (Figures \ref{fig:FCC179} and \ref{fig:FCC184}), though more significantly in the case of FCC184. Its relation between molecular gas surface density and star formation surface density becomes tighter in if we include composite spaxels, and its average depletion time decreases significantly (Figures \ref{fig:DT_SM}, \ref{fig:DT_SB}, and \ref{fig:DT_distance}). This suggests that while the AGN (and/or shocks/old stars) dominates the production of higher ionisation state lines such as [OIII]$\lambda 5007$, it does not contribute significant \Ha\ emission in the composite emission-line regions of this galaxy.}

\subsection{Spiral arm vs. interarm regions}
FCC179 is a large spiral galaxy (\S \ref{sub:FCC179}, Figure \ref{fig:FCC179}). From Figure \ref{subfig:DT_FCC179_SF_comp} it appears that it is undergoing ram pressure stripping, with both molecular gas and ionised gas being removed from its east side. Furthermore, depletion times trace the spiral arms in this figure, which have more efficient star formation (\DT $<$2 Gyr) than the interarm regions ($\sim$3 - 4 Gyr). \change{This implies that star formation in spiral arms is more efficient than in interarm regions, which is in line with our current understanding of spiral galaxies (e.g. \citealt{Elmegreen2011}).}

\section{Summary}
\label{sec:conclusions}
We study the resolved star formation relation in the Fornax cluster. CO is used as a tracer for molecular gas, and Balmer decrement corrected \Ha\ as a tracer for the star formation rates. Resolved maps are obtained from ALMA and MUSE observations, respectively. The resulting surface density maps are binned to the ALMA beam size using a flexible binning method, for which we use a grid that we repeatedly move by one pixel until all possible positions are covered. The result is a set of different possible (star formation and molecular gas) surface density maps for each galaxy, depending on the position of the grid. We combine the resulting maps into one star formation relation for each galaxy. Depletion time maps of the individual galaxies are obtained using the median gas and SFR surface densities of the possible maps in each pixel. We create two versions of each map and and \KS\ relation: one where only the star forming \Ha\ is taken into account according to a BPT classification, and one where \refrep{composite} \Ha\ is also considered. Besides star formation plots for each individual galaxy, we combine the possible values for each resolution element in the sample into one \KS\ relation. Results are compared with the relations found by \citet{Kennicutt1998} and \citet{Bigiel2008}. We also show galaxy-average depletion times as a function of stellar mass, stellar mass surface brightness, and distance to the cluster centre, and compare the results with similar relations for field galaxies from xCOLD GASS \citep{Saintonge2017}, ALLSMOG \citep{Cicone2017}, \refrep{the HRS \citep{Boselli2010,Boselli2014b}}, and \atlas3d\ \citep{Cappellari2011,Young2011}. Our main conclusions are as follows:

\begin{itemize}
\item \refrep{The overall star formation relation in the Fornax cluster is close to those found by K98 and B08, but lies towards the short-depletion time end of the B08 relation. This is mostly driven by dwarf galaxies with disturbed molecular gas reservoirs, although there are some observational differences between the studies that possibly contribute to this difference.}

\item Individual Fornax cluster galaxies lie in different locations on the star formation relation plane, their combined depletion times covering more than two orders of magnitude. \change{Most dwarf galaxies (with exception of FCC207, which has significantly longer depletion times than expected) and other galaxies close to the virial radius have depletion times shorter than the typical value of 1-2 Gyr for spiral field galaxies.}

\item Most galaxies exhibit significant scatter in their \KS\ relations. \change{Although some of this is due to edge-effects and the statistical nature of the relations shown, it also indicates a variety of depletion times on sub-galactic scales.} Depending on the galaxy, this scatter can be mostly towards either lower or higher depletion times, or both equally. \change{For some galaxies this is mostly the results of larger scale variations or gradients in depletion time (e.g. FCC090 and FCC207), in others it reflects small scale variations as a result of independent evolutionary lifecycles of clouds (e.g. \citealt{Schruba2010, Feldmann2011, KL14}).} FCC290, for example, an almost face-on flocculent spiral, shows the largest and most symmetric scatter. This reveals the spatial offsets between H{\sc i}{\sc i} regions and giant molecular clouds, \change{and indicates the time evolution of independent regions}. \change{We have applied the `uncertainty principle for star formation' to it, and find that it has relatively short molecular cloud lifetimes of $t_{\rm CO}=9.3_{-2.2}^{+2.6}~\myr$. Such short cloud lifetimes are typically found for low-mass galaxies with low molecular gas surface densities \citep{Chevance2020}, similar to that measured in FCC290, despite a relatively high total gas mass. At such low surface densities the cloud lifetime is regulated by internal dynamical processes rather than galactic dynamics, resulting in relatively short cloud lifetimes. The characteristic distance $\lambda \sim 300$ pc between individual regions is quite large in FCC290 compared to $\lambda \sim 100 - 250$ pc in similar galaxies in \citealt{Chevance2020}. This could be a result of the cluster environment.}

\item Fornax galaxies appear to follow the shallow trend in depletion time with stellar mass seen in the field, possibly with somewhat more extreme values at the low and high mass ends, although statistics are poor. At the lower mass end of the distribution this could be explained by the dwarf galaxies in the cluster being more susceptible to effects from the cluster environment, due to their shallow potentials. The increased depletion times observed at the higher mass end are likely due to the sample being limited to massive early-types here. \refrep{The HRS, which consists of spiral galaxies, and includes galaxies in the Virgo cluster, shows a spread in depletion times at the higher-mass end (see Appendix \ref{app:virgo} for a version of this Figure showing cluster galaxies exclusively).}

\item The same low-mass galaxies are outliers if we look at depletion times as a function of stellar mass surface density (defined here as 0.5 $M_\odot / (\pi R_e^2)$), which can be interpreted as a proxy for galaxy morphology. However, since some low-mass galaxies are dwarf ellipticals, and some higher-mass galaxies are spirals, these outliers are no longer at the extremes of the distribution, and therefore there is a weaker trend with SMSD than with stellar mass. Therefore, stellar mass appears to be a better predictor of galaxy depletion times, especially in clusters, where lower-mass galaxies are more susceptible to environmental effects decreasing their depletion times.

\item There is a relatively clear correlation between median depletion time and the (projected) distance between a galaxy and the cluster centre. It is, however, likely that this trend is (partly) caused by mass/morphological type rather than the distance itself. Massive ellipticals tend to be located in the cluster centre, and lower mass galaxies, such as spirals and dwarfs, are typically located further out. In order to break the degeneracy between these two variables, we need improved statistics and/or simulations.

\item The dwarf galaxy FCC090 has a large molecular gas tail, caused by either RPS or a tidal interaction. A similar though slightly offset tail is seen in the ionised gas. A starburst is likely taking place in its stellar body as it is falling into the cluster: depletion times there lie around $\sim$0.5 Gyr rather than the usual 1-2 Gyr. Depletion times in its gas tail, however, are much larger, typically around 10 times those in the galaxy body, or longer. This is in line with results from \citet{Moretti2018, Moretti2019, Ramatsoku2019, Jachym2019}.
\end{itemize}

\noindent In conclusion, the star formation relation in the Fornax cluster is close to the ``traditional'' ones (e.g. from K98 and B08), but with slightly decreased depletion times. This is mainly driven by dwarf galaxies with disturbed molecular gas reservoirs and galaxies possibly undergoing starbursts on their first infall into the cluster.

\section*{Acknowledgements}
NZ acknowledges support from the European Research Council (ERC) in the form of Consolidator Grant CosmicDust (ERC-2014-CoG-647939).

TAD acknowledges support from the Science and Technology Facilities Council through grant ST/S00033X/1.

MC and JMDK gratefully acknowledge funding from the Deutsche Forschungsgemeinschaft (DFG) through an Emmy Noether Research Group (grant number KR4801/1-1) and the DFG Sachbeihilfe (grant number KR4801/2-1). JMDK gratefully acknowledges funding from the European Research Council (ERC) under the European Union's Horizon 2020 research and innovation programme via the ERC Starting Grant MUSTANG (grant agreement number 714907).

IDL gratefully acknowledges the support of the Research Foundation Flanders (FWO).

EMC is supported by MIUR grant PRIN 2017 20173ML3WW\_001 and by Padua University grants DOR1715817/17, DOR1885254/18, and DOR1935272/19.

The National Radio Astronomy Observatory is a facility of the National Science Foundation operated under cooperative agreement by Associated Universities, Inc.

This paper makes use of the following ALMA data: ADS/JAO.ALMA\#2015.1.00497.S. ALMA is a partnership of ESO (representing its member states), NSF (USA) and NINS (Japan), together with NRC (Canada), MOST and ASIAA (Taiwan), and KASI (Republic of Korea), in cooperation with the Republic of Chile. The Joint ALMA Observatory is operated by ESO, AUI/NRAO and NAOJ.

This research has made use of the services of the ESO Science Archive Facility, in particular observations collected at the European Southern Observatory under ESO programmes 098.B-0239, 094.B-0576, 097.B-0761, and 096.B-0063.

This publication makes use of data products from the Wide-field Infrared Survey Explorer, which is a joint project of the University of California, Los Angeles, and the Jet Propulsion Laboratory/California Institute of Technology, funded by the National Aeronautics and Space Administration.

\refrep{This research has made use of data from HRS project. HRS is a Herschel Key Programme utilising Guaranteed Time from the SPIRE instrument team, ESAC scientists and a mission scientist.}

\refrep{The HRS data was accessed through the Herschel Database in Marseille (HeDaM - http://hedam.lam.fr) operated by CeSAM and hosted by the Laboratoire d'Astrophysique de Marseille.}

This research made use of Astropy,\footnote{http://www.astropy.org} a community-developed core Python package for Astronomy \citep{Astropy2013, Astropy2018}.

This research made use of astrodendro, a Python package to compute dendrograms of Astronomical data (http://www.dendrograms.org/).

This research made use of APLpy, an open-source plotting package for Python \citep{Robitaille2012}.

This research has made use of the NASA/IPAC Extragalactic Database (NED), which is operated by the Jet Propulsion Laboratory, California Institute of Technology, under contract with the National Aeronautics and Space Administration.




\bibliographystyle{mnras}
\bibliography{References} 



\appendix

\section{Comparison with the Virgo cluster}
\label{app:virgo}

\begin{figure*}

	\centering

	\hspace{-0mm}  \subfloat[\label{subfig:SM_clusters}]
	{\includegraphics[height=0.5\textwidth, width=0.7\textwidth]{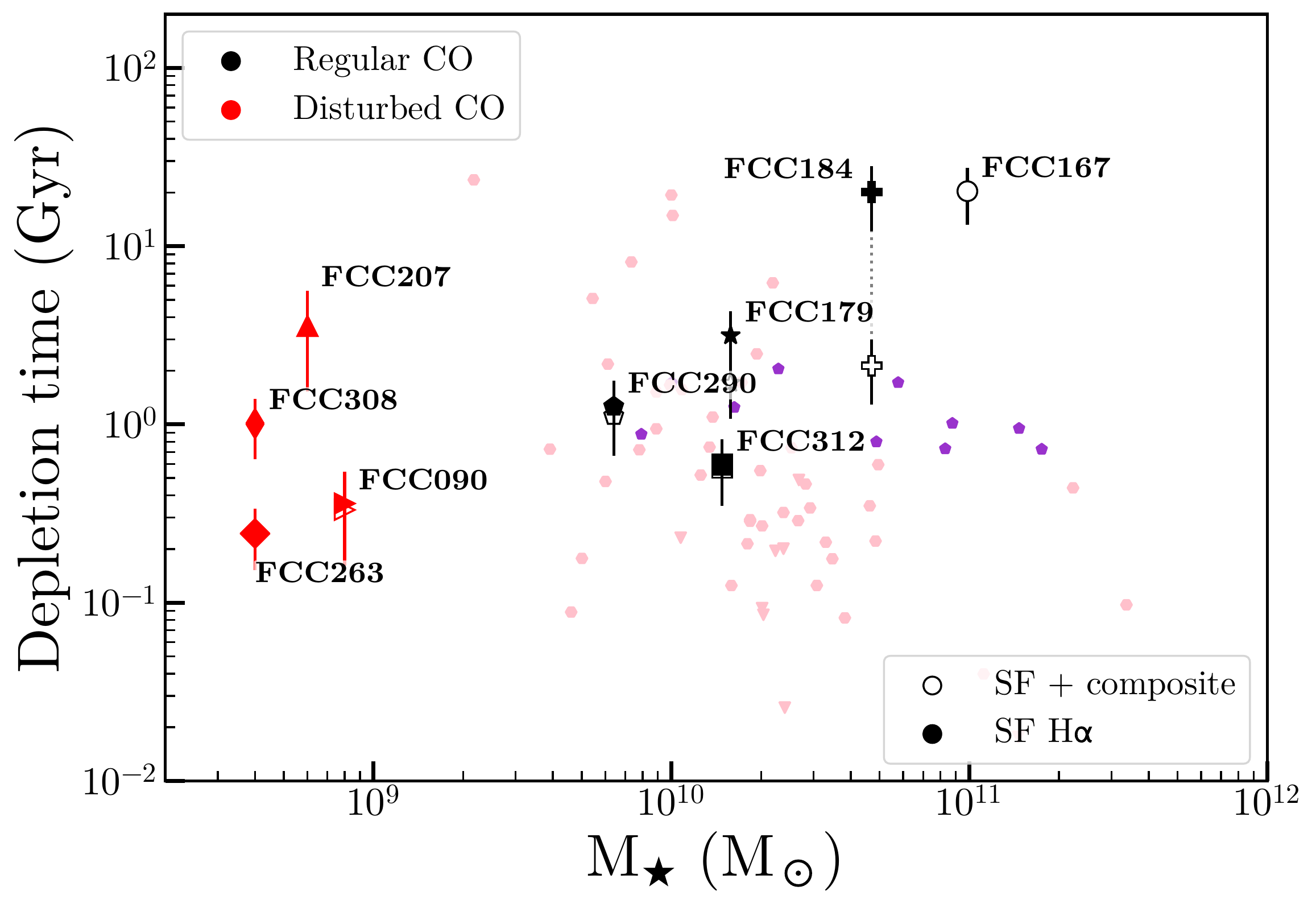}}	
	\caption{Similar to Figure \ref{fig:DT_SM}, but with data from the Virgo cluster only. Purple markers are ETGs from \atlas3d, and pink markers are galaxies from the HRS. High-mass Fornax galaxies have long depletion times compared to these samples (especially the HRS), probably because they are ETGs, while the HRS consists of spiral galaxies only. The low-mass end of the HRS agrees well with the Fornax data, both show a spread in depletion times here.}
	\label{fig:clusters_only}	
\end{figure*}

\section{SF plots, \DT\ maps, and discussion of individual galaxies}
\label{app:SFandDT}

\subsection{FCC90}
\label{sub:FCC90}
FCC90 (MCG-06-08-024) is a dwarf elliptical galaxy located (in projection) close to the virial radius. It has a molecular gas tail extending beyond its stellar body, either caused by RPS or by a tidal interaction. The ionised gas also shows a tail in the same location, but pointed in a slightly different direction. Depletion times in the galaxy body are significantly shorter than the ``normal'' 1-2 Gyr for nearby spiral galaxies, and are distributed around $\sim$0.5 Gyr instead. In the gas tail, on the other hand, depletion times are $\gtrsim10$ times longer. This is in agreement with what is found in other gas tails caused by RPS \citep{Moretti2018, Ramatsoku2019, Jachym2019}.

\begin{figure*}

	\centering

	\hspace{-0mm}  \subfloat[\label{subfig:KS_FCC090}]
	{\includegraphics[height=0.42\textwidth, width=0.44\textwidth]{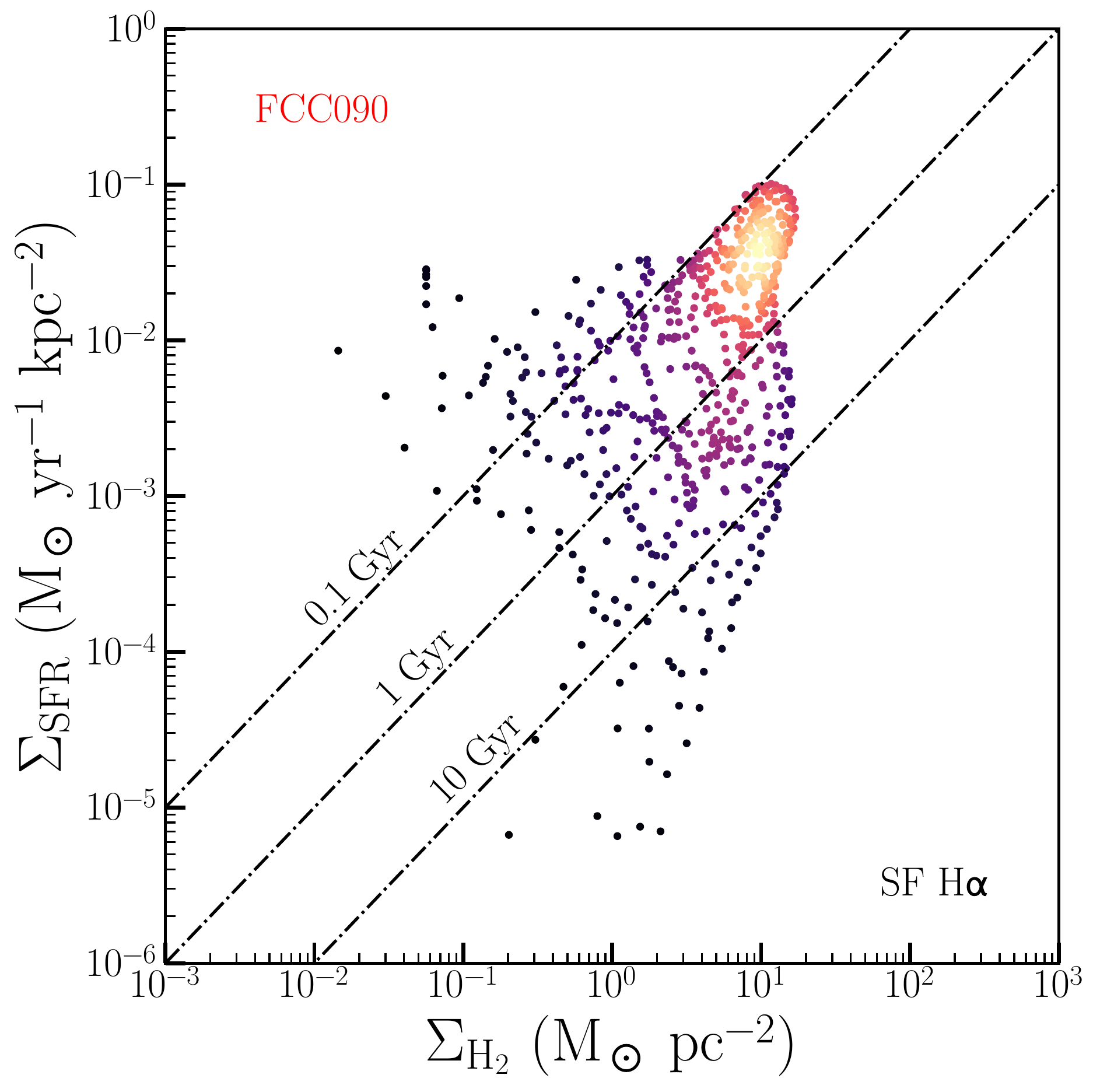}}	\hspace{13mm}
	\subfloat[\label{subfig:KS_FCC090_SF_comp}]
	{\includegraphics[height=0.42\textwidth, width=0.42\textwidth]{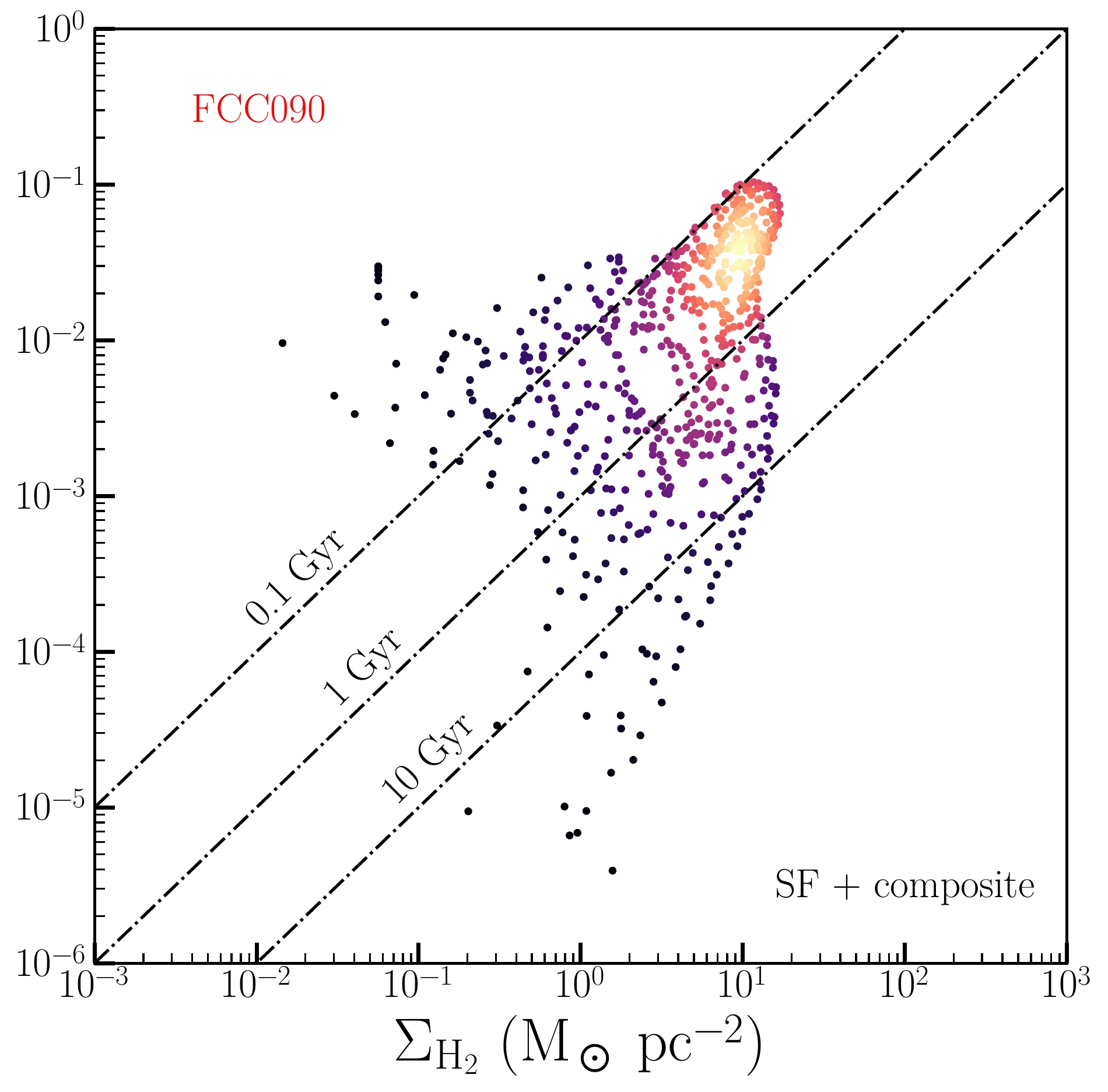}}	
	
	\subfloat[\label{subfig:DT_FCC090}]
	{\includegraphics[height=0.42\textwidth, width=0.47\textwidth]{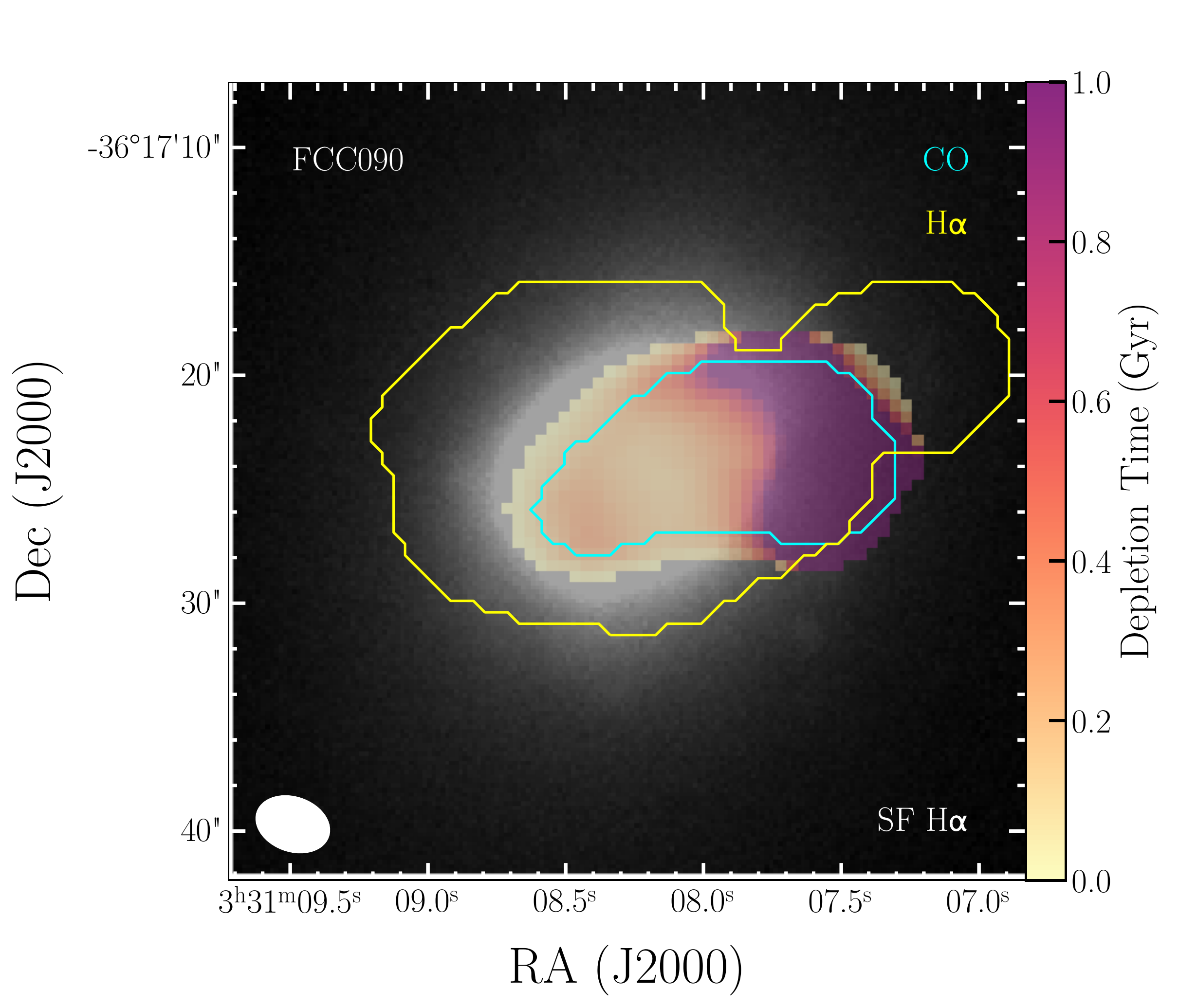}} \hspace{6mm}
	\subfloat[\label{subfig:DT_FCC090_SF_comp}]
	{\includegraphics[height=0.42\textwidth, width=0.47\textwidth]{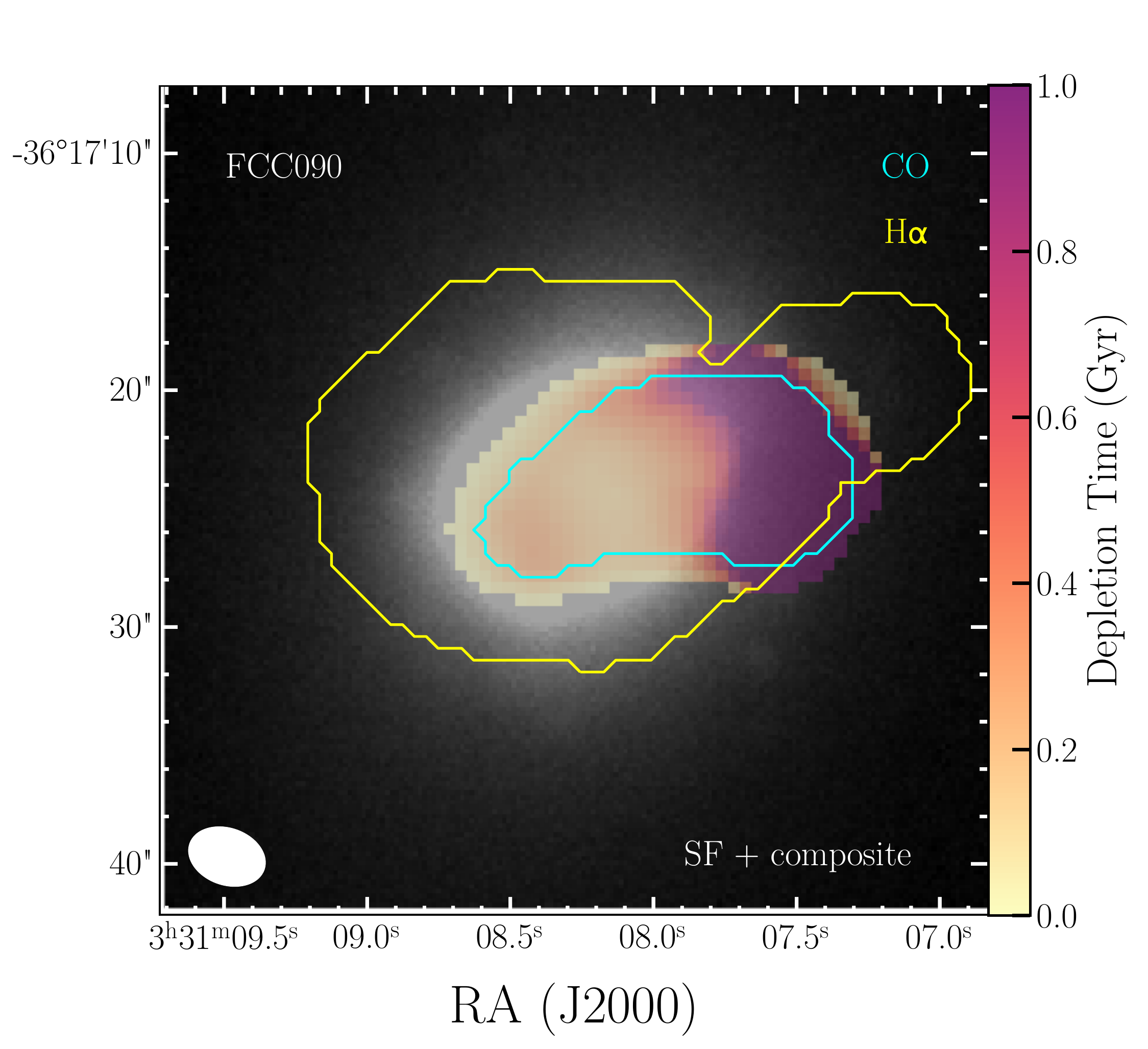}}	
	
	\caption{\textbf{Top row}: Relation between the molecular gas and star formation rate surface densities for FCC90. The colours of the markers indicate the density around the data point in this plane, determined using a Gaussian kernel density estimation. Lines of constant depletion times (0.1, 1, and 10 Gyr) are shown. The galaxy name is indicated in the upper left corner, and whether composite \Ha\ or only the star forming \Ha\ is used is indicated in the lower right corner. FCC90 deviates from the ``standard'' H$_2$ depletion time of 1-2 Gyr \refrep{often found in field galaxies (e.g. \citealt{Leroy2008}, \citealt{Saintonge2012}, Figure \ref{fig:DT_SM})} by a factor 5-10.
	\textbf{Bottom row}: depletion time maps of FCC90. Maps were made both taking into account composite and star forming \Ha\ (right column) and considering only purely star forming \Ha\ (left column). Whether composite \Ha\ or only the star forming \Ha\ is considered is indicated in the lower right corner. The maps are overplotted on optical (\textit{g}-band) images from the Fornax Deep Survey (FDS, \citealt{Iodice2016,Iodice2017, Venhola2017,Venhola2018}, Peletier et al., in prep.). The extent of the CO emission is indicated with a cyan contour, and that of the \Ha\ emission with a yellow contour (both clipped at the 3$\sigma$ level). The beam of the ALMA observations is shown in the lower left corner, and the FCC number of the galaxy is indicated in the upper left corners. Depletion times are short in the galaxy's body ($<$ 0.5 Gyr), but long in the gas tail ($\gg$2 Gyr).}
	
	\label{fig:FCC90}

\end{figure*}

\subsection{FCC167}
\label{sub:FCC167}
FCC167 (NGC1380) is an early-type galaxy with a nested ISM, consisting of a molecular gas ring/disk located inside a dust lane. Ionised gas is present throughout the galaxy \citep{Viaene2019}. According to our BPT classification, no star forming \Ha\ is present in this galaxy. The majority of the pixels are in the ``composite'' area of the diagram. The ionised gas forms a tail at the south side of the galaxy, likely because of RPS or an ongoing tidal interaction. Depletion times are mostly around 1-2 Gyr, although there is some scatter towards shorter depletion times.

\begin{figure*}

	\centering

	\subfloat[\label{subfig:KS_FCC167_SF_comp}]
	{\includegraphics[height=0.4\textwidth, width=0.4\textwidth]{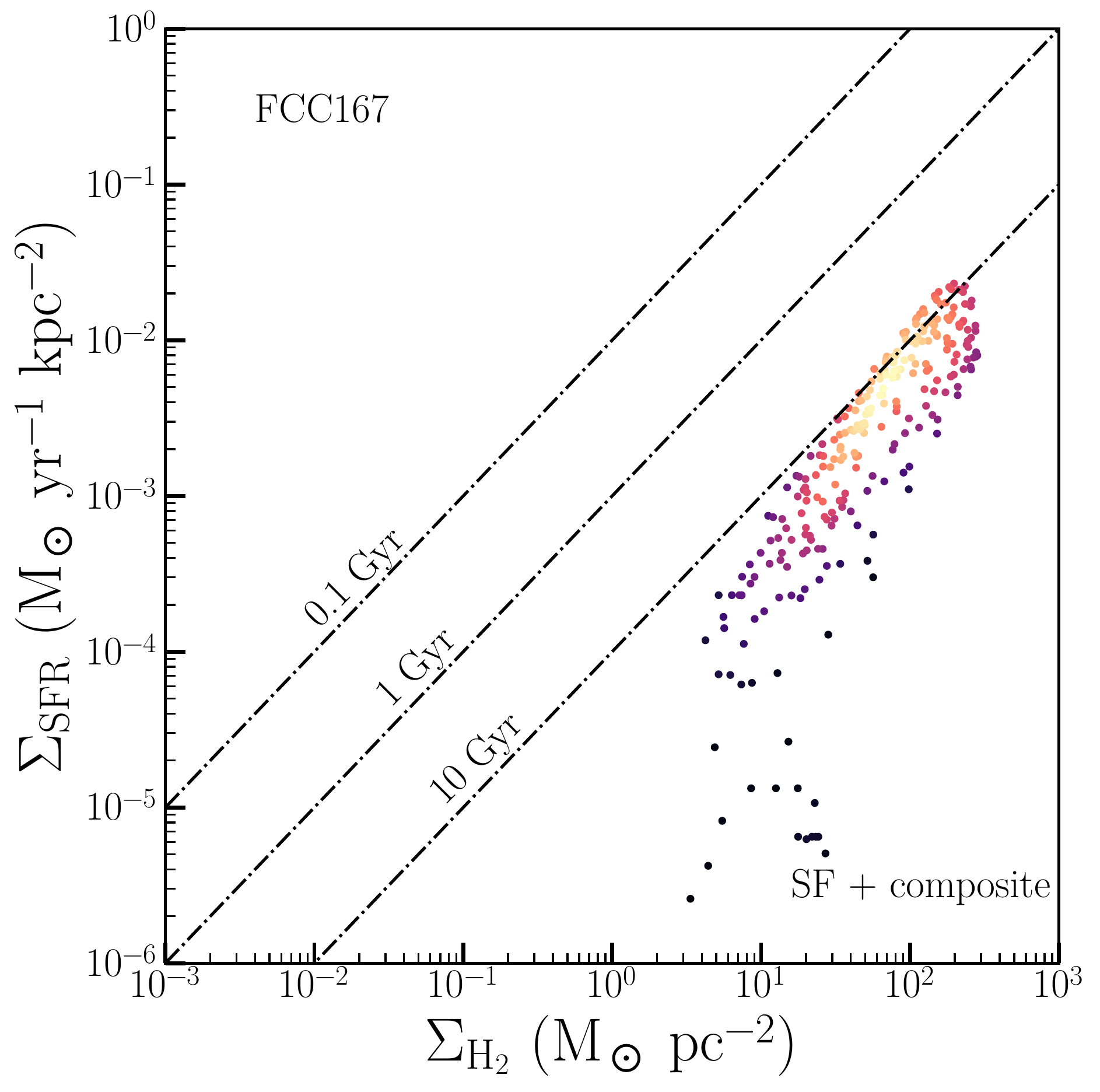}}	
	
		\subfloat[\label{subfig:DT_FCC167}]
	{\includegraphics[height=0.4\textwidth]{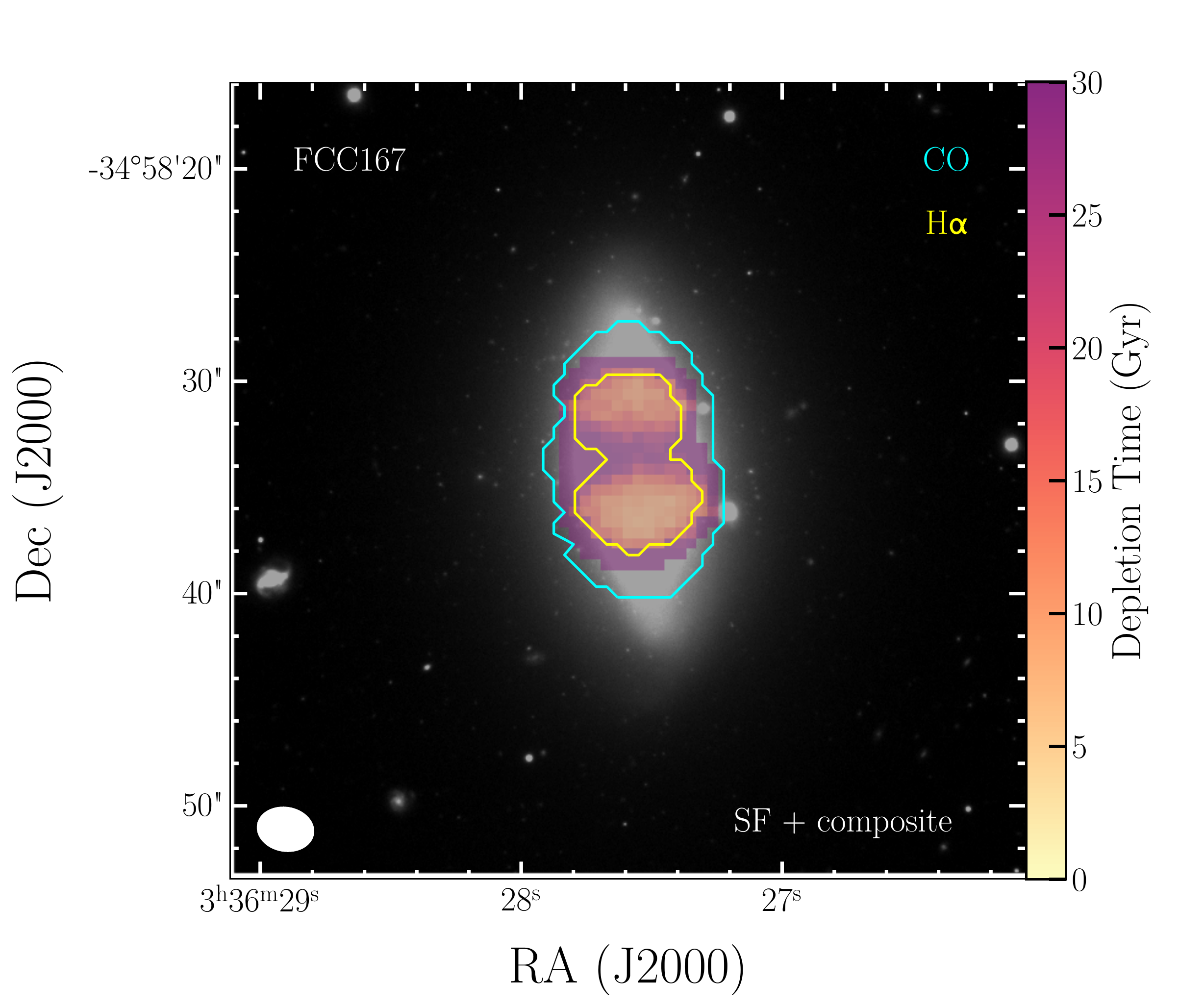}}	
	
	\caption{Similar to Figure \ref{fig:FCC90}, but for FCC167. FCC167 has no star forming \Ha, so only one (equivalent to the right column in Figure \ref{fig:FCC90}) column is shown. \refrep{Depletion times are long, significantly longer than 10 Gyr in almost all regions.}} 
	\label{fig:FCC167}

\end{figure*}

\subsection{FCC179}
\label{sub:FCC179}
\begin{figure*}

	\centering

	\hspace{-0mm}  \subfloat[\label{subfig:KS_FCC179}]
	{\includegraphics[height=0.42\textwidth, width=0.44\textwidth]{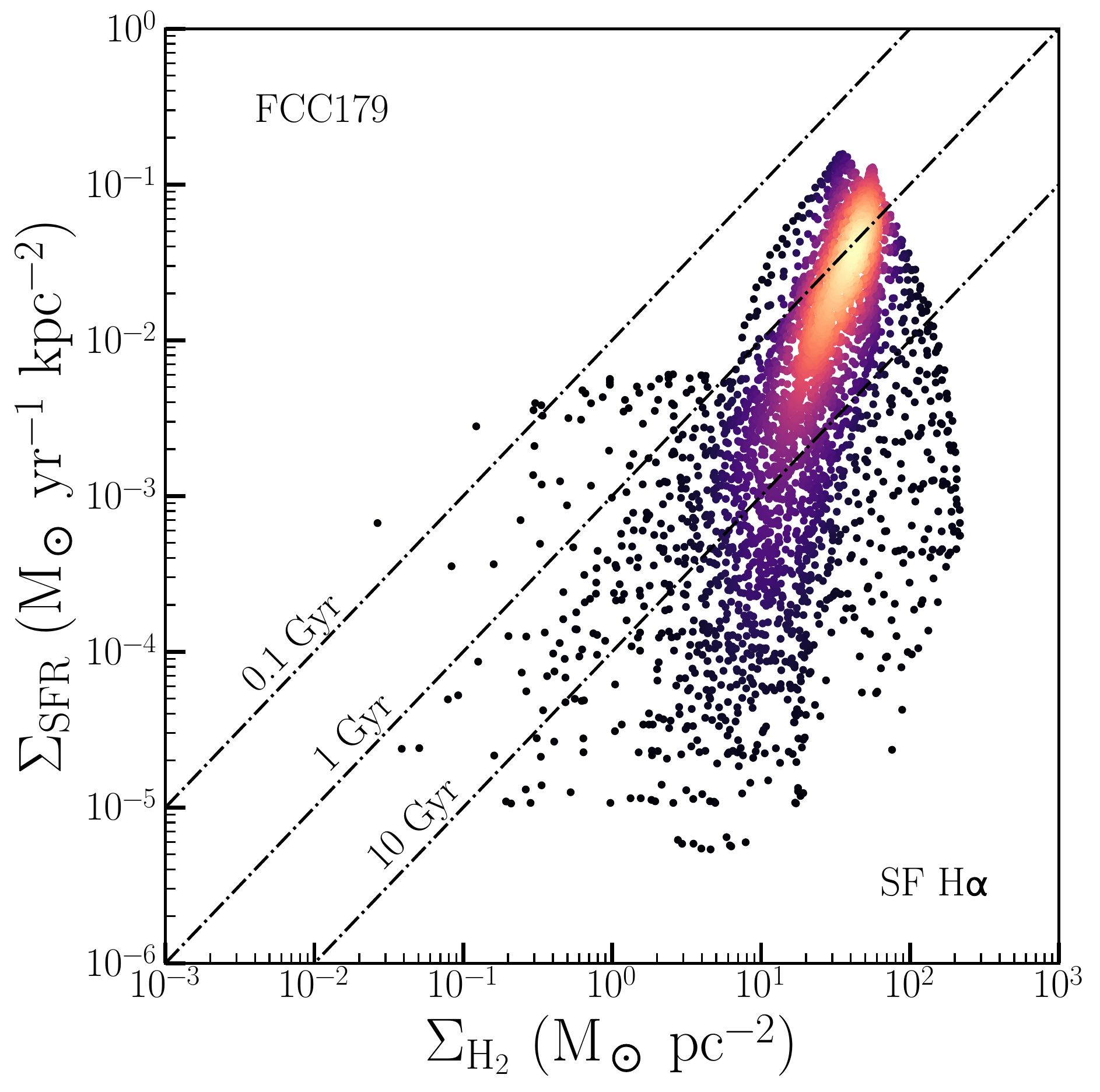}}	\hspace{13mm}
	\subfloat[\label{subfig:KS_FCC179_SF_comp}]
	{\includegraphics[height=0.42\textwidth, width=0.42\textwidth]{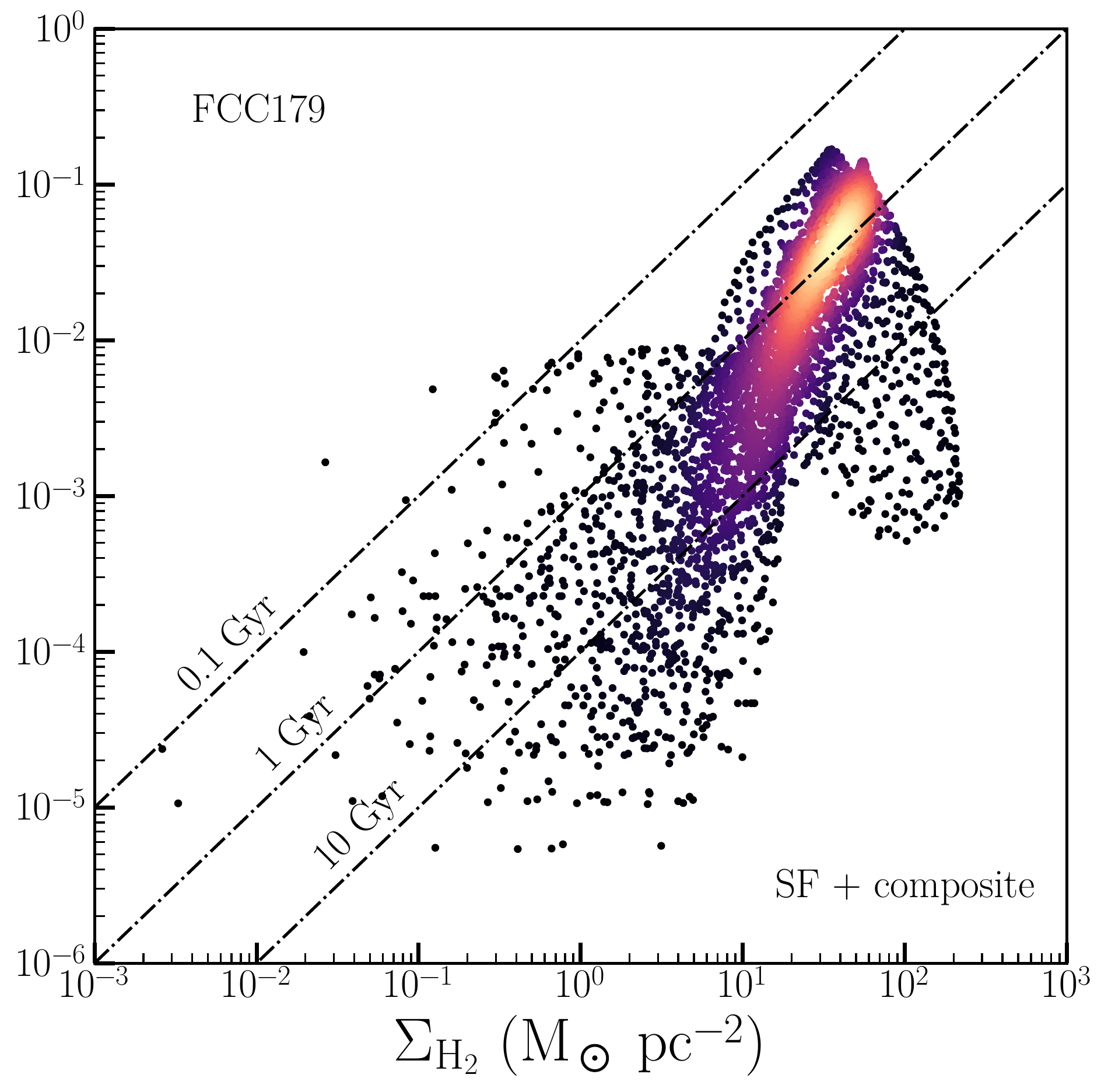}}	
	
	\subfloat[\label{subfig:DT_FCC179}]
	{\includegraphics[height=0.42\textwidth, width=0.47\textwidth]{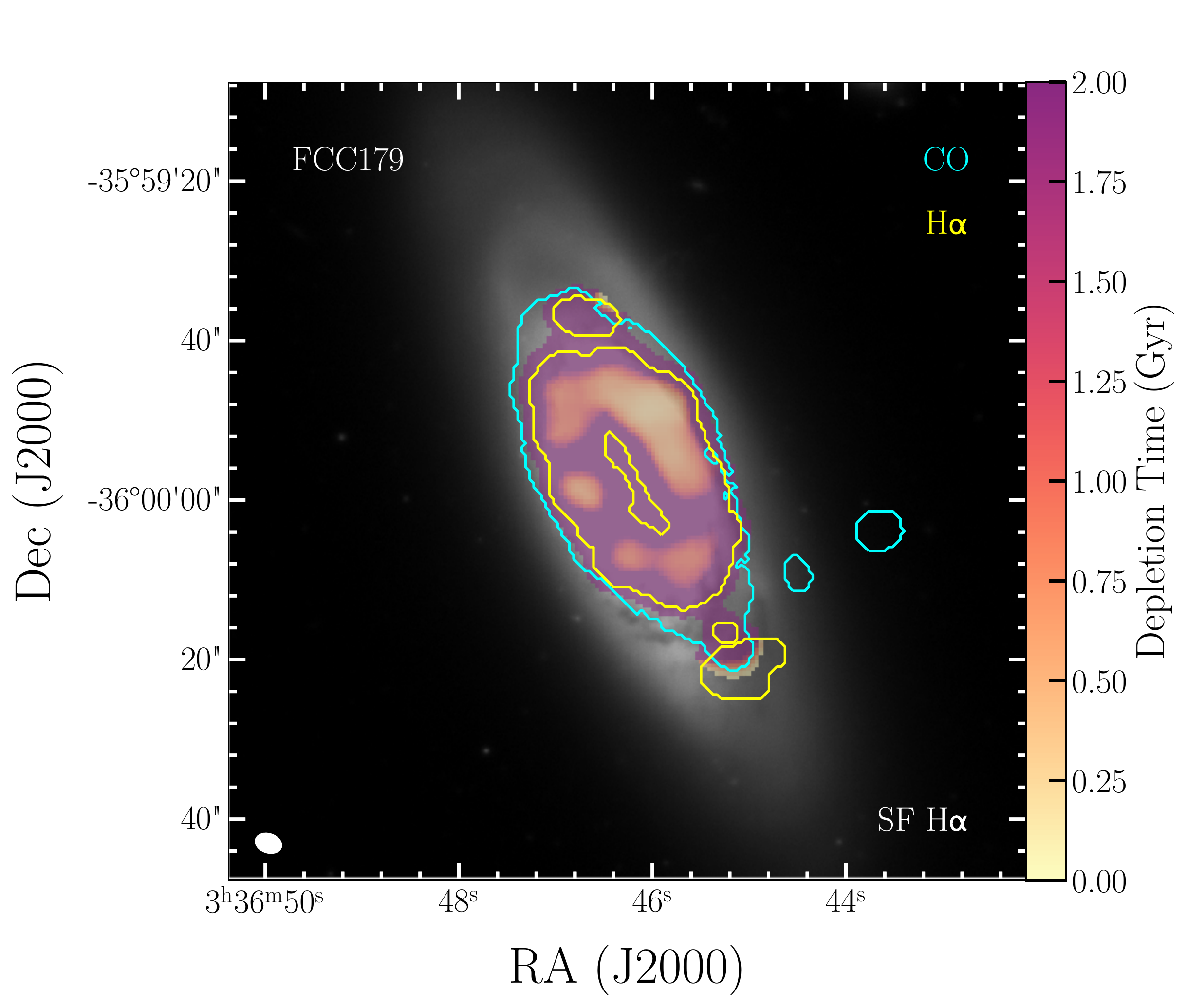}} \hspace{6mm}
	\subfloat[\label{subfig:DT_FCC179_SF_comp}]
	{\includegraphics[height=0.42\textwidth, width=0.47\textwidth]{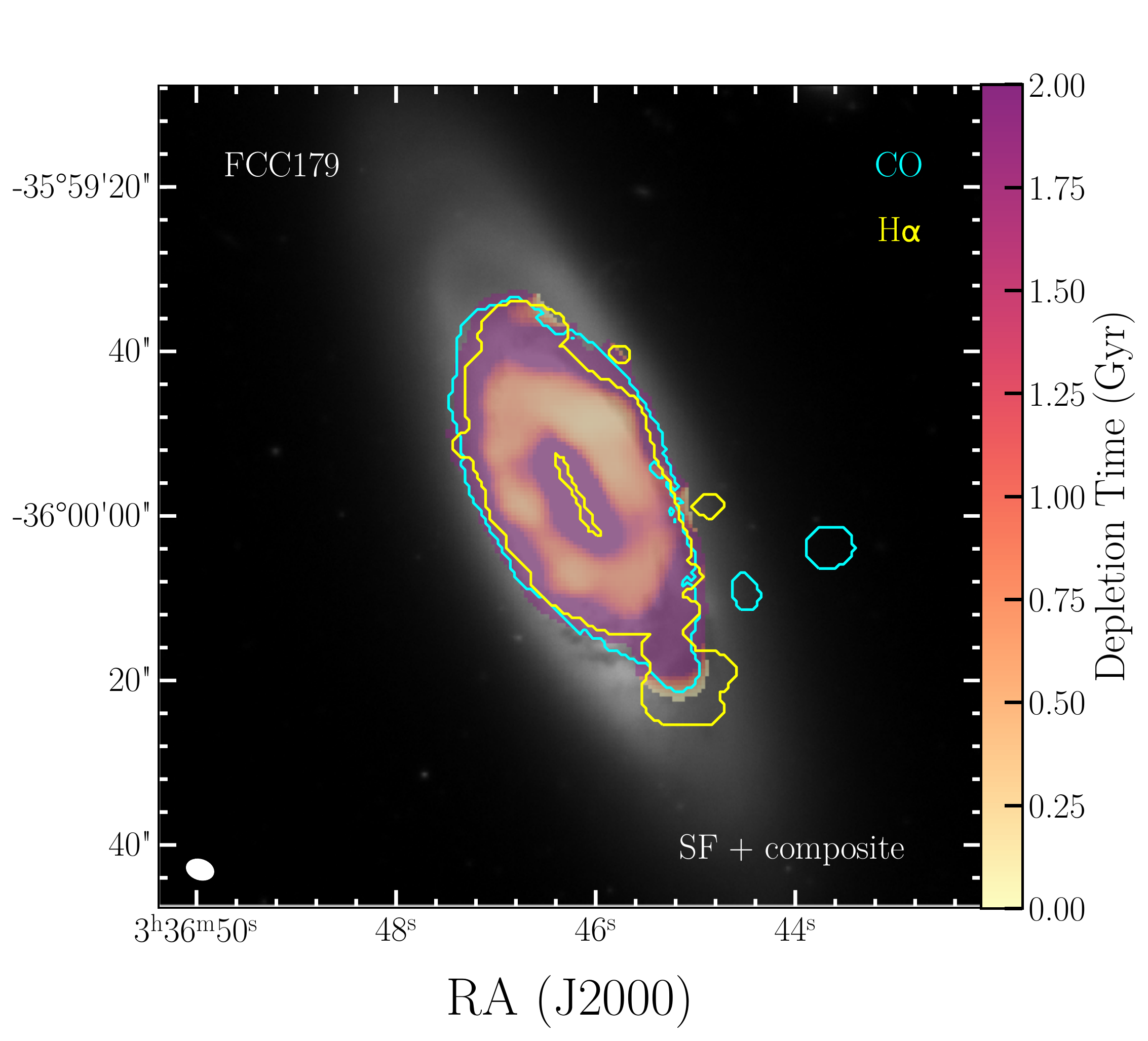}}	
	
	\caption{Similar to Figure \ref{fig:FCC90}, but for FCC179. The majority of the points are close to the 1 Gyr depletion time line, but the slope of the SF plot is quite steep, especially when only SF \Ha\ is considered. The latter forms a ring around the galactic centre.}
	\label{fig:FCC179}

\end{figure*}

FCC179 (NGC1386) is a barred spiral galaxy, that also harbours an AGN. This results in a significant difference between the left (only \Ha\ dominated by star formation) and right (composite \Ha considered also) columns in Figure \ref{fig:FCC179}. If only the star forming \Ha\ is taken into account, there is significant scatter in the star formation relation. The relationship is tighter when composite \Ha\ is considered. Depletion times are shorter in the spiral arms compared to the inter-arm regions. The shortest depletion times at high \Ha\ surface densities originate from the region around the AGN, and are therefore potentially contaminated. The distribution of the ionised gas at the east-side of the galaxy, as well as two separate blobs of molecular gas, could be a sign of ongoing ram pressure stripping.

\subsection{FCC184}
\label{sub:FCC184}
FCC184 (NGC1387) is a lenticular galaxy. It also harbours an AGN, and again, most of the pixels in the ionised gas map were classified as ``composite''. Depletion times in this galaxy also vary significantly depending on whether composite \Ha\ is included or not. If only star forming \Ha\ is considered, depletion times are long (up to tens of Gyrs) and show large variations. If composite \Ha\ is included, depletion times are a few Gyrs, showing an extremely tight relationship between the H$_2$ and SFR surface densities. 

\begin{figure*}

	\centering

	\hspace{-0mm}  \subfloat[\label{subfig:KS_FCC184}]
	{\includegraphics[height=0.42\textwidth, width=0.44\textwidth]{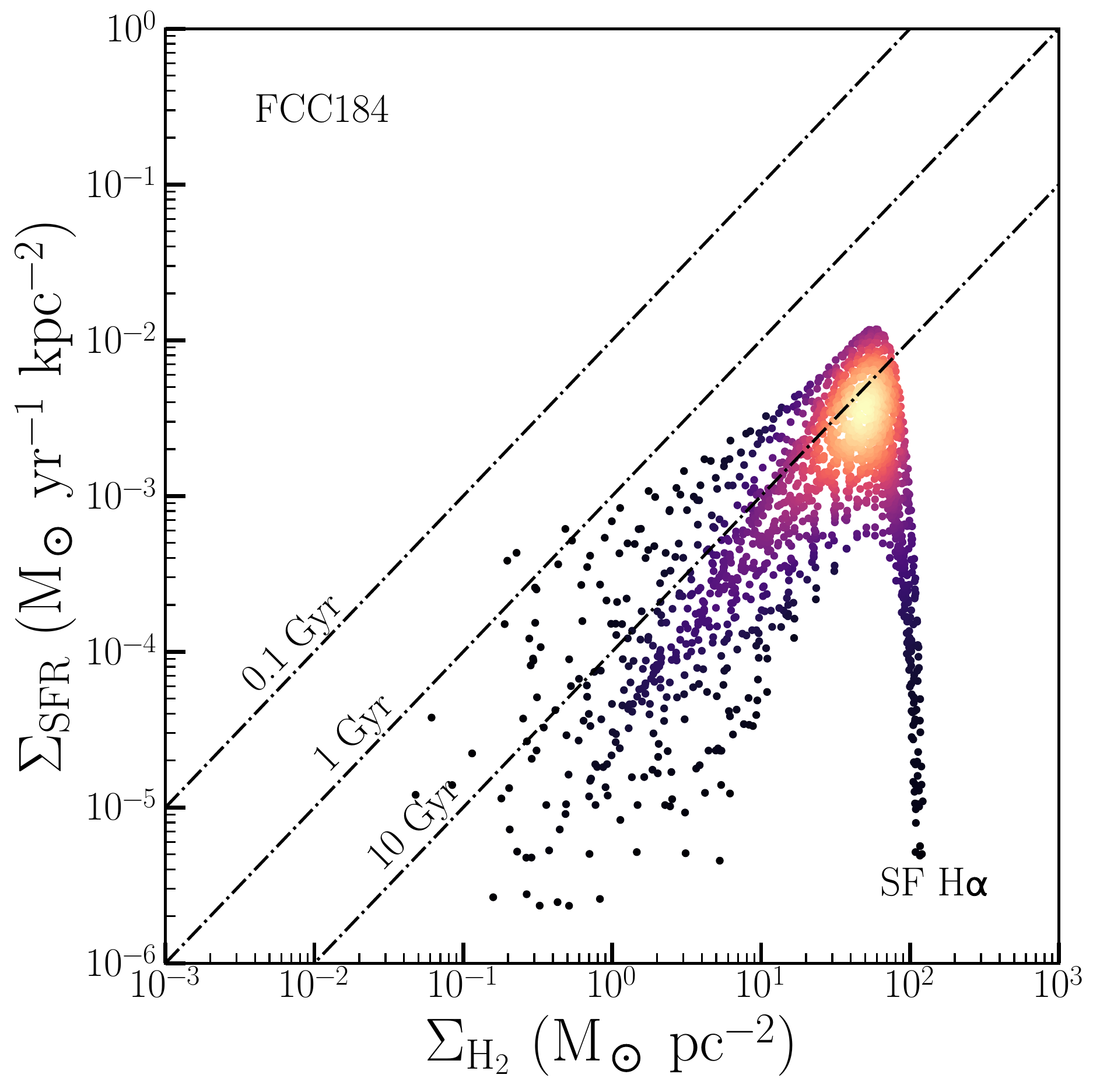}}	\hspace{13mm}
	\subfloat[\label{subfig:KS_FCC184_SF_comp}]
	{\includegraphics[height=0.42\textwidth, width=0.42\textwidth]{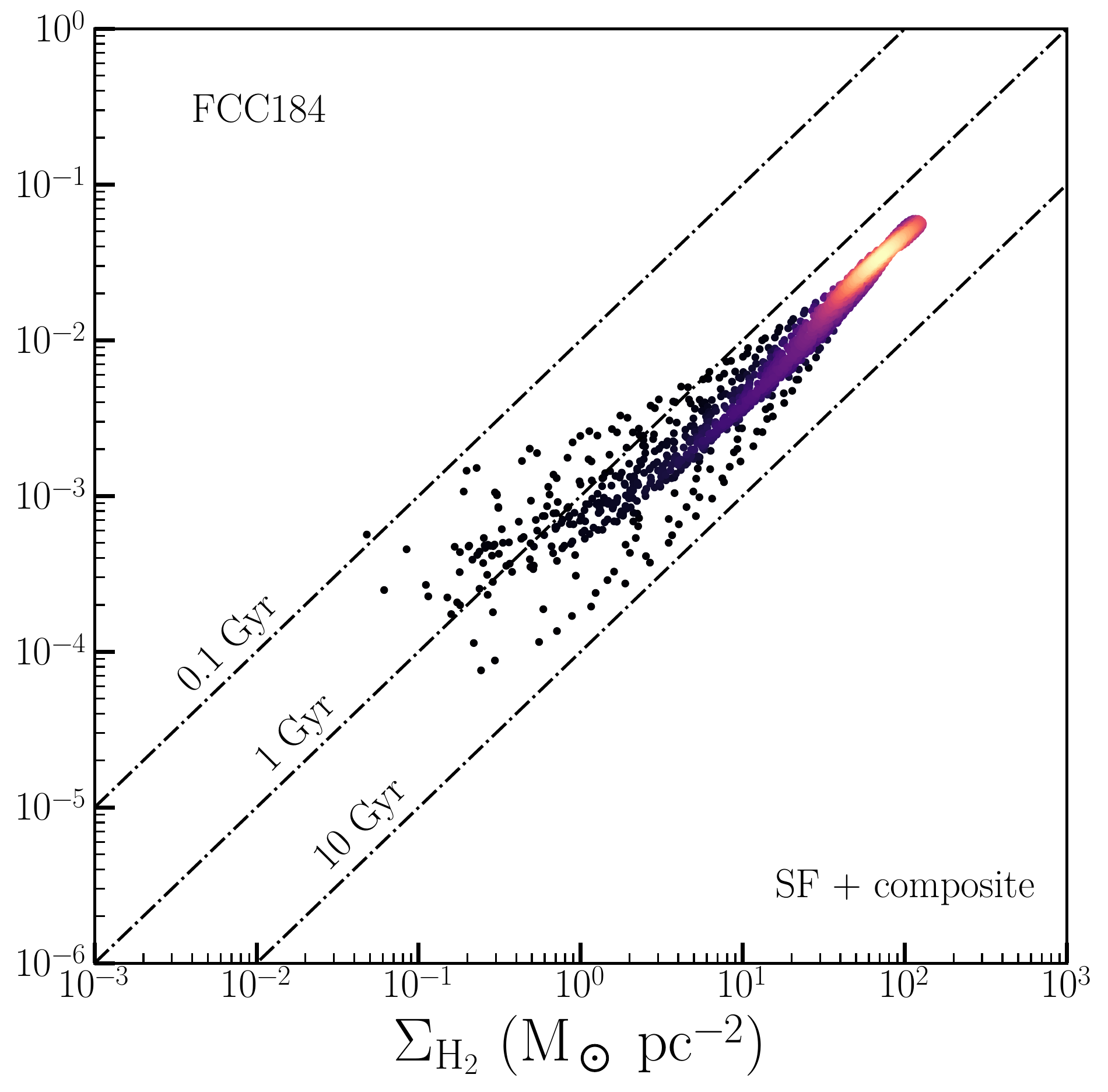}}	
	
	\subfloat[\label{subfig:DT_FCC184}]
	{\includegraphics[height=0.42\textwidth, width=0.47\textwidth]{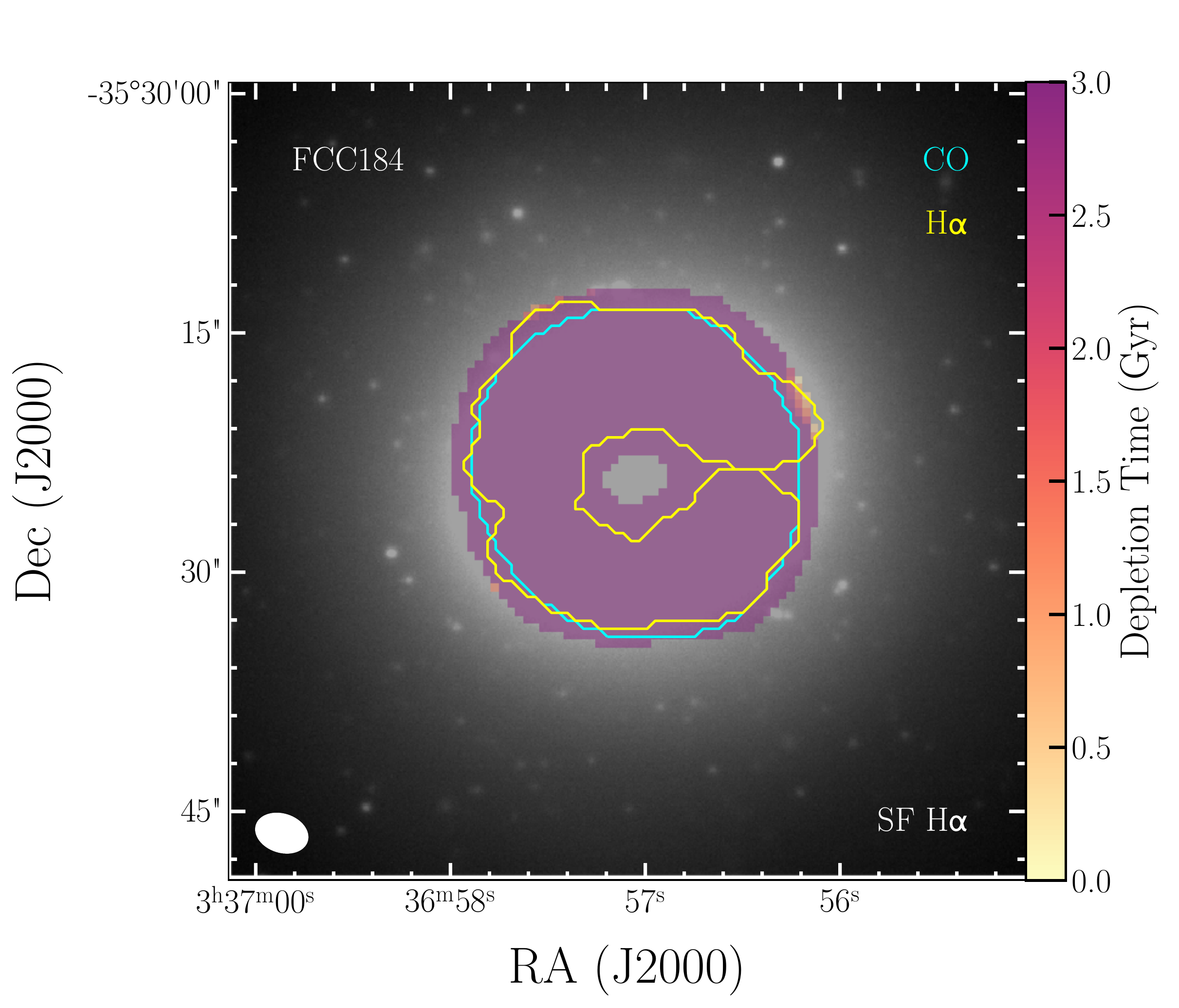}} \hspace{6mm}
	\subfloat[\label{subfig:DT_FCC184_SF_comp}]
	{\includegraphics[height=0.42\textwidth, width=0.47\textwidth]{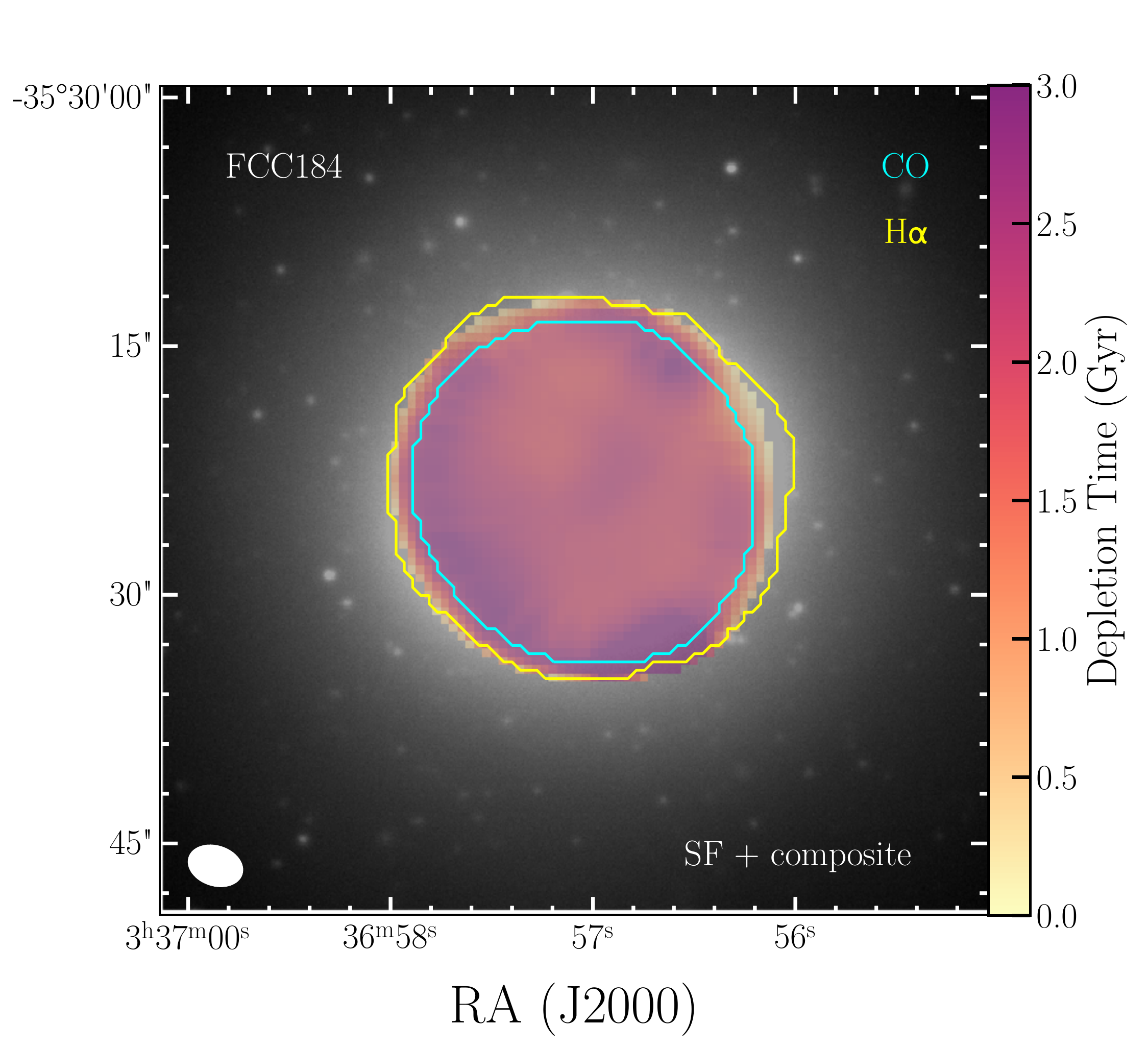}}
	
	\caption{Similar to Figure \ref{fig:FCC90}, but for FCC184. The SF plot for composite + star forming \Ha\ is very tight and around the expected 2 Gyr depletion time line. It is quite sparse for the star forming SF only, because that only forms a patchy ring around the galactic centre. Nearly all other \Ha\ in this galaxy is in the transition region of the BPT diagram, so in reality there is likely some more star forming \Ha, and the true SF relation plot is in between the left and right one.}
	\label{fig:FCC184}

\end{figure*}

\subsection{FCC207}
\label{sub:FCC207}
\begin{figure*}

	\centering

	\hspace{-0mm}  \subfloat[\label{subfig:KS_FCC207}]
	{\includegraphics[height=0.42\textwidth, width=0.44\textwidth]{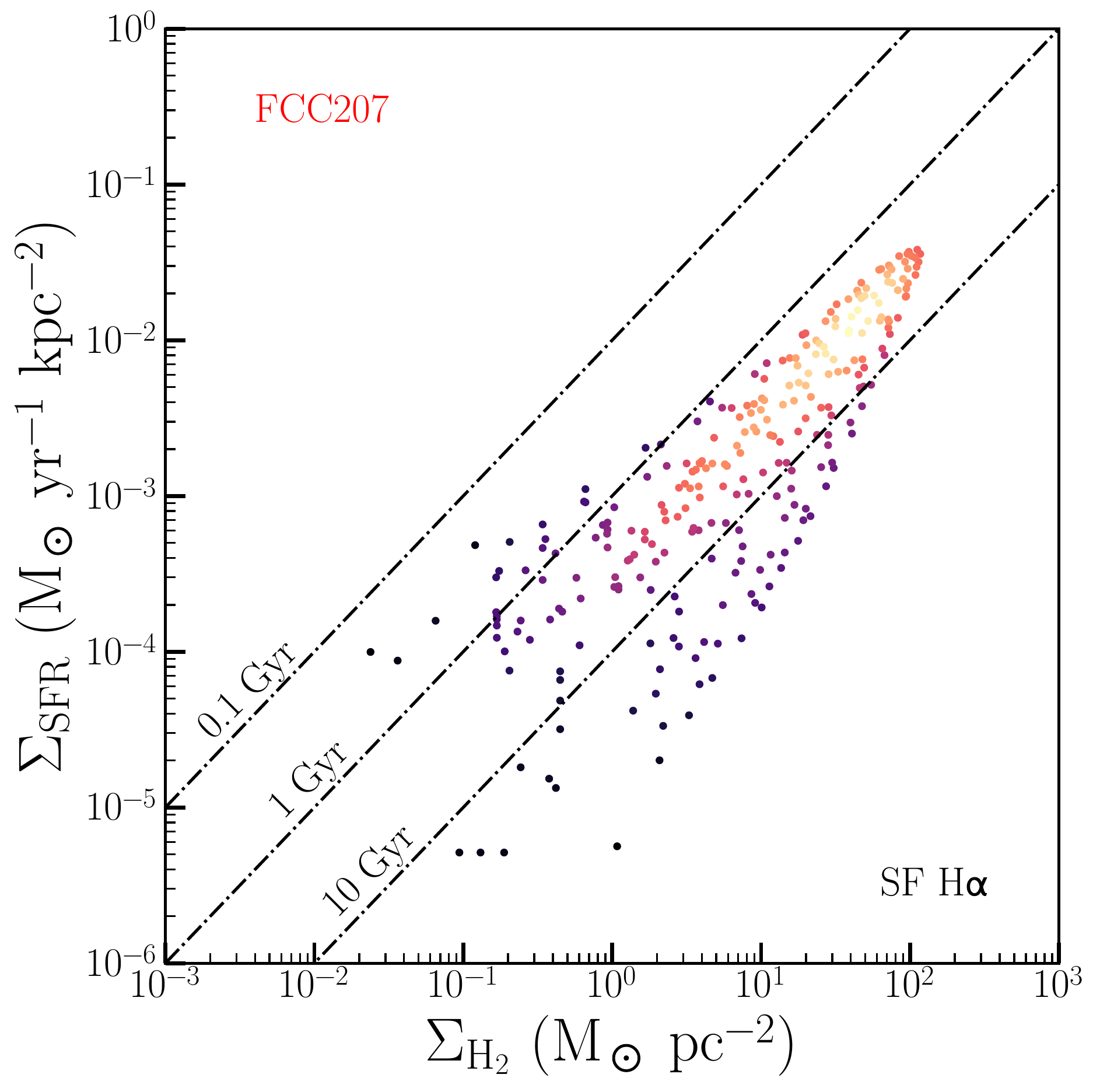}}	\hspace{13mm}
	\subfloat[\label{subfig:KS_FCC207_SF_comp}]
	{\includegraphics[height=0.42\textwidth, width=0.42\textwidth]{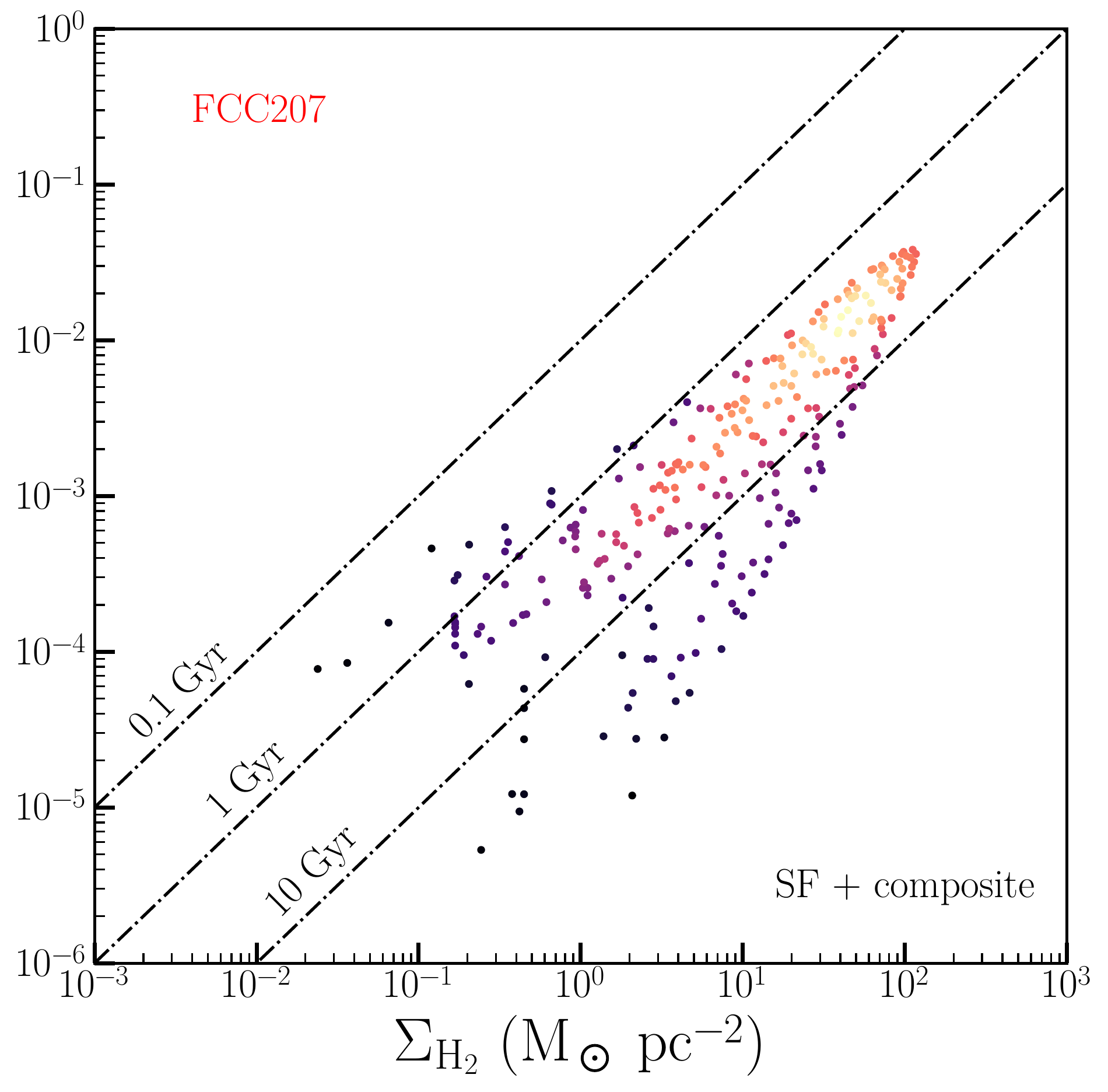}}	
	
	\subfloat[\label{subfig:DT_FCC207}]
	{\includegraphics[height=0.42\textwidth, width=0.47\textwidth]{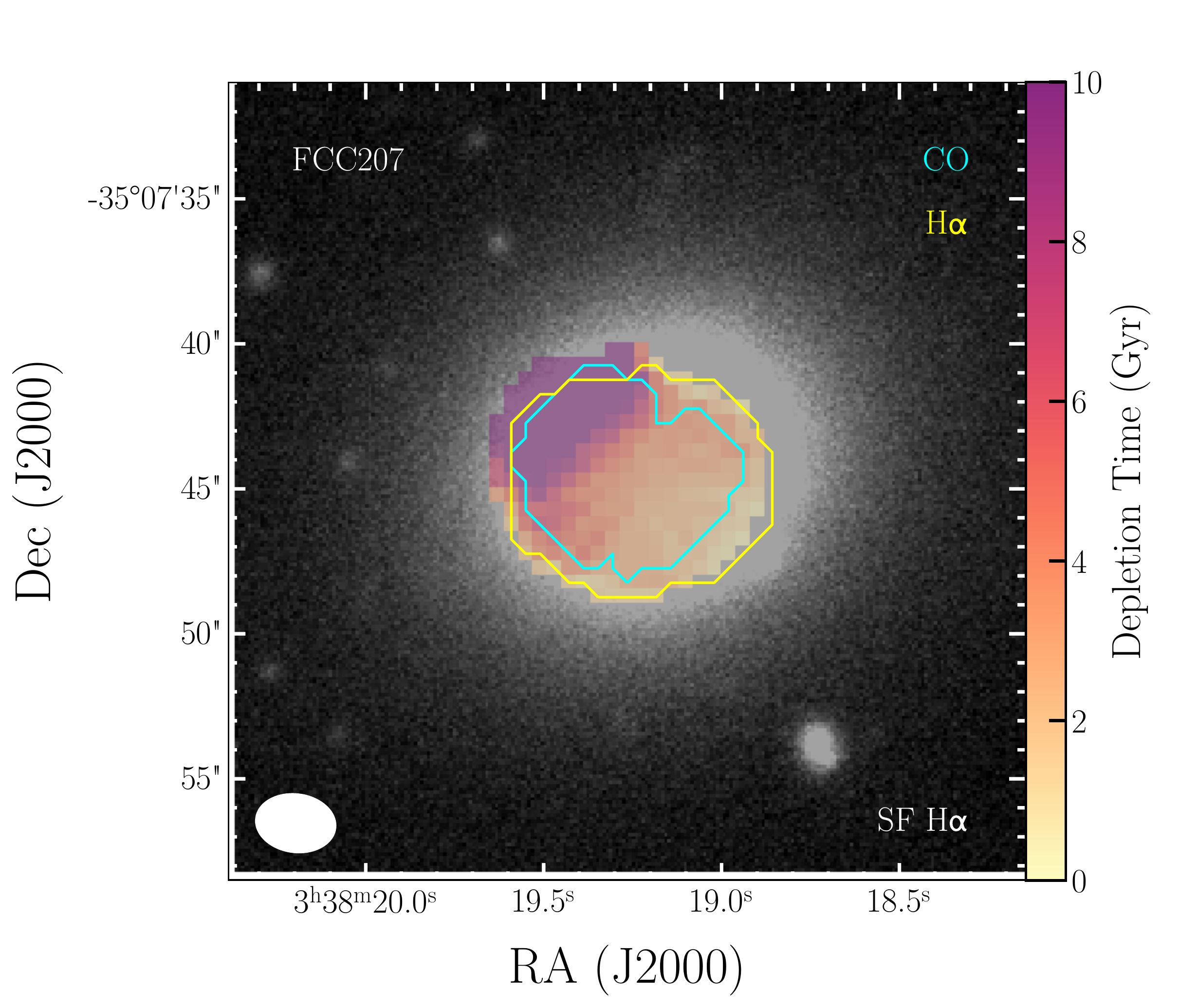}} \hspace{6mm}
	\subfloat[\label{subfig:DT_FCC207_SF_comp}]
	{\includegraphics[height=0.42\textwidth, width=0.47\textwidth]{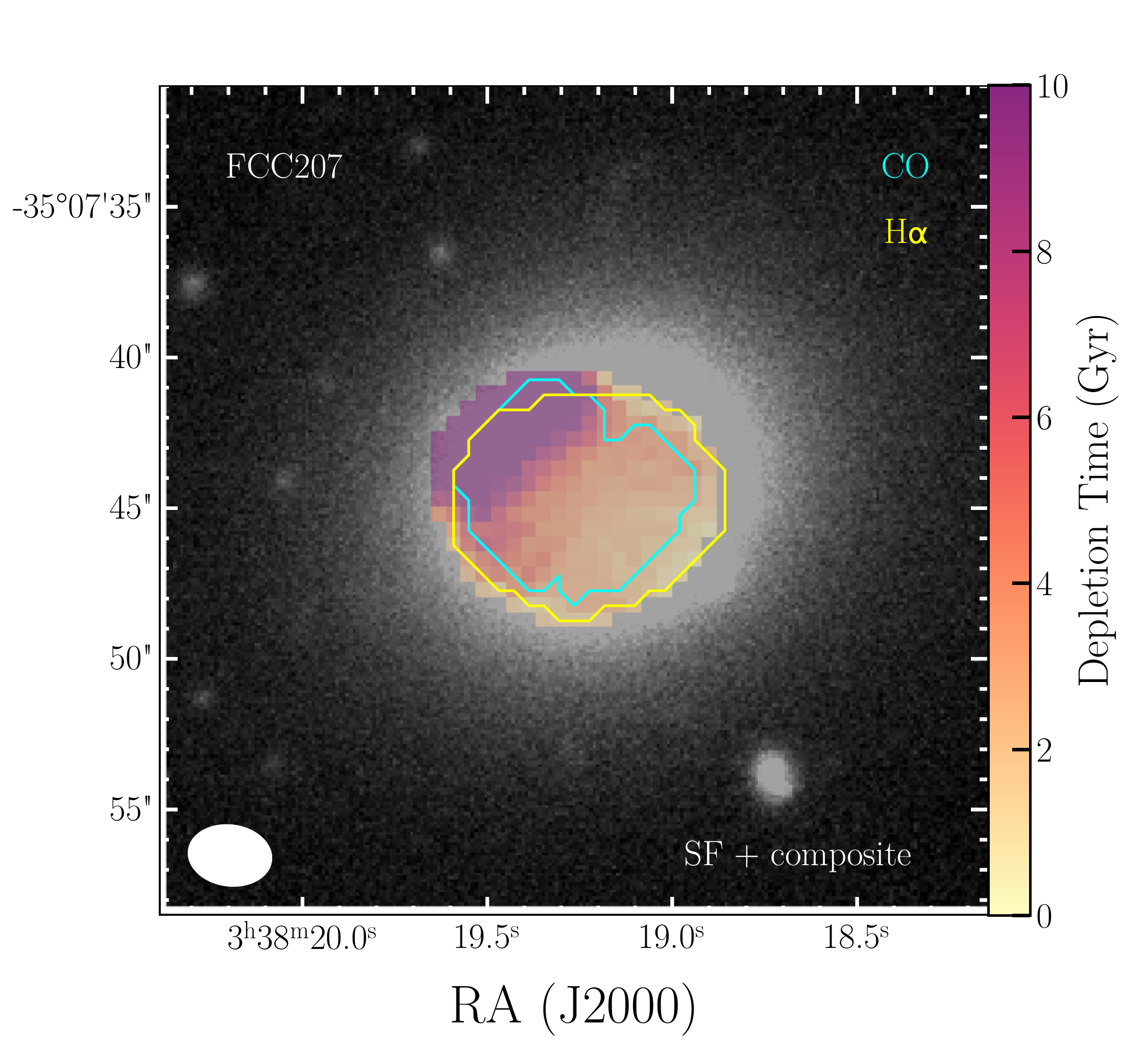}}	
	
	\caption{Similar to Figure \ref{fig:FCC90}, but for FCC207. \change{Although there is a gradient in depletion times, suggesting compression of the molecular gas as a result of the galaxy's first infall into the cluster, depletion times in FCC207 are long, varying between several Gyrs and several tens of Gyrs. It is possible that the slight starburst seen in other dwarf galaxies on their first infall has not yet started, or has already passed.}}
	\label{fig:FCC207}

\end{figure*}

FCC207 is a dwarf galaxy. It has relatively long depletion times, closer to 10 Gyr than to 1 Gyr. Molecular and ionised gas are present throughout the galaxy, but depletion times are significantly longer on the west-side compared to the east-side. FCC207 is the only dwarf galaxy with a disturbed molecular gas reservoir that has increased depletion times, rather than decreased. All \Ha\ in this galaxy is dominated by star formation, so the left and right columns (SF \Ha\ vs. composite + SF \Ha) in Figure \ref{fig:FCC207} are identical.

\subsection{FCC263}
\label{sub:FCC263}
FCC263 (PGC013571) is another dwarf galaxy with a disturbed molecular gas reservoir. Its molecular gas morphology and kinematics suggests that it either has a strong bar or has undergone a recent merger. The molecular gas is much more centrally distributed than the ionised gas, resulting in identical \KS\ relations in the left and right columns (the \Ha\ that is added in the right column does not overlap with any molecular gas). Depletion times in this galaxy are significantly shorter than 1 Gyr, and show significant scatter. There are various patches of shorter and longer depletion times.

\begin{figure*}

	\centering

	\hspace{-0mm}  \subfloat[\label{subfig:KS_FCC263}]
	{\includegraphics[height=0.42\textwidth, width=0.44\textwidth]{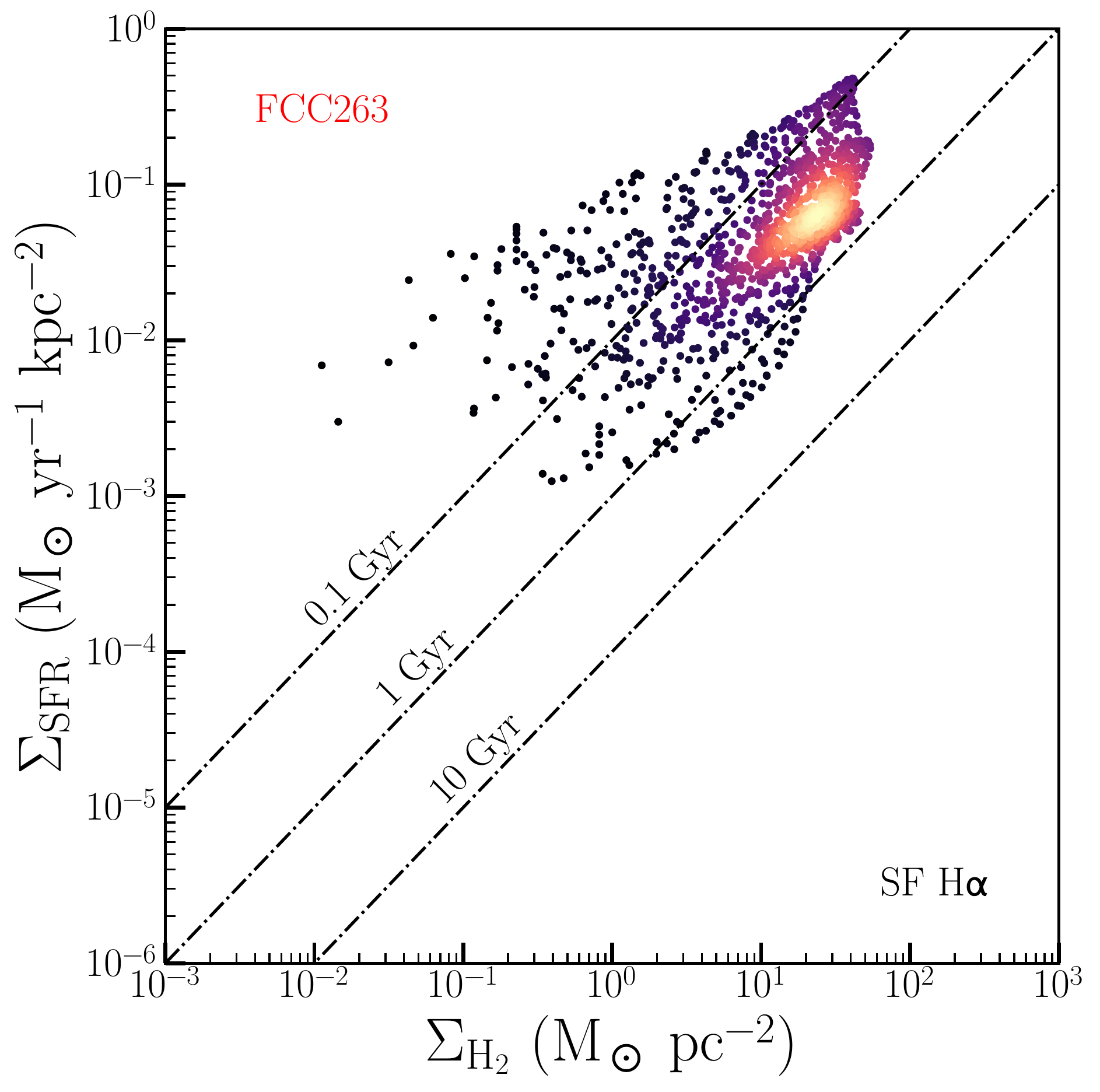}}	\hspace{13mm}
	\subfloat[\label{subfig:KS_FCC263_SF_comp}]
	{\includegraphics[height=0.42\textwidth, width=0.42\textwidth]{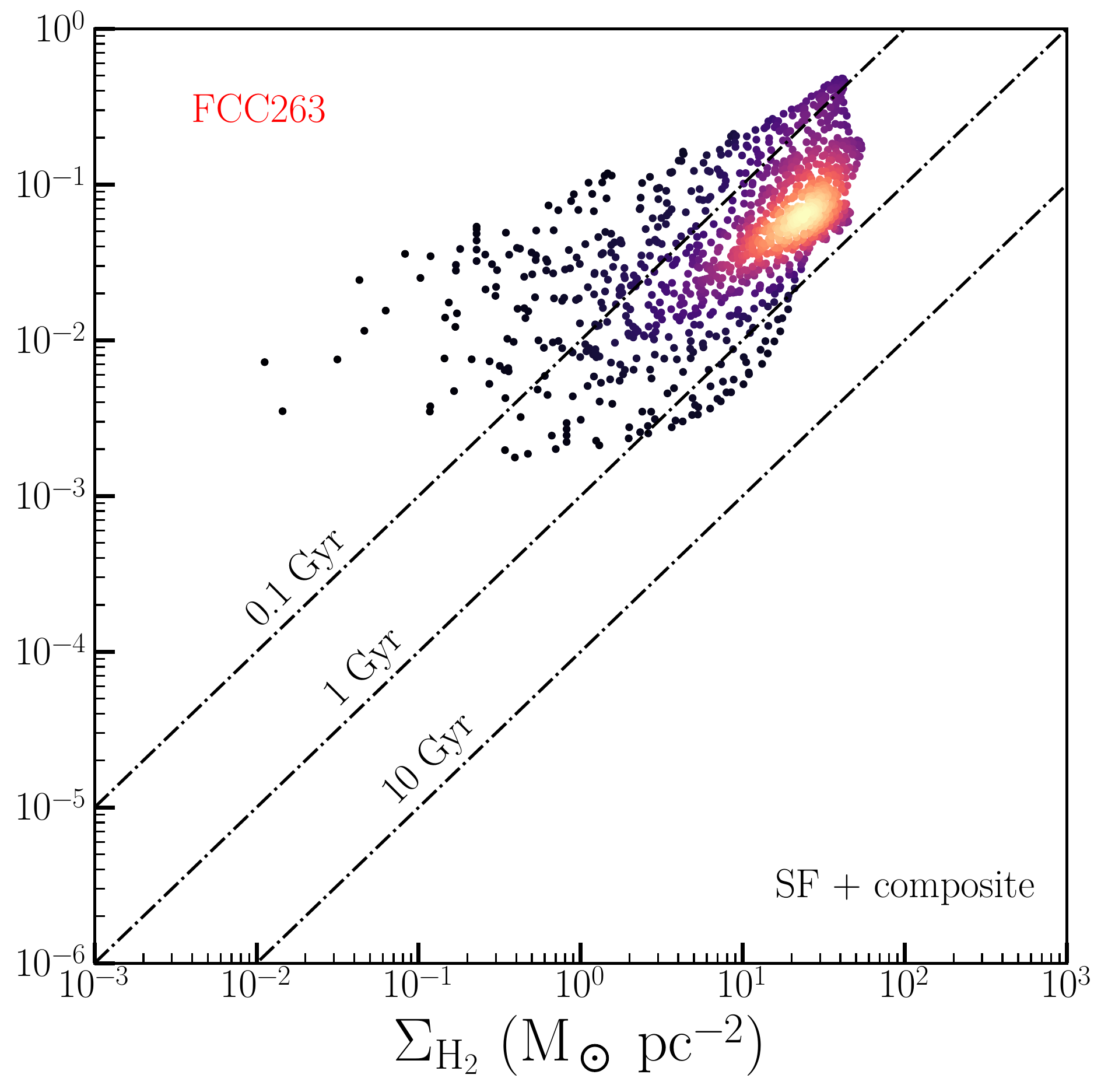}}	
	
	\subfloat[\label{subfig:DT_FCC263}]
	{\includegraphics[height=0.42\textwidth, width=0.47\textwidth]{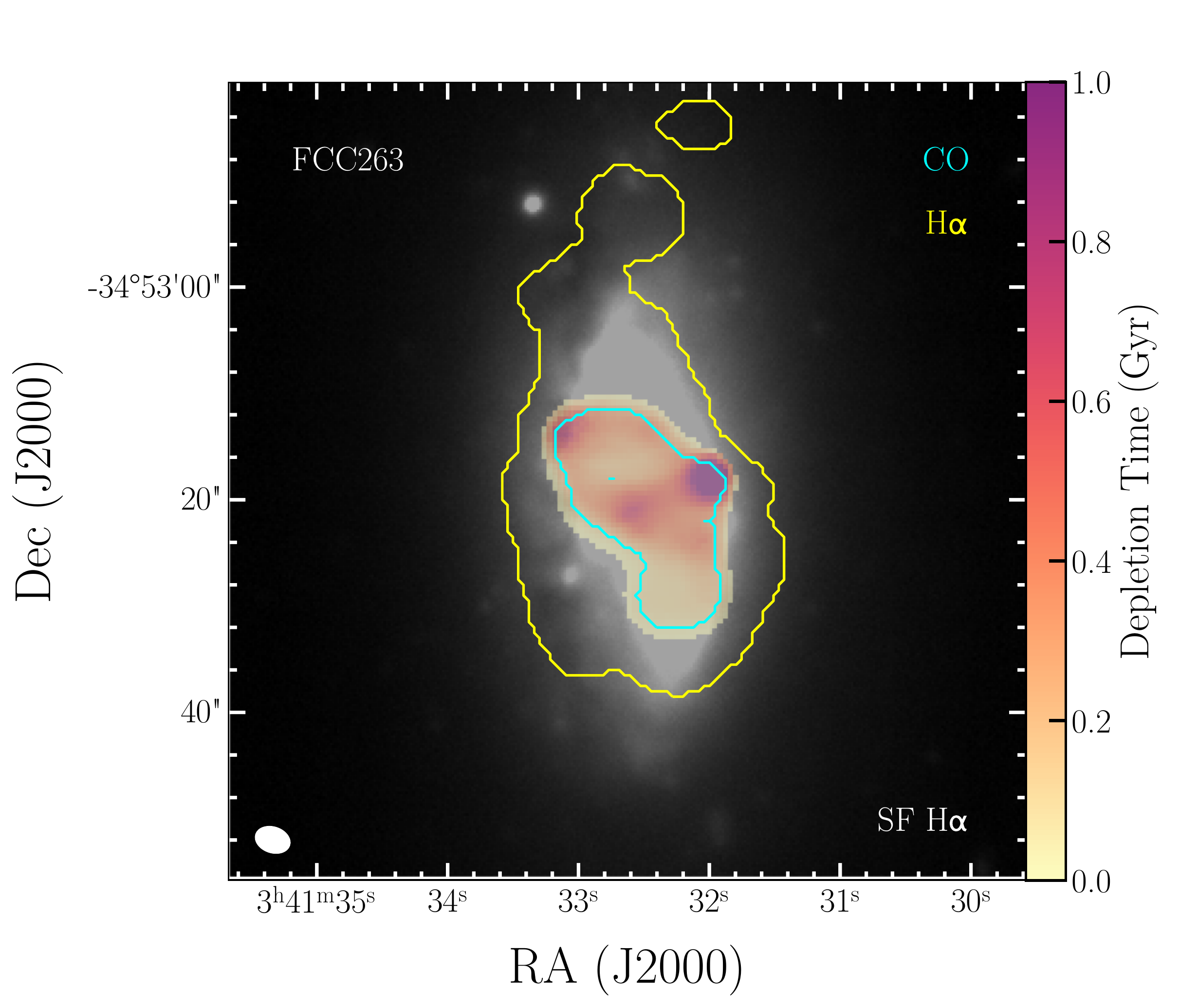}} \hspace{6mm}
	\subfloat[\label{subfig:DT_FCC263_SF_comp}]
	{\includegraphics[height=0.42\textwidth, width=0.47\textwidth]{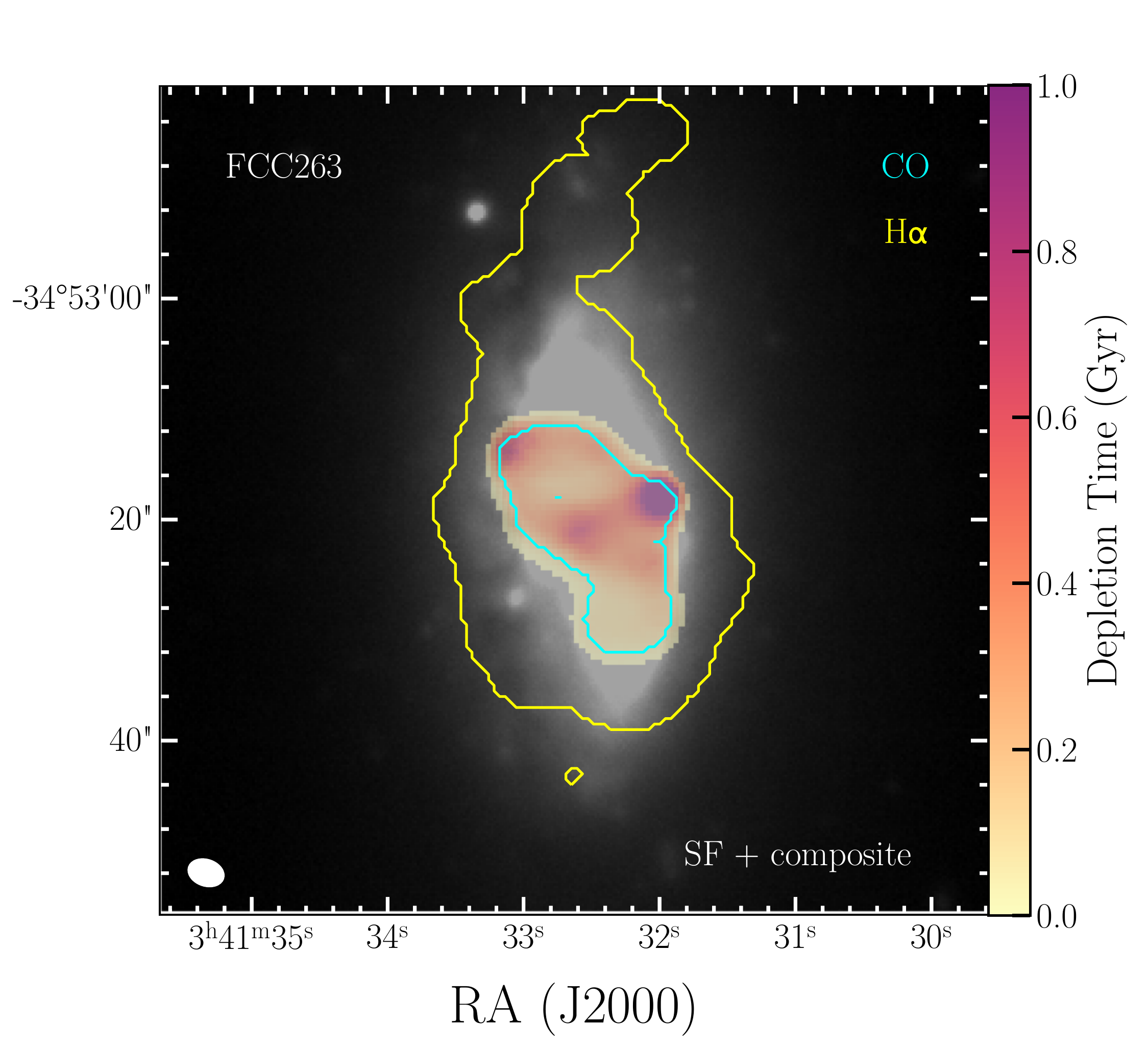}}
	
	\caption{Similar to Figure \ref{fig:FCC90}, but for FCC263. There is some spread in the SF relation, and depletion times are around 0.5 Gyr. The CO is more concentrated than the \Ha, and there are some regions where the depletion times are relatively long.}
	\label{Fig:FCC263}

\end{figure*}

\subsection{FCC290}
\label{sub:FCC290}
FCC290 (NGC1436) is a flocculent spiral that is almost face-on. This makes it an ideal candidate for comparing the molecular and ionised gas distributions, without suffering too much from inclination effects. Although not all of the galaxy is covered by the MUSE observations, the data should be representative of the entire galaxy. Although not all \Ha\ in this galaxy is star forming, the resulting \KS\ relations for both cases are very similar, with slightly more scatter in the star formation rate surface densities in the case where only star forming \Ha\ is taken into account. The spread in depletion times in this galaxy is large, with the densest area in the plot varying from less than a Gyr to up to 10 Gyr. It is strikingly symmetric in both directions, and the maps reveal both areas of increased and decreased depletion times. This suggests that the \Ha\ and CO do not overlap perfectly, their offset possibly revealing the time passed since the OB stars were actually formed (see \S \ref{sec:cloud_lifetimes}). The bulk of the depletion times is around the ``standard'' 1-2 Gyr.

\begin{figure*}

	\centering

	\hspace{-0mm}  \subfloat[\label{subfig:KS_FCC290}]
	{\includegraphics[height=0.42\textwidth, width=0.44\textwidth]{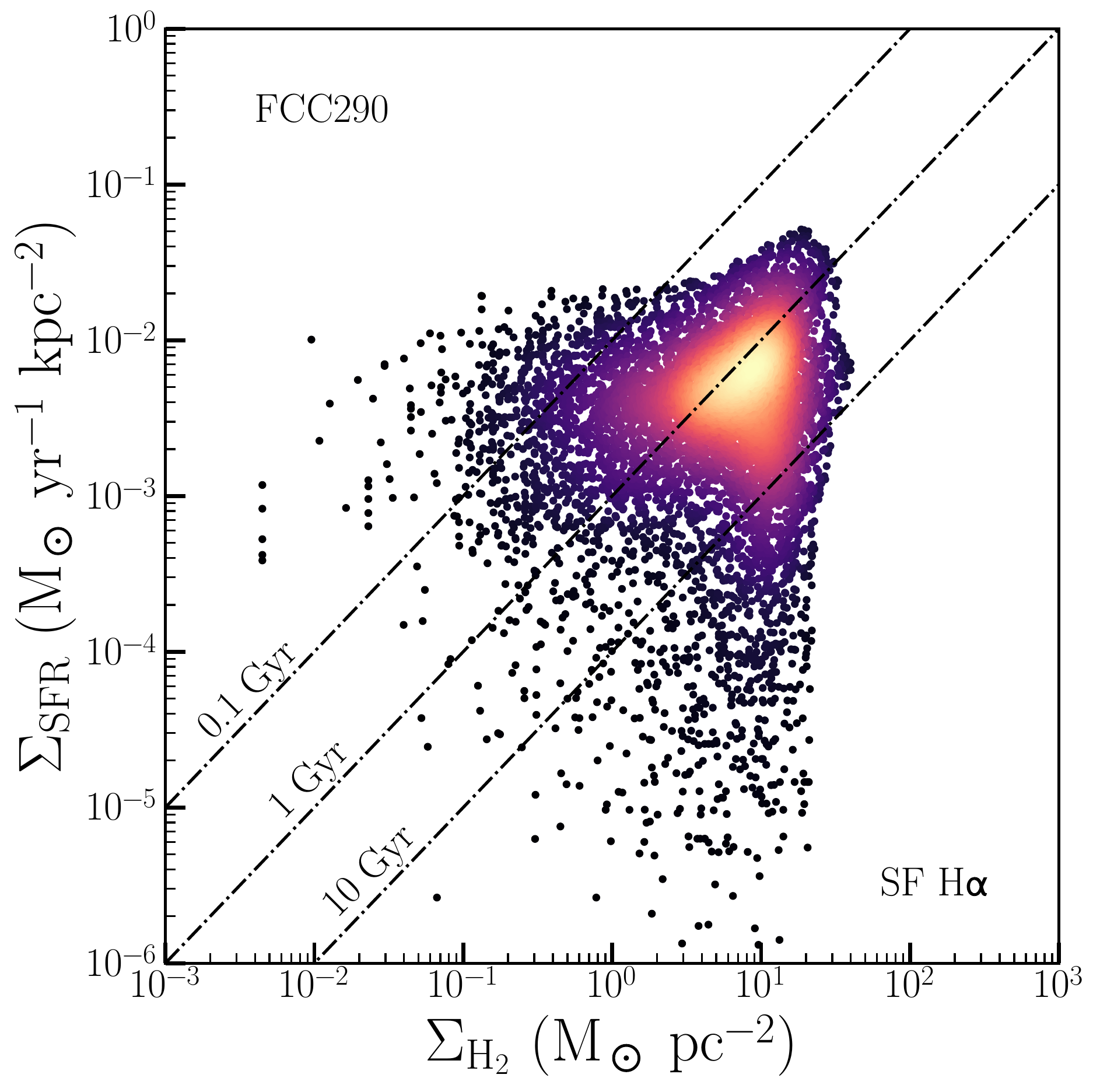}}	\hspace{13mm}
	\subfloat[\label{subfig:KS_FCC290_SF_comp}]
	{\includegraphics[height=0.42\textwidth, width=0.42\textwidth]{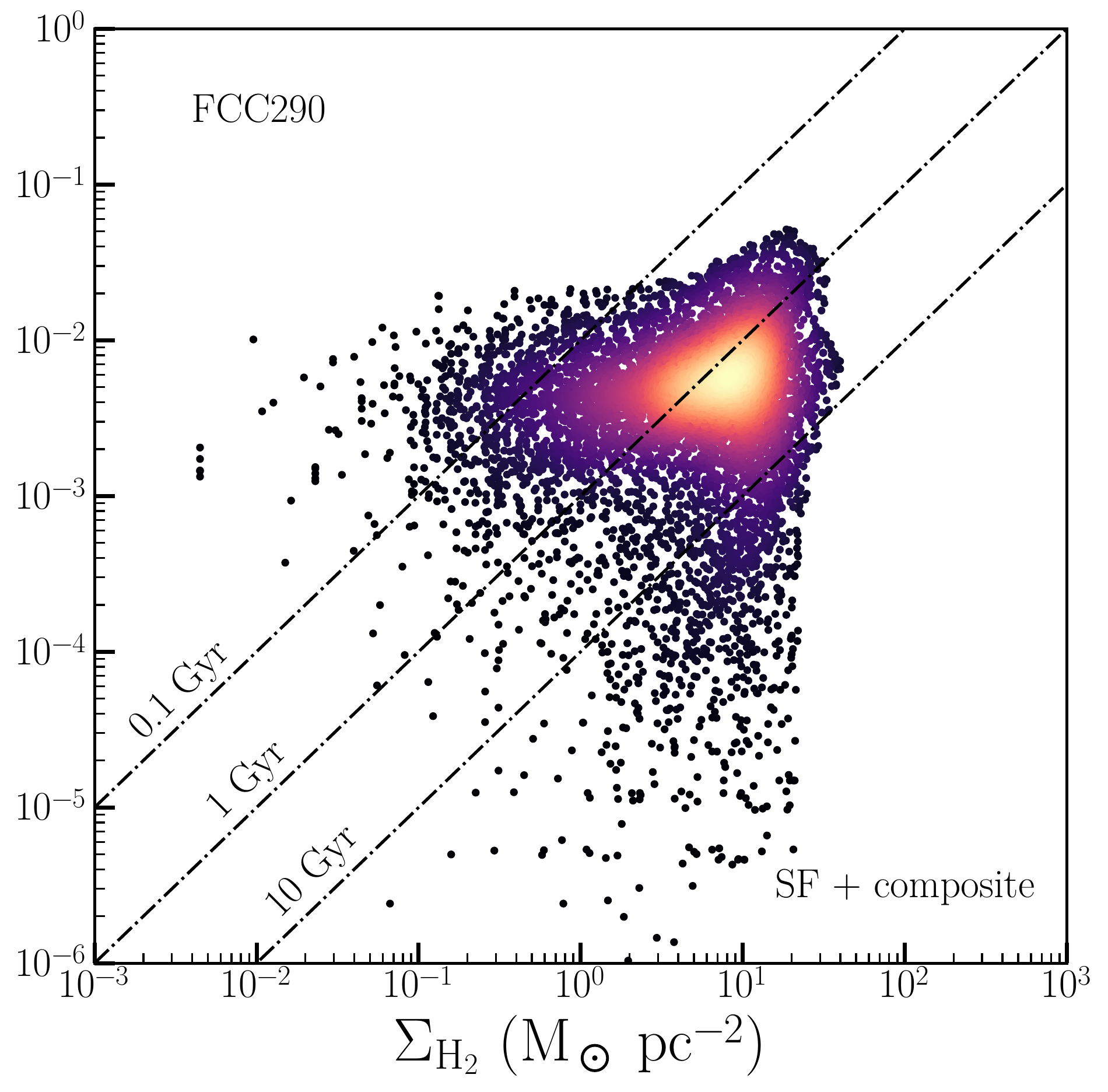}}	
	
	\subfloat[\label{subfig:DT_FCC290}]
	{\includegraphics[height=0.42\textwidth, width=0.47\textwidth]{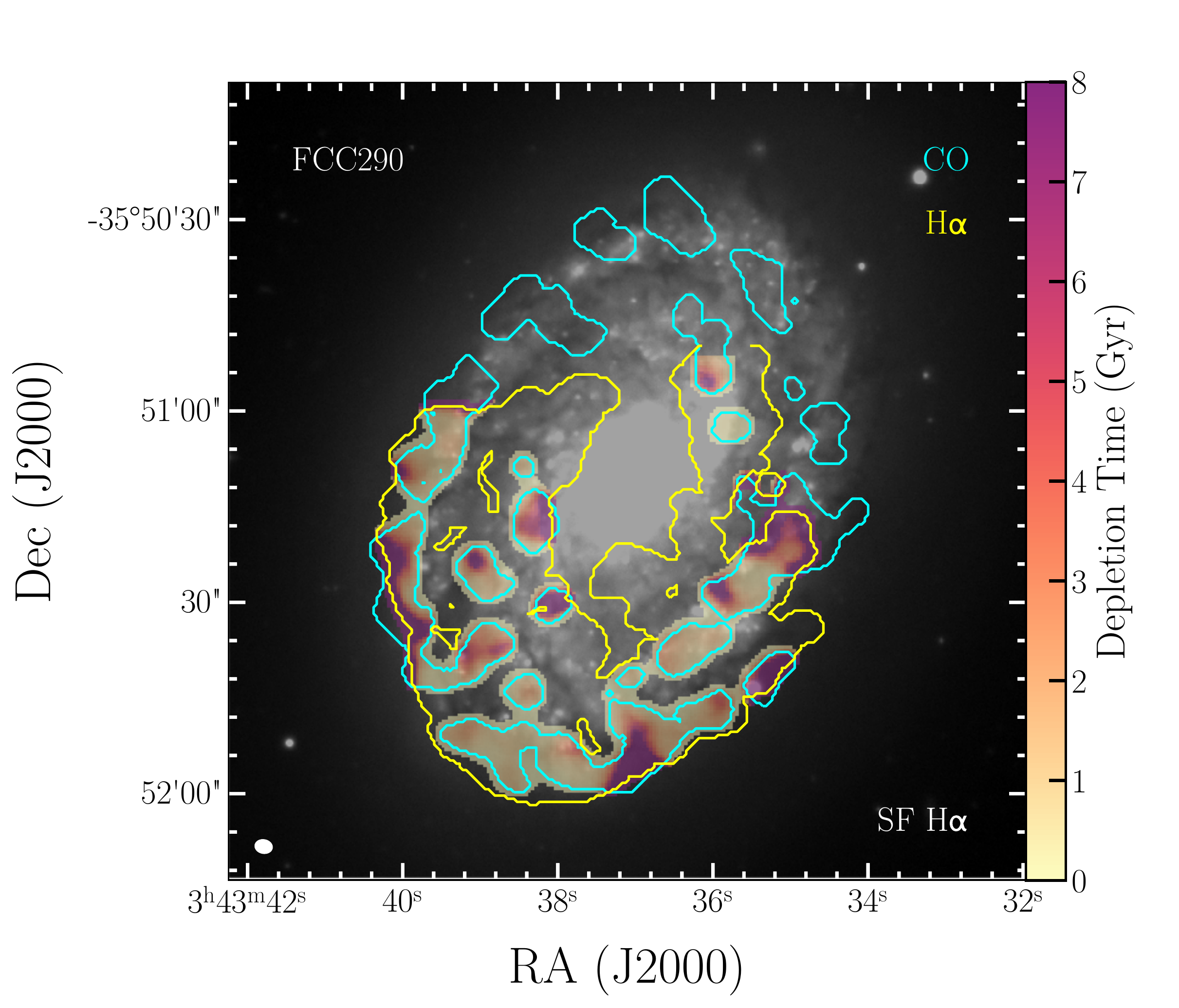}} \hspace{6mm}
	\subfloat[\label{subfig:DT_FCC290_SF_comp}]
	{\includegraphics[height=0.42\textwidth, width=0.47\textwidth]{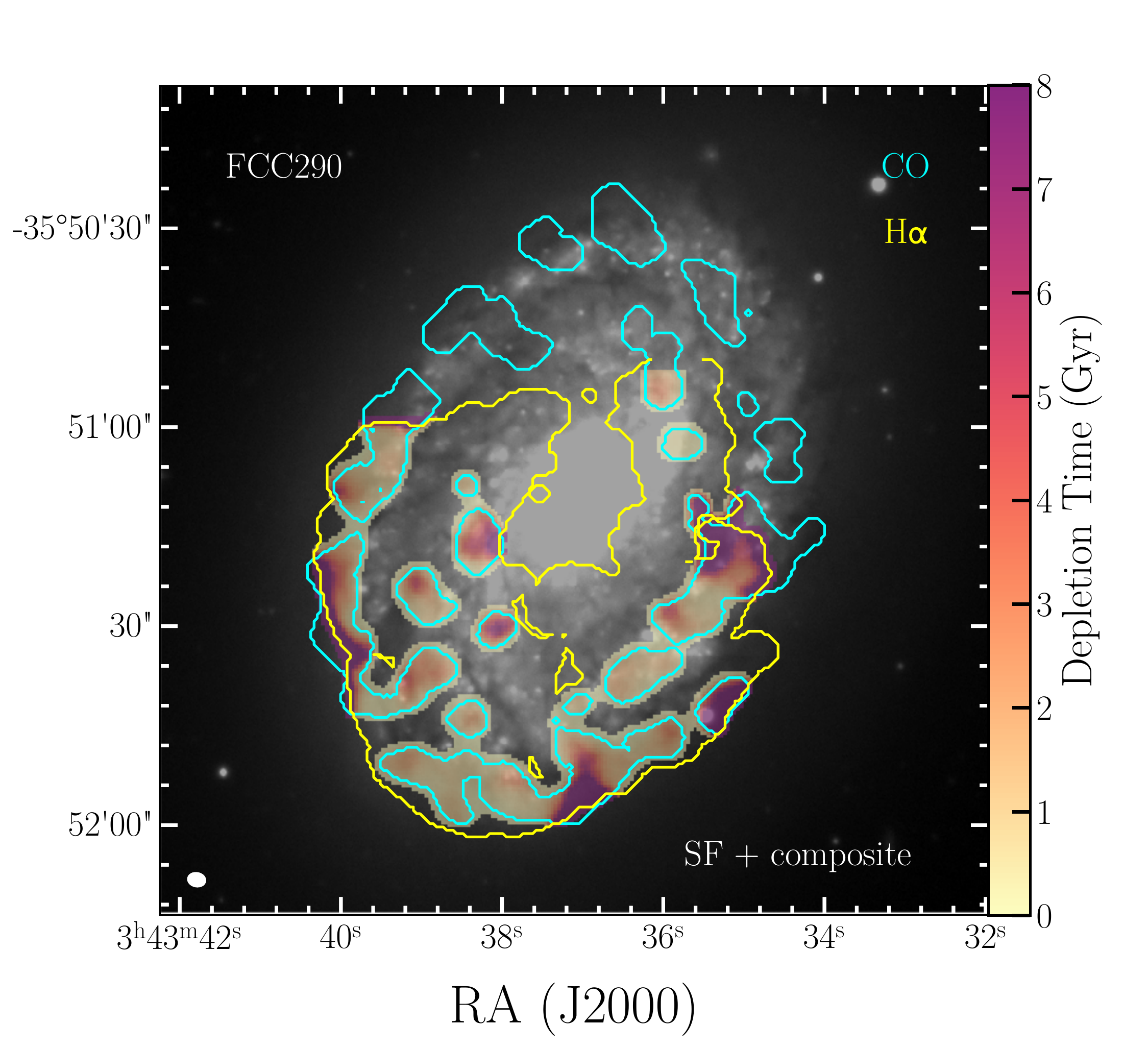}}		
	
	\caption{Similar to Figure \ref{fig:FCC90}, but for FCC290. There is some spread in the SF plot, but the majority of the points lie around the 2 Gyr line. The variation in depletion times can be seen in the maps. The MUSE field of view does not cover the entire galaxy, but the area covered should be quite representative of the entire object. There is a large spread in depletion times, that is quite symmetric around the 2 Gyr depletion time line.}
	\label{fig:FCC290}

\end{figure*}

\subsection{FCC308}
\label{sub:FCC308}
\begin{figure*}

	\centering

	\hspace{-0mm}  \subfloat[\label{subfig:KS_FCC308}]
	{\includegraphics[height=0.42\textwidth, width=0.44\textwidth]{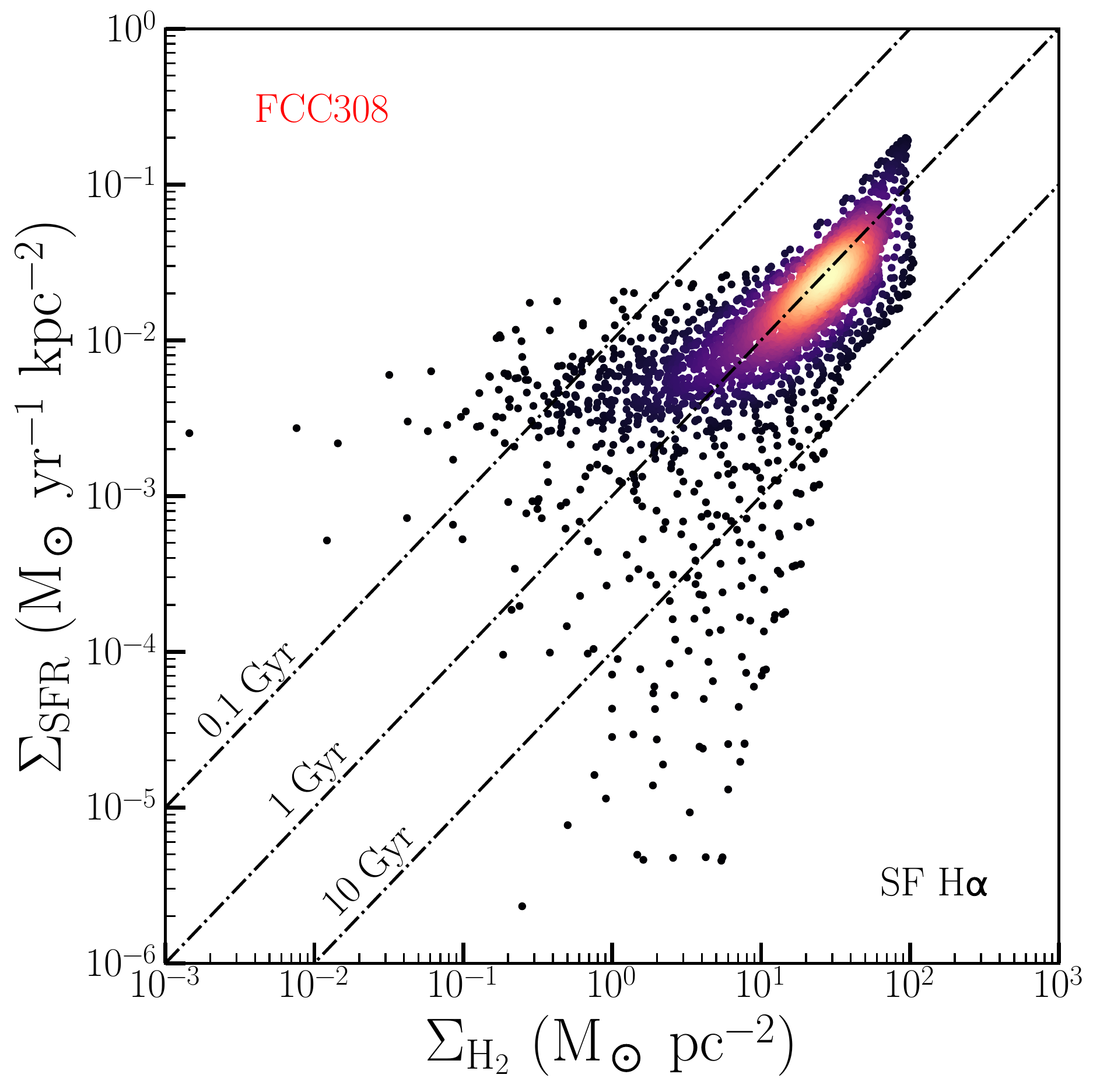}}	\hspace{13mm}
	\subfloat[\label{subfig:KS_FCC308_SF_comp}]
	{\includegraphics[height=0.42\textwidth, width=0.42\textwidth]{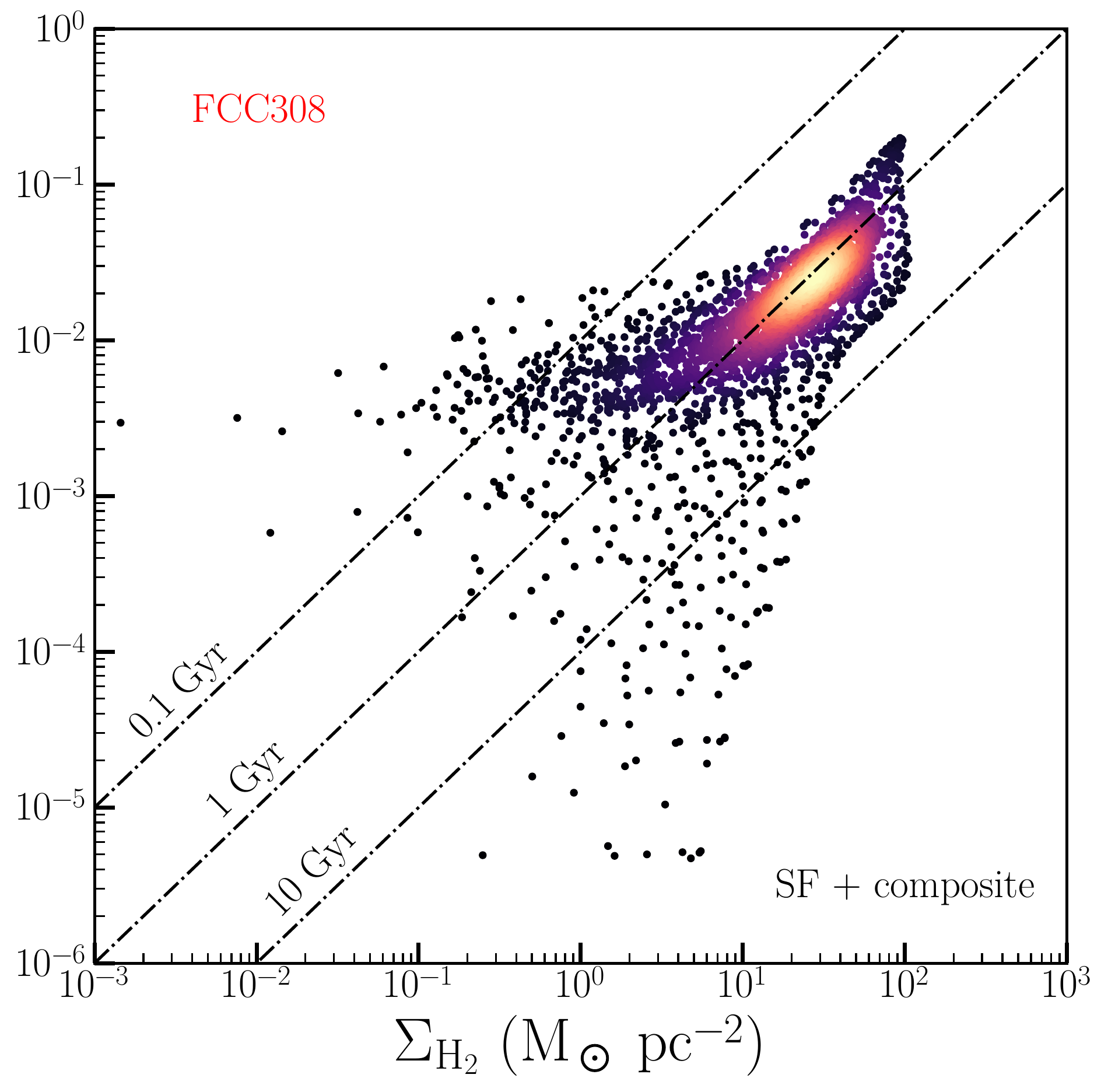}}	
	
	\subfloat[\label{subfig:DT_FCC308}]
	{\includegraphics[height=0.42\textwidth, width=0.47\textwidth]{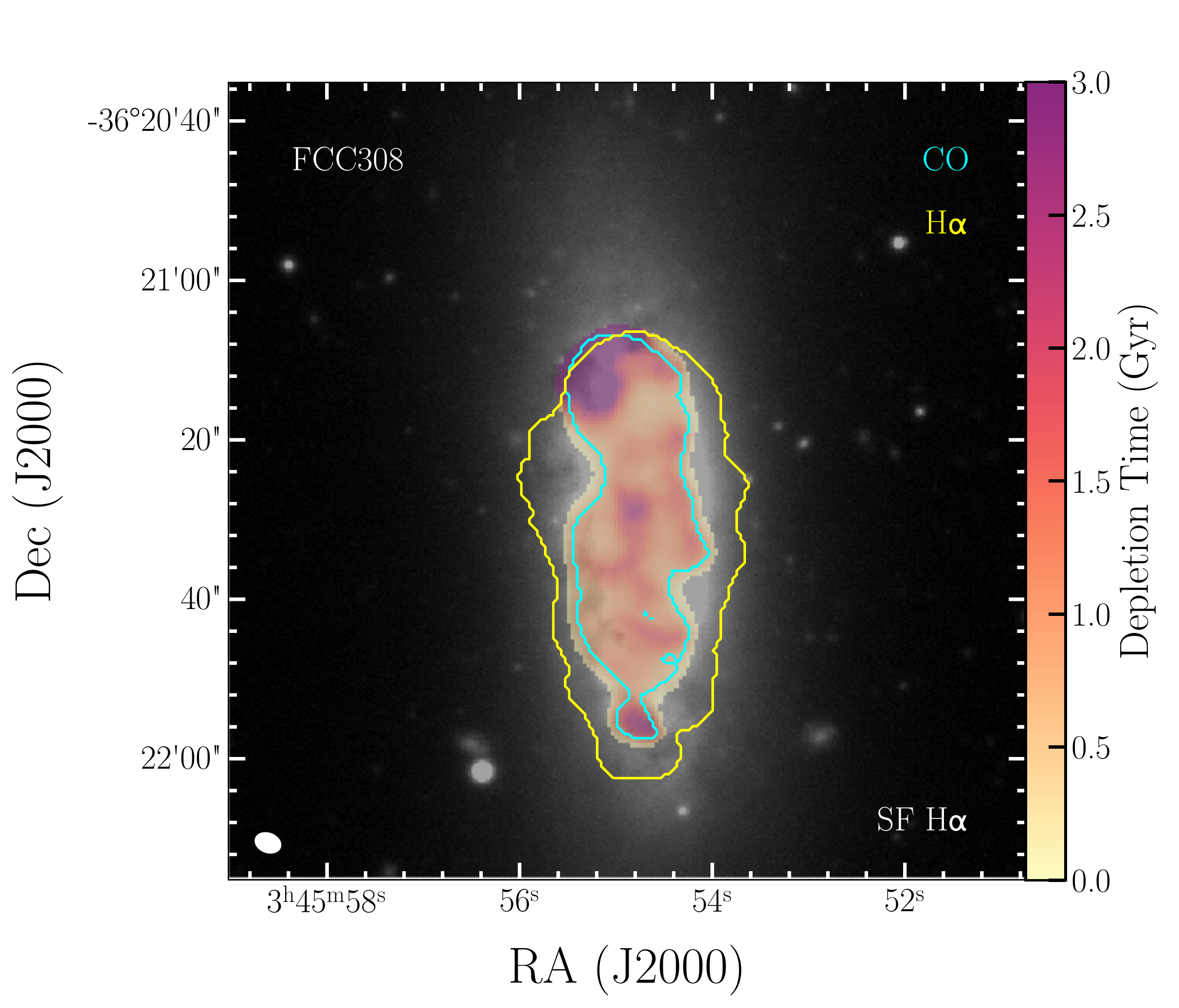}} \hspace{6mm}
	\subfloat[\label{subfig:DT_FCC308_SF_comp}]
	{\includegraphics[height=0.42\textwidth, width=0.47\textwidth]{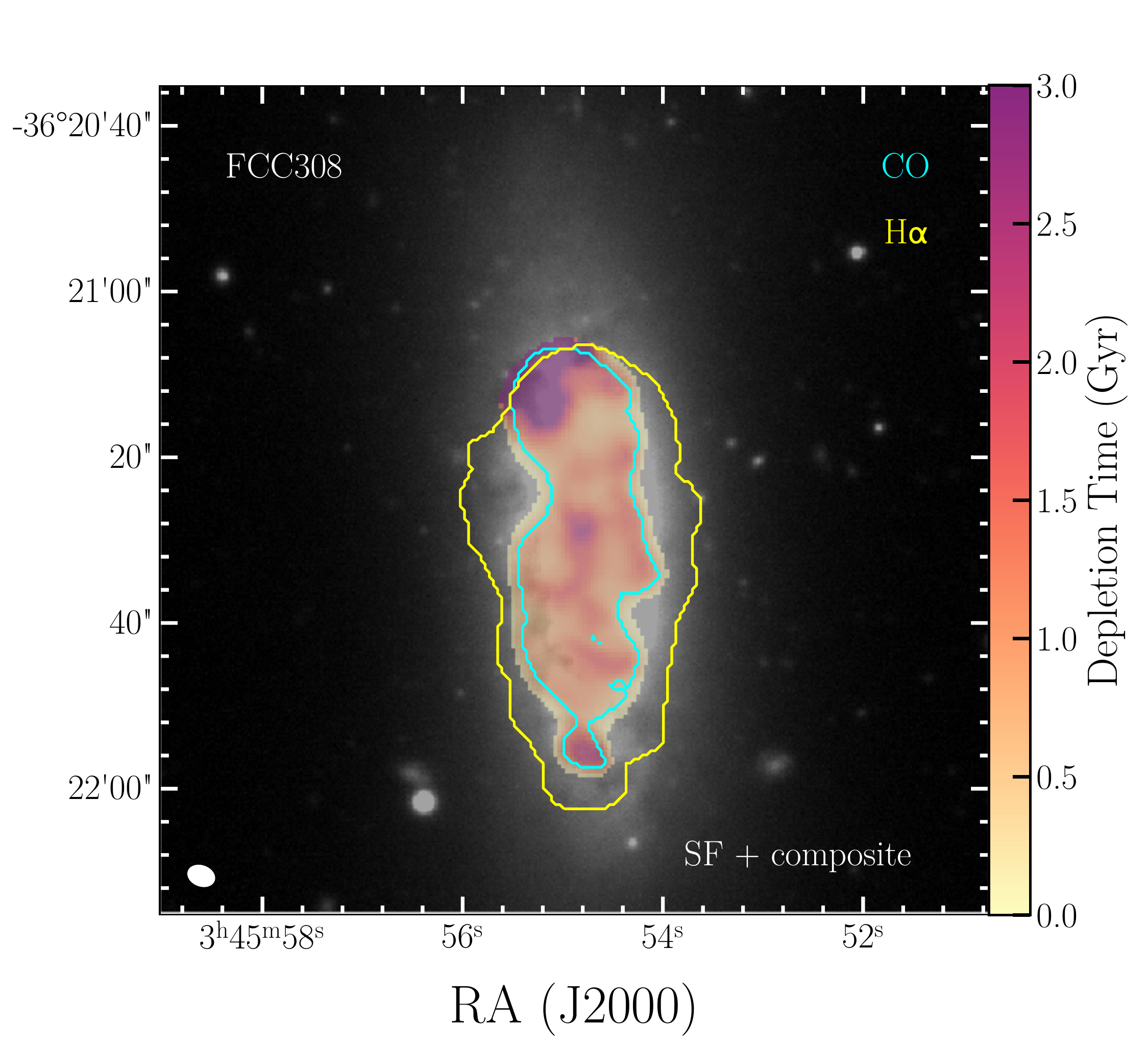}}			
	
	\caption{Similar to Figure \ref{fig:FCC90}, but for FCC308. Depletion times are relatively short, and vary slightly throughout the galaxy. There is a small patch in the north-west of the galaxy where CO is detected, but not \Ha.}
	\label{fig:FCC308}

\end{figure*}

FCC308 (NGC1437B) is a very blue galaxy with a disturbed molecular gas reservoir. The ionised gas is distributed somewhat more widely than the molecular gas, but generally follows the same distribution. As the \Ha\ that is not dominated by star formation does not overlap with any molecular gas, there is very little difference between the left and right columns in Figure \ref{fig:FCC308}. As \change{could be} expected from its colour, depletion times are short in this galaxy, well below 1 Gyr. There is some scatter in the \KS\ relation towards longer depletion times, mostly resulting from the north-west of the galaxy where the edges of the detected CO and \Ha\ emission are very close together. Throughout the rest of the galaxy there \change{also} is some variation in depletion times. 

\subsection{FCC312}
\label{sub:FCC312}
FCC312 (ESO358-G063) is a large spiral galaxy located around the virial radius of the cluster (in projection). There is some scatter in the depletion times due to small scale variations clearly visible in the maps. Depletion times are on the short side (mostly slightly shorter than 1 Gyr, but much shorter in some locations). Most of the \Ha\ is star forming and extends more widely than the molecular gas, resulting in very similar \KS\ relations \change{in the cases where composite \Ha\ is taken into account versus the star formation dominated \Ha\ only}. The exception is a slight ``chip'' in the north-west of the galaxy (\change{indicated with a blue/white arrow}); this is likely due mosaic effects in the MUSE data. FCC312 is possibly undergoing a slight starburst as it is entering the cluster.

\begin{figure*}

	\centering

	\hspace{-0mm}  \subfloat[\label{subfig:KS_FCC312}]
	{\includegraphics[height=0.42\textwidth, width=0.44\textwidth]{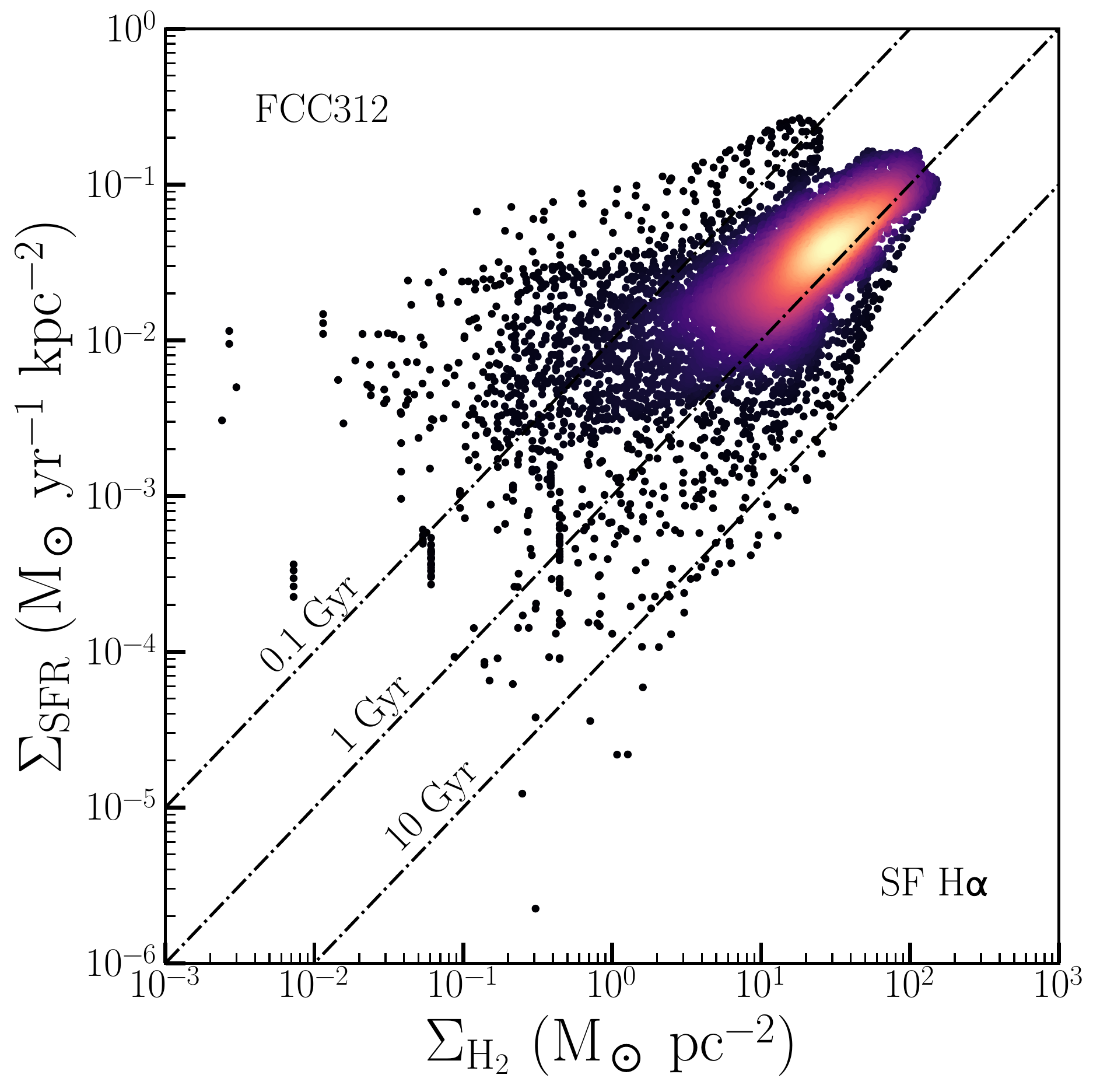}}	\hspace{13mm}
	\subfloat[\label{subfig:KS_FCC312_SF_comp}]
	{\includegraphics[height=0.42\textwidth, width=0.42\textwidth]{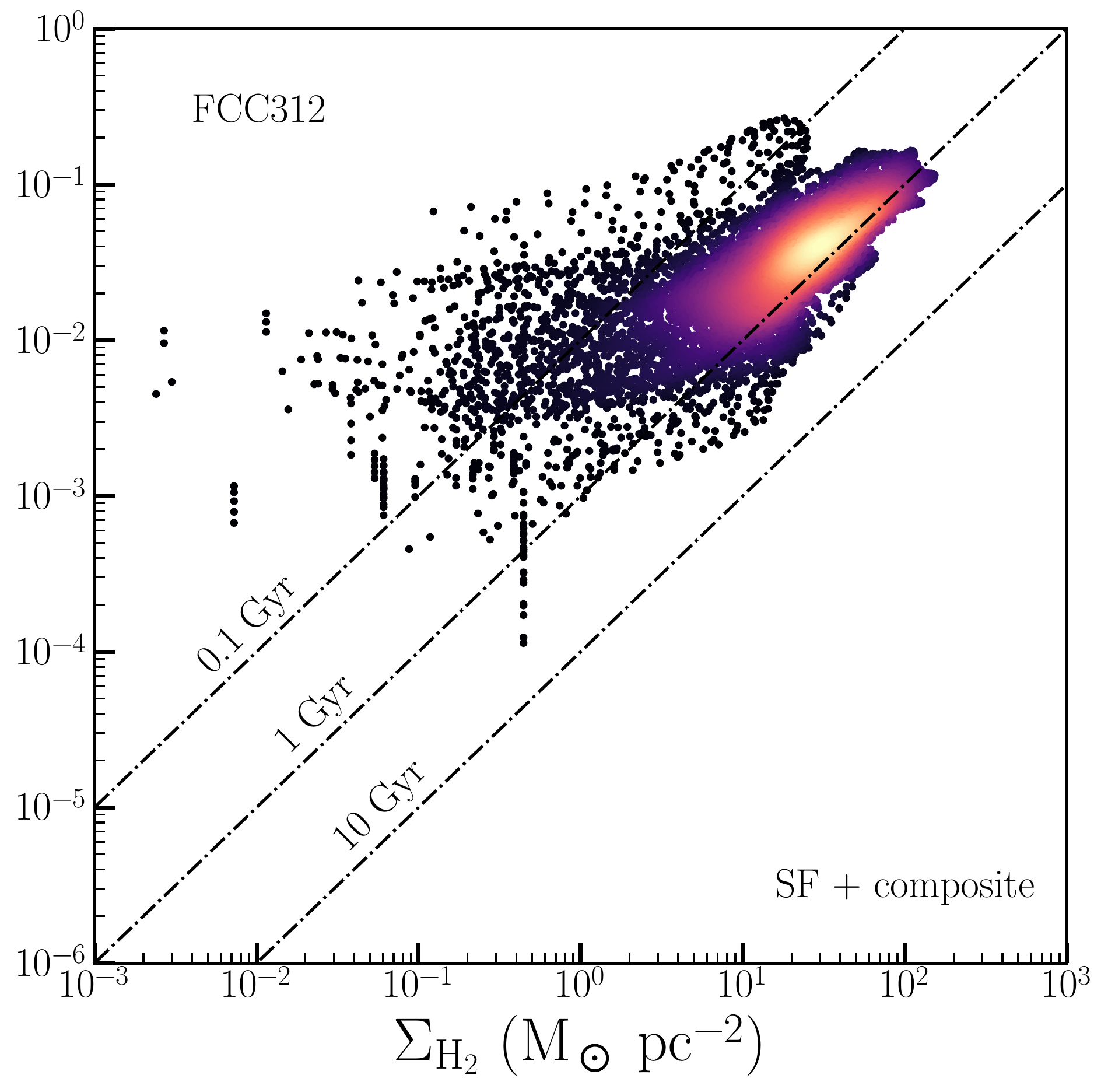}}	
	
	\subfloat[\label{subfig:DT_FCC312}]
	{\includegraphics[height=0.42\textwidth, width=0.47\textwidth]{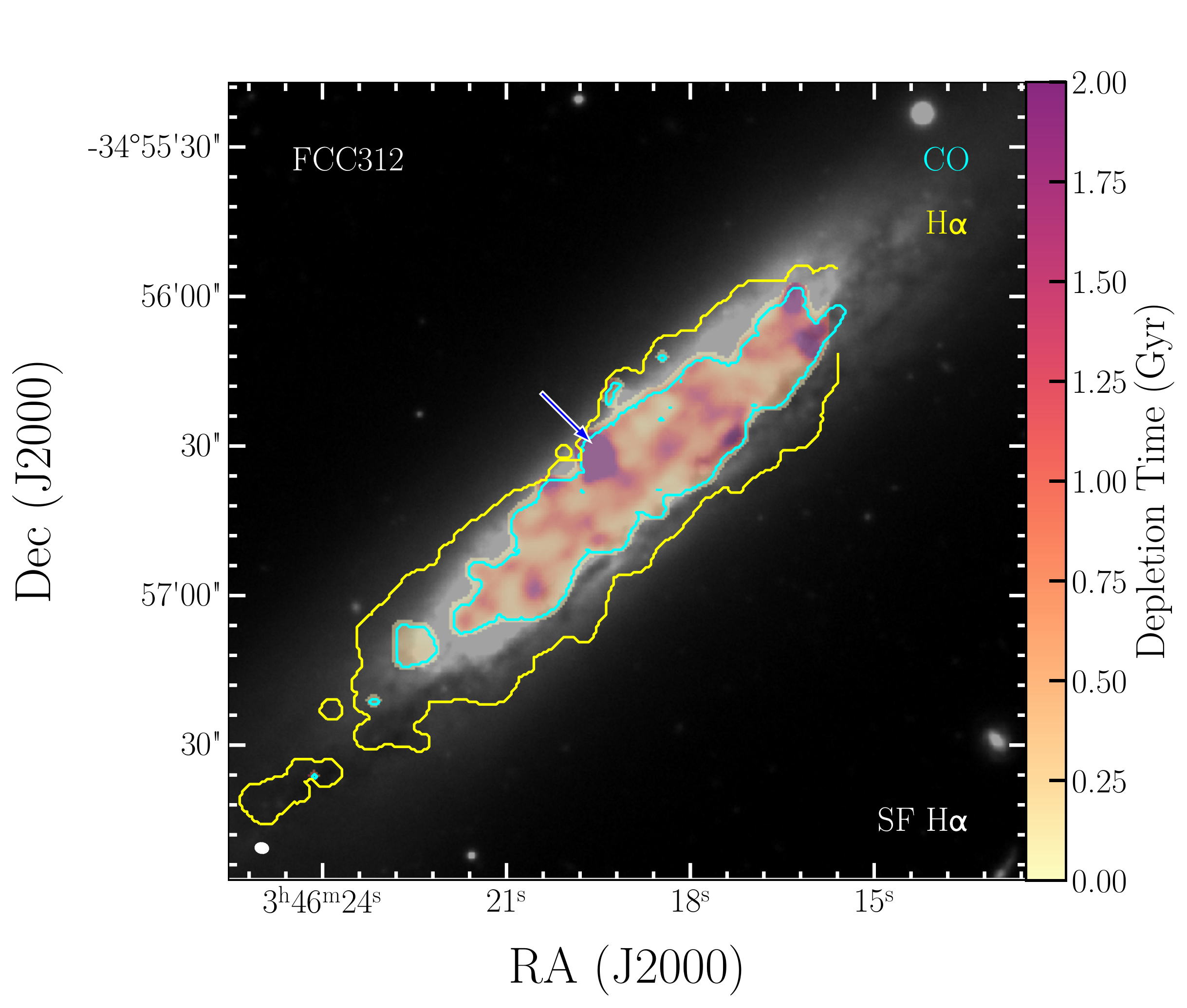}} \hspace{6mm}
	\subfloat[\label{subfig:DT_FCC312_SF_comp}]
	{\includegraphics[height=0.42\textwidth, width=0.47\textwidth]{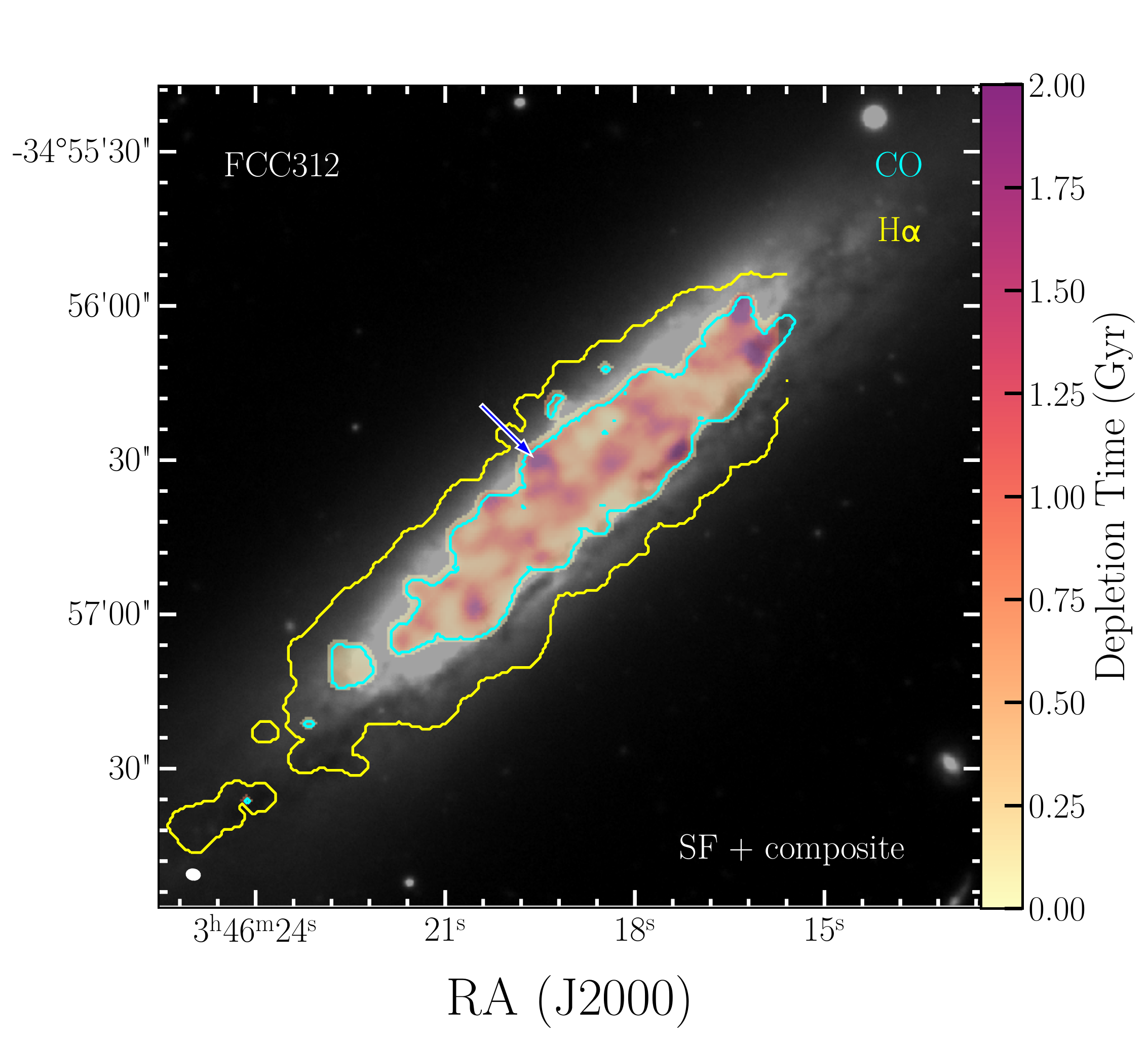}}	
	
	\caption{Similar to Figure \ref{fig:FCC90}, but for FCC312. \change{Depletion times in this galaxy are relatively short: most resolution elements have depletion times shorter than 1 Gyr. The long depletion time area in the lower-left panel, in the centre of the top half of the galaxy (indicated with a blue/white arrow), is due to mosaicking effects in the MUSE data\refrep{: a small fraction of the galaxy was not covered by the mosaic, resulting in an overestimation of the depletion time here.}}}
	\label{fig:FCC312}

\end{figure*}


\bsp	
\label{lastpage}
\end{document}